\documentclass[11pt]{article}
\usepackage[utf8]{inputenc}
\usepackage{verbatim,amsmath,amssymb}
\usepackage{epsfig,float,color}
\usepackage[table,x11names,dvipsnames]{xcolor}
\usepackage{geometry}
\usepackage{enumerate}
\usepackage{enumitem}
\usepackage{graphicx}
\usepackage{epstopdf}
\usepackage{setspace}
\usepackage{fancyhdr}
\usepackage{booktabs}
\usepackage{subcaption}
\usepackage[font=small]{caption} 
\usepackage{longtable}
\usepackage{tabularx}
\usepackage{tabulary}
\usepackage{natbib}
\usepackage{amsmath,amssymb,amsthm,mathrsfs,amsfonts,dsfont} 
\usepackage{hyperref}
\usepackage{framed}
\graphicspath{figs/}

\definecolor{darkblue}{rgb}{0,0,1}
\definecolor{col2}{rgb}{0,0,0}
\definecolor{col1}{rgb}{0,0,0}
\hypersetup{colorlinks=true, breaklinks=true, linkcolor=darkblue, menucolor=darkblue, citecolor=darkblue, urlcolor=darkblue}

\newcommand{\bitm}{\begin{itemize}}
\newcommand{\eitm}{\end{itemize}}
\newcommand{\bnumr}{\begin{enumerate}}
\newcommand{\enumr}{\end{enumerate}}

\newcommand{\mrT}{\mathrm{T}}

\newcommand {\eqb}[1]{\begin{equation}\begin{array}{#1}}
\newcommand {\eqe}{\end{array}\end{equation}}

\newcommand {\esb}[1]{\begin{equation*}\begin{array}{#1}}
\newcommand {\ese}{\end{array}\end{equation*}}
\newcommand {\ds}{\displaystyle}

\newcommand {\pa}[2]{\frac{\partial{#1}}{\partial{#2}}}

\newcommand {\dif}{\mathrm{d}}

\newcommand {\II}{{I\kern-.3em I}}
\newcommand {\III}{{I\kern-.3em I\kern-.3em I}}

\newcommand {\mrb}{\mathrm{b}}

\newcommand {\mrd}{\mathrm{d}}

\newcommand {\mrm}{\mathrm{m}}

\newcommand {\mrr}{\mathrm{r}}

\newcommand {\mf}{\mathbf{f}}
\newcommand {\mg}{\mathbf{g}}

\newcommand {\mq}{\mathbf{q}}

\newcommand {\muu}{\mathbf{u}}

\newcommand {\mx}{\mathbf{x}}

\newcommand {\ba}{\boldsymbol{a}}

\newcommand {\be}{\boldsymbol{e}}

\newcommand {\bn}{\boldsymbol{n}}

\newcommand {\bq}{\boldsymbol{q}}

\newcommand {\bu}{\boldsymbol{u}}

\newcommand {\bx}{\boldsymbol{x}}

\newcommand {\bxi}{\mbox{\boldmath$\xi$}}

\newcommand {\mB}{\mathbf{B}}
\newcommand {\mC}{\mathbf{C}}

\newcommand {\mH}{\mathbf{H}}

\newcommand {\mJ}{\mathbf{J}}
\newcommand {\mK}{\mathbf{K}}

\newcommand {\mN}{\mathbf{N}}

\newcommand {\mR}{\mathbf{R}}
\newcommand {\mS}{\mathbf{S}}

\newcommand {\mU}{\mathbf{U}}

\newcommand {\mX}{\mathbf{X}}

\newcommand {\bX}{\boldsymbol{X}}

\newcommand {\eps}{\varepsilon}

\newcommand {\bsig}{\mbox{\boldmath$\sigma$}}

\newcommand {\IR}{{\rm\kern.24em
   \vrule width.02em height1.53ex depth-.05ex
   \kern-.3em R}}
\newcommand {\ic}{{\rm\kern.20em
   \vrule width.02em height1.0ex depth-.05ex
   \kern-.22em c}}
\newcommand {\ia}{{\rm\kern.20em
   \vrule width.02em height1.05ex depth-.0ex
   \kern-.25em a}}
\newcommand {\IC}{{\rm\kern.24em
   \vrule width.02em height1.4ex depth-.05ex
   \kern-.26em C}}
\newcommand {\ID}{{\rm\kern.34em
   \vrule width.02em height1.5ex depth-.05ex
   \kern-.36em D}}
\newcommand {\IS}{{\rm\kern.24em
   \vrule width.02em height1.6ex depth.05ex
   \kern-.26em S}}
\newcommand {\IT}{{\rm\kern.50em
   \vrule width.02em height1.55ex depth-.05ex
   \kern-.52em T}}

\newcommand {\IE}{{\rm\kern.24em
   \vrule width.02em height1.55ex depth-.05ex
   \kern-.33em E}}
\newcommand {\IEa}{{\rm\kern.24em
   \vrule width.02em height1.55ex depth-.05ex
   \kern-.33em E}^{1}_{ijkl}}
\newcommand {\IEb}{{\rm\kern.24em
   \vrule width.02em height1.55ex depth-.05ex
   \kern-.33em E}^{2}_{ijkl}}

\newcommand {\sN}{\mathcal{N}}

\newcommand {\sP}{\mathcal{P}}

\newcommand {\sS}{\mathcal{S}}

\newcommand {\sV}{\mathcal{V}}

\newcommand {\Ass}[2]{\kern 0.9ex \vrule width0.45em height0.2ex depth0ex \kern -2.1ex \bigwedge_{#1}^{#2}}
\newcommand {\ASS}[2]{\kern 1.45ex \vrule width0.5em height0.2ex depth0ex \kern -2.65ex \bigwedge_{#1}^{#2}}

\begin{document}
%
%
%
%
%
%
%
%
%-------------------------------------- ABSTRACT --------------------------------------
\begin{center}
\Large{\bf{Nonlinear material identification of heterogeneous isogeometric Kirchhoff-Love shells}} \\[7mm]
\end{center}
\begin{center}
\renewcommand{\thefootnote}{\fnsymbol{footnote}}
\large{Bartosz Borzeszkowski$^{\,\mathrm{a}}$, Izabela Lubowiecka$^{\,\mathrm{a}}$, Roger A. Sauer$^{\,\mathrm{a,b,c,}}$\footnote{corresponding author, email: sauer@aices.rwth-aachen.de}}\\\vspace{4mm}
\small{$^\mathrm{a}$\textit{Gda\'nsk University of Technology, Faculty of Civil and Environmental Engineering,\\ul. Narutowicza 11/12, 80-233 Gda\'nsk, Poland}}\\\vspace{2mm}
\small{$^\mathrm{b}$\textit{Aachen Institute for Advanced Study in Computational Engineering Science (AICES), RWTH Aachen
University, Templergraben 55, 52056 Aachen, Germany}}\\\vspace{2mm}
\small{$^\mathrm{c}$\textit{Department of Mechanical Engineering, Indian Institute of Technology, Kanpur, UP 208016, India}}\\\vspace{4.5mm}
\small{Published\footnote{This pdf is the personal version of an article whose journal version is available at \href{https://doi.org/10.1016/j.cma.2021.114442}{https:/\!/sciencedirect.com}} 
in \textit{Comput.~Methods Appl.~Mech.~Eng.}, \href{https://doi.org/10.1016/j.cma.2021.114442}{DOI: 10.1016/j.cma.2021.114442} \\
Submitted on 11 August 2021; Revised on 11 November 2021; Accepted on 5 December 2021} 
\end{center}
\vspace{-4.5mm}
\setcounter{footnote}{0}
\renewcommand{\thefootnote}{\arabic{footnote}}
\rule{\linewidth}{.15mm}
{\bf Abstract:}
This work presents a Finite Element Model Updating inverse methodology for reconstructing heterogeneous material distributions based on an efficient isogeometric shell formulation. 
It uses nonlinear hyperelastic material models suitable for describing incompressible material behavior as well as initially curved shells. The material distribution is discretized by bilinear elements such that the nodal values are the design variables to be identified. 
Independent FE analysis and material discretization, as well as flexible incorporation of experimental data, offer high robustness and control. 
Three elementary test cases and one application example, which exhibit large deformations and different challenges, are considered: uniaxial tension, pure bending, sheet inflation, and abdominal wall pressurization. 
Experiment-like results are generated from high-resolution simulations with the subsequent addition of up to 4\% noise. Local optimization based on the trust-region approach is used. 
The results show that with a sufficient number of experimental measurements, design variables and analysis elements, the algorithm is capable to reconstruct material distributions with high precision even in the presence of large noise. 
The proposed formulation is very general, facilitating its extension to other material models, optimization algorithms and meshing approaches. 
Adapted material discretizations allow for an efficient and accurate reconstruction of material discontinuities by avoiding overfitting due to superfluous design variables. 
For increased computational efficiency, the analytical sensitivities and Jacobians are provided. 


\vspace{-5mm}
\rule{\linewidth}{.15mm}
\vspace{-8mm}
%
%
%-------------------------------------- SYMBOLS --------------------------------------
\section*{List of important symbols}
\vspace{-5mm}
\begin{center}
\begin{longtable}{ll}
\addtocounter{table}{-1}
$\boldsymbol{a}_\alpha$ & \quad  covariant tangent vectors of surface $\mathcal{S}$ at point $\boldsymbol{x}$; $\alpha=1,2$\\
$\boldsymbol{A}_\alpha$ &\quad covariant tangent vectors of surface $\mathcal{S}_0$ at point $\boldsymbol{X}$; $\alpha=1,2$\\
$\boldsymbol{a}^\alpha$ &\quad  contravariant tangent vectors of surface $\mathcal{S}$ at point $\boldsymbol{x}$; $\alpha=1,2$\\
$\boldsymbol{A}^\alpha$ &\quad  contravariant tangent vectors of surface $\mathcal{S}_0$ at point $\boldsymbol{X}$; $\alpha=1,2$\\
$\boldsymbol{a}_{\alpha,\beta}$ & \quad parametric derivative of $\boldsymbol{a}_{\alpha}$ w.r.t. $\xi^\beta$\\
$a_{\alpha\beta}$ &\quad  covariant metric components of surface $\mathcal{S}$ at point $\boldsymbol{x}$\\
$A_{\alpha\beta}$ &\quad  covariant metric components of surface $\mathcal{S}_0$ at point $\boldsymbol{X}$\\
$\mB^e$ & \quad matrix of the coefficients of the Bernstein polynomials for element $\Omega^e$\\
$b_{\alpha\beta}$ &\quad  covariant curvature tensor components of surface $\mathcal{S}$ at point $\boldsymbol{x}$\\
$B_{\alpha\beta}$ &\quad covariant curvature tensor components of surface $\mathcal{S}_0$ at point $\boldsymbol{X}$\\
$c$ & \quad bending stiffness\\
$\mC^e_\alpha$ &\quad B\'ezier extraction operator for element $\Omega^e$ in direction $\xi_\alpha$\\
$\Gamma_{\alpha\beta}^\gamma$ & \quad Christoffel symbols of the second kind of surface $\sS$\\
$\mathrm{d}a$ &\quad differential area element on $\mathcal{S}$\\
$\mathrm{d}A$ &\quad differential area element on $\mathcal{S}_0$\\
$\delta\,...$ & \quad variation of\,... \\
$d$ & \quad number of displacement dofs per analysis node ($=3$ in 3D)\\
$\bar d$ & \quad number of material dofs per material node ($=1$ in Secs.~\ref{sec:uni} \& \ref{sec:purebending}, otherwise $=2$)\\
$\delta_{\mathrm{max}}$ & \quad maximum of the relative error between reference and estimated parameters\\
$\delta_{\mathrm{ave}}$ & \quad average of the relative error between reference and estimated parameters\\
$e$ &\quad index numbering of elements\\
$E$ & \quad Young's modulus\\
$\eps_{\alpha\beta}$ & \quad covariant components of the membrane strain tensor\\
$e^\alpha$ & \quad offset between $\Omega_0^e$ and $\bar{\Omega}_0^e$ coordinate centers\\
$...^{\mathrm{exp}}$  &\quad  corresponding quantity of the experimental grid; e.g. $\boldsymbol{x}_I^{\mathrm{exp}}$\\
$f$ &\quad objective function\\
$\boldsymbol{f}$ &\quad  `body' force acting on $\mathcal{S}$\\
$\boldsymbol{F}$ &\quad  surface deformation gradient\\
$\boldsymbol{f}_{\!0}$ &\quad constant surface force (due to dead loading)\\
$\textbf{f}^{\,e}_\bullet$ &\quad  finite element force vector of element $\Omega^e$\\
$\textbf{g}$ &\quad gradient of the objective function $f$\\
$G$ &\quad expression for the weak form\\
$G^e$ &\quad  contribution to $G$ from finite element $\Omega^e$\\
$G_{\mathrm{ext}}$ &\quad external virtual work\\
$G_{\mathrm{int}}$ &\quad internal virtual work\\
$...^h$ &\quad ... approximated by finite elements\\
$\textbf{H}$ &\quad Hessian of the objective function $f$\\
$I$ &\quad  index numbering of finite \textcolor{col2}{and material} element nodes, and experimental points\\
$J$ &\quad  area change between $\mathcal{S}_0$ and $\mathcal{S}$\\
$\textbf{J}$ &\quad Jacobian of the residual $\bar{\textbf{U}}_\mrr$\\
\textcolor{col1}{$\kappa_{\alpha\beta}$} &\quad  \textcolor{col1}{covariant components of the relative curvature tensor}\\
\textcolor{col2}{$\textbf{K}$} &\quad  \textcolor{col2}{global finite element tangent matrix}\\
$\textbf{k}^e$ &\quad \textcolor{col2}{elemental} finite element tangent matrix associated with $\textbf{f}^{\,e}$\\
\textcolor{col1}{$\Lambda$} &\quad \textcolor{col1}{2D Lam{\'e} parameter}\\
\textcolor{col1}{$\Tilde{\Lambda}$} &\quad \textcolor{col1}{3D Lam{\'e} parameter}\\
$m_\tau,m_\nu$ & \quad bending moment components acting at $\bx \in \partial\sS$\\
$M^{\alpha\beta}$ & \quad contravariant bending moment components\\
$M_0^{\alpha\beta}$ & \quad $=JM^{\alpha\beta}$\\
$\mu$ &\quad \textcolor{col1}{2D surface shear modulus}\\
\textcolor{col1}{$\Tilde{\mu}$} &\quad \textcolor{col1}{3D shear modulus}\\
$\boldsymbol{n}$ &\quad surface normal of $\mathcal{S}$ at $\boldsymbol{x}$\\
$\boldsymbol{N}$ &\quad surface normal of $\mathcal{S}_0$ at $\boldsymbol{X}$\\
$\mathcal{N}$ &\quad trust region\\
$\textbf{N}$ &\quad array of the shape functions for element $\Omega^e$\\
$\bar{\textbf{N}}$ &\quad array of the shape functions for element $\bar{\Omega}^e$\\
$N_I$ &\quad displacement shape function of finite element node $I$\\
$\bar{N}_I$ &\quad shape function of material element node $I$\\
$n_\mathrm{el}$ &\quad total number of FE used to discretize $\mathcal{S}$\\
$n_\mathrm{ll}$ &\quad number of load levels considered in the experiment\\
$\bar{n}_\mathrm{el}$ &\quad total number of material elements used to discretize $q$\\
$n_\mathrm{exp}$ & \quad number of sampled experimental points for all $n_\mathrm{ll}$ load levels\\
$n_\mathrm{no}$ & \quad  number of FE \textcolor{col2}{analysis} nodes\\
$\bar{n}_\mathrm{no}$  & \quad number of material nodes\\
$n_e$ & \quad number of nodes of $\Omega^e$\\
$\bar{n}_e$ &\quad number of nodes of $\bar{\Omega}^e$\\
\textcolor{col2}{$n_{\mathrm{var}}$} &\quad \textcolor{col2}{number of design variables}\\
\textcolor{col1}{$\nu$} &\quad \textcolor{col1}{Poisson's ratio}\\
$\boldsymbol{\nu}$ &\quad normal vector on boundary $\partial\sS$\\
$p$ &\quad external pressure normal to $\sS$\\
$\mathcal{P}$ &\quad parametric domain spanned by $\xi^1$ and $\xi^2$\\
$q$ & \quad material parameter field\\
$\textbf{q}$  & \quad stacked array of all nodal $\textbf{q}^e$ in the system\\
$\textbf{q}^e$  & \quad stacked array of all nodal $q_I$ of material element $\bar{\Omega}^e$\\
$q_I$ & \quad material parameter value at material node $I$\\
$\textbf{q}_\mathrm{opt}$  & \quad solution of the inverse problem\\
$\textbf{q}_0$  & \quad initial estimate for the optimization\\
$\mR$ &\quad vector of reaction forces\\
$\mS$ &\quad sensitivity matrix\\
$\mathcal{S}$ &\quad current configuration of the surface\\
$\mathcal{S}_0$ &\quad initial configuration of the surface\\
$\textbf{s}_k$ &\quad trial step at \textcolor{col2}{inverse} iteration $k$\\
$\boldsymbol{\sigma}$ &\quad surface Cauchy stress tensor of $\mathcal{S}$ at point $\boldsymbol{x}$\\
$\sigma^{\alpha\beta}$ &\quad contravariant in-plane stress components\\
$\boldsymbol{t}^\alpha$ &\quad traction vector acting on the surface $\bot\ \boldsymbol{a}^\alpha$\\
$\boldsymbol{t}$ &\quad effective traction acting on boundary $\partial\sS$\\
$T$ &\quad reference thickness of $\mathcal{S}$\\
$\boldsymbol{\tau}$ &\quad surface Kirchhoff stress tensor of $\sS$ at $\boldsymbol{x}$\\
$\tau^{\alpha\beta}$ &\quad $=J\sigma^{\alpha\beta}$ Kirchhoff membrane stress components\\
$\boldsymbol{\varphi}$ &\quad deformation map of surface $\mathcal{S}$\\
$\boldsymbol{u}$ & \quad displacement field of $\sS$ at $\boldsymbol{x}$\\
$\boldsymbol{\overline{u}}$ &\quad prescribed boundary displacements on Dirichlet boundary $\partial_u\mathcal{S}$\\
$\textbf{u}_\bullet^\bullet $ &\quad discrete displacements; $\bullet$ takes the same options as in $\textbf{x}_\bullet^\bullet$\\
$\textbf{U}_{\mathrm{exp}}$ &\quad stacked vector of $n_\mathrm{exp}$ experimentally measured displacements $\boldsymbol{u}_I$ at surface point $\boldsymbol{x}_I$\\
$\textbf{U}_{\rm FE}$ &\quad vector of FE surface displacements $\boldsymbol{u}^h$ interpolated at all $n_{\rm exp}$ experimental points $\boldsymbol{x}_I^{\rm exp}$\\
$\bar{\textbf{U}}_\mrr$ &\quad residual vector\\
$W$ &\quad 2D hyperelastic stored surface energy density\\
$\mathcal{V}$ &\quad space of admissible functions\\
$\boldsymbol{x}$ & \quad current position on surface \textcolor{col2}{$\sS$}\\
$\boldsymbol{X}$ &\quad initial position on surface $\sS_0$\\
$\textbf{x}_I$ & \quad current position of FE node (control point); may not lie on $\sS^h$\\
$\textbf{X}_I$ & \quad initial position of FE node (control point); may not lie on $\sS^h_0$\\
$\boldsymbol{x}_I$ &\quad $= \sum_J N_J(\bxi_I) \mx_J$; closest surface point of $\mx_I$\\
$\boldsymbol{X}_I$ &\quad $= \sum_J N_J(\bxi_I) \mX_J$; closest surface point of $\mX_I$\\
$\textbf{x}^e$ &\quad stacked array of all nodal $\textbf{x}_I$ of finite element $\Omega^e$\\
$\textbf{X}^e$ &\quad stacked array of all nodal $\textbf{X}_I$ of finite element $\Omega^e$\\
$\textbf{x}$ &\quad stacked array of all nodal $\textbf{x}_I$ in the system\\
$\textbf{X}$ &\quad stacked array of all nodal $\textbf{X}_I$ in the system\\
$\xi^{\alpha}$ &\quad convective surface coordinates; $\alpha=1,2$\\
$\boldsymbol{\xi}$ &\quad $=[\xi^1,\xi^2]$\\
$\Omega^e$ &\quad finite element $e$ in the current configuration\\
$\Omega^e_0$ &\quad finite element $e$ in the initial configuration\\
$\Omega^e_\square$ &\quad finite element $e$ in the parametric domain $\mathcal{P}$\\
$\bar{...}$  & \quad corresponding quantity of the material reconstruction mesh; e.g. $\bar\Omega^e$, $\bar x_I$, $\bar{\textbf{x}}^e$
\end{longtable}
\end{center}
%
%
%
%
%
%
%-------------------------------------- INTRODUCTION --------------------------------------
\vspace*{-4em}
\section{Introduction}\label{s:intro}
Material modeling and design are rapidly advancing in many fields of engineering. This creates a demand for the a priori knowledge of material properties. Many modern materials, including textiles, concrete, composites and biological materials, are characterized by heterogeneity. This local change of the material properties can result from microstructure, imperfections, damage, or manufacturing processes. The characterization of heterogeneous materials is often inaccessible through standard testing methods \citep{wineman1979material,pierron2021towards}. This is a particular difficulty in soft biological materials, since they present a unique set of challenges to experimental material identification procedures, such as their small size and delicate
structure, the restricted access to representative samples, the difficult mounting of specimens, the provision of physiological conditions, as well as the obtainment of formal, ethical and medical consent \citep{evans2017can}. Therefore, nondestructive \textit{in vivo} experiments, accompanied by numerical models and \textit{in silico} inverse identification offer a path to overcome these challenges. Seminal work in this direction has been done on skin tissue \citep{vossen1994mixed}, heart muscles \citep{moulton1995inverse}, aortic aneurysms \citep{raghavan2000toward}, tympanic membranes \citep{aernouts2011elastic}, and abdominal walls \citep{simon2017towards}.\par
Various inverse identification methods have been developed in the past in conjunction with digital image correlation systems (DIC) that provide full-field data, e.g. see \cite{avril2008overview}. The first class of inverse identification methods are \textit{direct inverse} methods. They are based on the principle that material properties can be expressed explicitly in terms of the strain and stress components. Due to their high efficiency, these methods are favored in patient-specific rupture risk assessment, especially for vascular disorders, which are characterized by localized changes in wall composition and structure \citep{bersi2016novel}. An example for a direct inverse method, is the Pointwise Identification Method \citep{zhao2009pointwise}. It assumes that the local material properties follow directly from the local stress-strain data. It has been used to identify heterogeneous, anisotropic properties of planar soft tissues \citep{zhao2011identifying,genovese2014digital,davis2015pointwise}. Another direct inverse method is Local Extensional Stiffness Identification (LESI) proposed by \cite{farzaneh2019inverse,farzaneh2019identifying}. It is based on the local membrane equilibrium equations, and has been applied to the identification of regional elastic properties. The Virtual Field Method (VFM) proposed by \cite{pierron2012virtual}, a popular direct inverse method based on the principle of virtual work, has been developed and used by many researchers for the identification of material parameters in homogeneous and heterogeneous linear elasticity \citep{avril2004sensitivity,avril2007general}, hyperelasticity \citep{avril2010anisotropic,bersi2016novel,marek2017sensitivity}, plasticity \citep{pierron2010extension,martins2018comparison} and incompressible elasticity \citep{mei2019improving}.\par
The second class of inverse identification methods are \textit{iterative inverse} methods, such as the Finite Element Model Updating method (FEMU) \citep{kavanagh1971finite}, which minimizes the discrepancy between experimental measurements and finite element model predictions in a global least-squares sense. It is a well established technique associated with high robustness and low sensitivity to measurement noise, capable to model complex mechanical tests and structures \citep{goenezen2012linear}. However, it requires a priori knowledge of boundary conditions and, due to its iterative character, can be computationally expensive. FEMU has been used for the identification of homogeneous hyperelastic solids \citep{iding1974identification}, membranes \citep{kyriacou1997inverse}, nonlinear viscoelastic continua \citep{kauer2002inverse}, linear elastic continua \citep{oberai2003solution}, as well as various anisotropic hyperelasticity models \citep{genovese2006mechanical,bischoff2009quantifying,badel2012mechanical,wittek2013vivo}. Although FEMU was successfully used to identify homogeneous material models, to the best of our knowledge only few attempts have been made to deal with the heterogeneous distribution of material parameters. \cite{seshaiyer2003sub} assumed homogeneity in sub-domains to identify material parameters based on the Neo-Hookean, Mooney-Rivlin and Fung models. A similar approach was used by \cite{khalil2006combined} to identify the elastic properties of vascular tissues. \cite{kroon2008estimation,kroon2009elastic} and \cite{kroon2010efficient} applied FEMU to determine element-wise constant material distributions of anisotropic nonlinear membranes. This formulation was extended by \cite{kroon2010numerical} to more general material distributions. The work by Kroon seems to be the most recent on FEMU for heterogeneous materials.\par
Many soft materials consist of thin, surface-like structures that can be efficiently described by \textit{rotation-free} shell and membrane models, especially in the context of Isogeometric analysis (IGA) \citep{hughes2005isogeometric}. Such models require no rotational degrees of freedom. Further, compared to classical finite element methods (FEM), IGA exhibits higher accuracy and robustness per degree of freedom in many areas of computational mechanics \citep{de2014isogeometric,nguyen2015isogeometric,schillinger2018isogeometric}. This is due to the fact that IGA discretizations can provide smoothness of any order across element boundaries, while no Gibbs oscillations appear in high order elements \citep{hughes2005isogeometric}. Furthermore, IGA can be integrated straightforwardly into existing FE software using the B\'ezier extraction operator \citep{borden2011isogeometric,scott2011isogeometric}. Due to its accurate yet efficient geometrical description with relatively few elements, IGA has become a particularly advantageous computational tool for shell structures. \par
A number of shell and membrane formulations using IGA have been proposed for both linear and nonlinear deformation regimes, see e.g. \cite{kiendl2009isogeometric, nguyen2011rotation, benson2011large, sauer2014computational, tepole2015isogeometric, guo2015weak, kiendl2015isogeometric}. IGA has been applied to inverse problems such as shape optimization \citep{wall2008isogeometric,manh2011isogeometric,kiendl2014isogeometric}, topology optimization \citep{seo2010isogeometric,dede2012isogeometric,wang2018structural}, load reconstruction \citep{vu2018nurbs,vu2019nurbs} and material identification \citep{dufour2015shape,do2019isogeometric}. To the best of our knowledge, \cite{tepole2015isogeometric} are the first to consider isogeometric Kirchhoff–Love shells for biological materials. In their approach, numerical integration through the shell thickness is used to obtain shell material models. On the other hand, \citet{roohbakhshan2017efficient} propose an analytical integration approach for biomaterial models to obtain direct surface models for Kirchhoff–Love shells. The application of isogeometric shell formulations to the forward and inverse simulation of biological materials can be expected to play an important role in the future. \par
In this paper, we propose a FEMU framework based on direct isogeometric shell formulations and gradient-based optimization, aimed at identifying the heterogeneous distribution of material properties. Large deformation incompressible isotropic material behavior is considered, since it is a basis for many soft material and biological tissue models. Our framework is most closely related to the approach by \cite{kroon2010numerical}, which we extend here by adding isogeometric shell formulations and FE mesh-independent heterogeneity descriptions. The latter allow for more flexibility and efficiency in the inverse analysis. In particular, the proposed use of low order Lagrange interpolation is better suited for capturing material discontinuities than the high order IGA discretization used for rotation-free shell analysis. To the best of our knowledge, such a flexible meshing approach has not been considered in IGA-based inverse analysis before. The proposed inverse method contains the following features: 
\begin{itemize}[noitemsep,topsep=0pt]
    \item Isogeometric shell FE formulation based on separate membrane and bending contributions derived from analytical thickness integration.
    \item Material discretization capable of capturing general material distributions independently from the FE analysis mesh.
    \item General Finite Element Model Updating inverse framework capable of reconstructing distributed constitutive parameters. 
    \item Analytical sensitivities and Jacobians w.r.t.~the design variables.
    \item Discussion of various error sources and strategies to reduce their influence.
    \item Systematic investigation of the influence of noise on the nonlinear material behavior.
\end{itemize}
\par
The remainder of this paper is organized as follows: In Sec.~\ref{s:shelltheory} rotation-free thin shell theory is summarized. Sec.~\ref{s:FE} presents the finite element discretization of the shell equations and the distributed material parameter field. A general framework for inverse analysis is proposed in Sec.~\ref{s:inverse}, which is followed by several numerical examples in Sec.~\ref{s:examples} to illustrate the capability of the identification protocol. The paper concludes with Sec.~\ref{s:conclusions}.
%
%
%
%
%
%
%-------------------------------------- SHELL THEORY --------------------------------------
\section{Thin shell theory}\label{s:shelltheory}
This section briefly summarizes the nonlinear theory of rotation-free Kirchhoff-Love shells in the
framework of curvilinear coordinates. The formulation admits arbitrary hyperelastic material laws with general decomposition into bending and membrane contributions. A more detailed presentation can be found in \citet{sauer2018computational}. 
\subsection{Surface description and kinematics}
The shell surface, denoted $\mathcal{S}$, is characterized by the parametric description
\begin{equation}\label{eq:covariant}
	\boldsymbol{x}=\boldsymbol{x}(\xi^{\alpha}), \quad \alpha=1, 2\ ,
\end{equation}
where $\xi^{\alpha}$ are \emph{curvilinear} coordinates associated with a 2D parameter domain $\mathcal{P}$. In the following, lower case symbols are used to denote kinematical quantities in the current configuration $\mathcal{S}$, while upper case symbols are used for the reference configuration $\mathcal{S}_0$. The tangent vectors to coordinate $\xi^{\alpha}$ at point $\boldsymbol{x} \in \mathcal{S}$ (corresponding to $\boldsymbol{X} \in \mathcal{S}_0$) are given by
\begin{equation}\label{eq:covariant2}
	\boldsymbol{a}_{\alpha}=\frac{\partial \boldsymbol{x}}{\partial \xi^{\alpha}}\,,\qquad \boldsymbol{A}_{\alpha}=\frac{\partial \boldsymbol{X}}{\partial \xi^{\alpha}}\ ,
\end{equation}
which form a basis. It is characterized by the surface metric, that has the covariant components
\begin{equation}\label{eq:metric_basis}
	a_{\alpha\beta}=\boldsymbol{a}_{\alpha}\cdot \boldsymbol{a}_{\beta}\,,\qquad A_{\alpha\beta}=\boldsymbol{A}_{\alpha}\cdot \boldsymbol{A}_{\beta} \ .
\end{equation}
Then the contravariant surface metric $[a^{\alpha\beta}]=[a_{\alpha\beta}]^{-1}$ and $[A^{\alpha\beta}]=[A_{\alpha\beta}]^{-1}$ can be evaluated, such that the contravariant base vectors are determined by\,\footnote{with summation from 1 to 2 implied over repeated Greek indices.}
\begin{equation}
	\boldsymbol{a}^{\alpha}=a^{\alpha\beta}\boldsymbol{a}_{\beta}\,,\qquad \boldsymbol{A}^{\alpha}=A^{\alpha\beta}\boldsymbol{A}_{\beta}\,.
\end{equation}
The normal unit vector to surface $\mathcal{S}$ can be obtained as
\begin{equation}
	\boldsymbol{n}=\frac{\boldsymbol{a}_{1}\times\boldsymbol{a}_{2}}{\|\boldsymbol{a}_{1}\times\boldsymbol{a}_{2}\| }\,,\qquad \boldsymbol{N}=\frac{\boldsymbol{A}_{1}\times\boldsymbol{A}_{2}}{\|\boldsymbol{A}_{1}\times\boldsymbol{A}_{2}\| }\ .
\end{equation}
Based on the second parametric derivative $\boldsymbol{a}_{\alpha,\beta}=\partial\boldsymbol{a}_{\alpha}/\partial\xi^\beta$ and $\boldsymbol{A}_{\alpha,\beta}=\partial\boldsymbol{A}_{\alpha}/\partial\xi^\beta$, the covariant curvature tensor components are given by
\begin{equation}
    b_{\alpha\beta}=\boldsymbol{a}_{\alpha,\beta}\cdot\boldsymbol{n}\,,\qquad B_{\alpha\beta}=\boldsymbol{A}_{\alpha,\beta}\cdot\boldsymbol{N}\,.
\end{equation}
The deformation map between $\mathcal{S}_0$ and $\mathcal{S}$, denoted $\boldsymbol{x} =\boldsymbol{\varphi}(\boldsymbol{X})$, is characterized by the surface deformation gradient
\begin{equation}\label{defgrad}
\boldsymbol{F}=\boldsymbol{a}_\alpha\otimes \boldsymbol{A}^\alpha
\end{equation}
and the surface stretch
\begin{equation}\label{eq:surfstre}
	J=\frac{\sqrt{\textrm{det}[a_{\alpha\beta}]}}{\sqrt{\textrm{det}[A_{\alpha\beta}]}}\ .
\end{equation}
\subsection{Weak form}
 The weak form of the shell, 
 \begin{equation}\label{eq:weakform}
    G_{\mathrm{in}}+G_{\mathrm{int}}-G_{\mathrm{ext}}=0\qquad \forall\ \delta\bx \in \mathcal{V},
\end{equation}
contains contributions from inertia as well as internal and external virtual work. 
The first is zero for quasi-static problems. 
The internal virtual work is given by
\begin{equation}\label{eq:intvirtw}
    G_{\mathrm{int}}=\int_{\mathcal{S}_0}\frac{1}{2}\delta a_{\alpha\beta}\,\tau^{\alpha\beta} \mathrm{d}A+\int_{\mathcal{S}_0}\delta b_{\alpha\beta}\,  M_{0}^{\alpha\beta}\mathrm{d}A\,,
\end{equation}
where $\delta \boldsymbol{x}\in\sV$ denotes a kinematically admissible variation of the position vector $\bx$, while $\tau^{\alpha\beta}$ are the membrane stress components and $M_{0}^{\alpha\beta}$ are the bending moment components that can also be associated with the bending stresses. 
It is emphasized that $\tau^{\alpha\beta}$ and $M_0^{\alpha\beta}$ are not tensors, only tensor components. 
The full stress and moment tensors can be constructed by multiplying by the basis vectors from Eq.~\eqref{eq:covariant2}, but this is unnecessary for evaluating weak form \eqref{eq:intvirtw}.
Both $\tau^{\alpha\beta}$ and $M_0^{\alpha\beta}$ are expressed w.r.t.~reference surface configuration $\mathcal{S}_0$. 
The stress components w.r.t.~the current configuration are $\sigma^{\alpha\beta} = \tau^{\alpha\beta}/J$ and $M^{\alpha\beta} = M_0^{\alpha\beta}/J$, where $J$ is the surface stretch from Eq.~\eqref{eq:surfstre}. 
Further, $\delta a_{\alpha\beta}$ and $\delta b_{\alpha\beta}$ are the variations of the covariant surface metric and curvature tensor components, respectively. \\
The external work is given by
\begin{equation}
    G_{\mathrm{ext}}=\int_{\mathcal{S}} \delta\bx \cdot \boldsymbol{f}\, \mathrm{d}a+\int_{\partial_{t}\mathcal{S}}\delta\bx\cdot\boldsymbol{t}\,\mathrm{d}s +\int_{\partial_{m}\mathcal{S}}\delta \bn \cdot m_{\tau}\, \boldsymbol{\nu}\ \mathrm{d}s\ ,
\end{equation}
where $\boldsymbol{f} = \boldsymbol{f}_{\!0} + p\,\bn\ $ is a prescribed body force on $\mathcal{S}$ containing the dead load $\boldsymbol{f}_{\!0}$ and the external pressure $p$. Further, $\boldsymbol{t}$, $m_{\tau}$ and $m_{\nu}$ are distributed forces and moments prescribed along the edges, and $\boldsymbol{\nu}=\nu_\alpha \boldsymbol{a}^{\alpha}$ is the normal to edge $\partial_{m}\mathcal{S}$, where the bending moment $m_{\tau}$ is applied. In the following, the inertia term $G_{\mathrm{in}}$ is neglected. Due to its nonlinear character, the weak form needs to be linearized. Details on the linearization can be found in \citet{sauer2015theoretical}.
%
%%------------------------------------------------------
%
\subsection{Constitution}\label{sec:shellconstitution}
The constitutive relations for the shell can be either obtained via projection of 3D material laws onto the two-dimensional manifold \citep{roohbakhshan2016}, or directly derived for surfaces. 
In the case of hyperelasticity the latter approach starts from a surface strain energy density function of the form
\begin{equation}
    W=W(a_{\alpha\beta},b_{\alpha\beta})=W_{\mrm}(a_{\alpha\beta})+W_{\mrb}(b_{\alpha\beta},a_{\alpha\beta})\,,
\end{equation}
where $W_{\mrm}$ is the membrane part that depends on the surface metric $a_{\alpha\beta}$, and $W_{\mrb}$ is the bending part that predominantly depends on the curvature tensor $b_{\alpha\beta}$ \citep{roohbakhshan2017efficient}. Therefore, different constitutive models can be assigned for membrane deformation and bending. 
Given the total strain energy 
\eqb{l}
\Pi_\mathrm{int} = \ds\int_{\sS_0} W\,\dif A\,,
\eqe
the internal virtual work in \eqref{eq:intvirtw} follows from the variation
\eqb{l}
G_\mathrm{int} = \delta\Pi_\mathrm{int} = \ds\int_{\sS_0} \delta W\,\dif A\,,
\eqe
since
\eqb{l}
\delta W = \ds\frac{1}{2} \tau^{\alpha\beta} \delta a_{\alpha\beta} + M_0^{\alpha\beta} \delta b_{\alpha\beta} 
\eqe
for the stress and bending moment components
\begin{equation}\label{tauforNH}
\tau^{\alpha \beta}=2\frac{\partial W}{\partial a_{\alpha\beta}}\,,\qquad M^{\alpha \beta}_0=\frac{\partial W}{\partial b_{\alpha \beta}}\,,
\end{equation}
that can be formally introduced through Cauchy's theorem \citep{sauer2015theoretical}. \\
Introducing the strains and relative curvatures
\eqb{l}
\eps_{\alpha\beta} := \frac{1}{2}(a_{\alpha\beta} - A_{\alpha\beta})\,,\qquad
\kappa_{\alpha\beta} := b_{\alpha\beta} - B_{\alpha\beta}\,,
\eqe
one can also consider strain energy functions in the form $W = W(\eps_{\alpha\beta},\kappa_{\alpha\beta})$ and use
\eqb{l}
\tau^{\alpha\beta} = \ds\pa{W}{\eps_{\alpha\beta}}\,,\qquad
M_0^{\alpha\beta} = \ds\pa{W}{\kappa_{\alpha\beta}}\,,
\eqe
since $\delta \eps_{\alpha\beta} = \delta a_{\alpha\beta}/2$ and $\delta\kappa_{\alpha\beta} = \delta b_{\alpha\beta}$.
\subsubsection{Initially planar shells}
For initially planar shells, a simple choice is the 2-(material-)parameter formulation based on the Canham bending model \citep{canham1970minimum} and the nonlinear incompressible Neo-Hookean membrane model \citep{sauer2014computational}. This results in the membrane stress\,\footnote{\,There is also a stress contribution coming from the Canham bending model \citep{sauer2015theoretical}, but it is negligible for thin shells.}
\begin{equation}\label{eq:tauneohooke}
    \tau^{\alpha\beta}=\mu\left(A^{\alpha\beta}-\frac{a^{\alpha\beta}}{J^2}\right)\,,
\end{equation}
and the bending moment
\begin{equation}\label{eq:momentcanham}
    M^{\alpha\beta}_0=c\, J\, b^{\alpha\beta}\,,
\end{equation}
where the two material parameters $\mu$ and $c$, characterize the in-plane shear stiffness and the out-of-plane bending stiffness, respectively. According to the Canham model, the bending moment is linear w.r.t.~the curvature component $b^{\alpha\beta}$, however, $b^{\alpha\beta}$ is nonlinear w.r.t.~$b_{\alpha\beta}$, which implies a nonlinear bending model with non-constant bending stiffness. The corresponding material tangents are given in~\citet{sauer2015theoretical}. Since Eqs.~\eqref{eq:tauneohooke} and \eqref{eq:momentcanham} result from a surface strain energy, they require no further thickness integration. Instead, thickness integration is inherent to \eqref{eq:tauneohooke} and \eqref{eq:momentcanham}, so that the unit of $\mu$~(and likewise $\tau^{\alpha\beta}$) is [N/m], while the unit of $c$ is [Nm] (and the unit of $M_0^{\alpha\beta}$ is [Nm/m]).\footnote{\,In case basis $\ba_\alpha$ has no units. On the other hand, if $\ba_\alpha$ has units of length, $\tau^{\alpha\beta}$ and $M^{\alpha\beta}_0$ adjust accordingly.} In principle $\mu$ and $c$ can be treated fully independent from each other. But given the shell thickness $T$, $c$ and $\mu$ can both be related to the material parameters of 3D elasticity, e.g.~Young's modulus $E$ and Poisson's ratio $\nu$, e.g. see \cite{duong2017new}. 
\subsubsection{Initially curved shells}\label{sec:koiter}
For initially curved shells, a simple choice is the Koiter model \citep{ciarlet2005introduction,steigmann2013koiter}
\begin{equation}\label{eq:tauKoiter}
    \tau^{\alpha\beta} =c^{\alpha\beta\gamma\delta} \eps_{\gamma\delta}\,,
\end{equation}
\begin{equation}\label{eq:MKoiter}
    M^{\alpha\beta}_0=\frac{T^2}{12}c^{\alpha\beta\gamma\delta} \kappa_{\gamma\delta}\,,
\end{equation}
with
\begin{gather}
\begin{aligned}\label{eq:cabgd}
    c^{\alpha\beta\gamma\delta}&:=\Lambda\, \mathrm{I}^{\,\alpha\beta\gamma\delta}_{\mathrm{dil}}+2\,\mu\,\mathrm{I}^{\,\alpha\beta\gamma\delta}_{\mathrm{dev}},\\
    \mathrm{I}^{\,\alpha\beta\gamma\delta}_{\mathrm{dil}}&:=A^{\alpha\beta}A^{\gamma\delta},\\
    \mathrm{I}^{\,\alpha\beta\gamma\delta}_{\mathrm{dev}}&:=\frac{1}{2}\big(A^{\alpha\gamma}A^{\beta\delta}+A^{\alpha\delta}A^{\beta\gamma}\big),
\end{aligned}
\end{gather}
where $\mu$ and $\Lambda$ are the 2D Lam{\'e} parameters. They can be obtained, e.g.,~by analytical integration of the 3D Saint Venant-Kirchhoff model over the shell thickness $T$. This gives
\begin{equation}
    \mu = T\,\Tilde{\mu}\,,\qquad \Lambda=T\frac{2\,\Tilde{\Lambda}\,\Tilde{\mu}}{\Tilde{\Lambda}+2\,\Tilde{\mu}}\,,
\end{equation}
where $\Tilde{\Lambda}$ and $\Tilde{\mu}$ are the Lam{\'e} parameters in 3D elasticity. They are related to Young's modulus $E$ and Poisson's ratio $\nu$ by 
\begin{equation}
    \Tilde{\mu}=\frac{E}{2(\nu+1)}\,,\quad 
    \Tilde{\Lambda}=\frac{2\,\Tilde{\mu}\,\nu}{(1-2\,\nu)}\,.
\end{equation}
For incompressible materials $\nu=0.5$, leading to
\begin{equation}\label{eq:muandlam}
    \mu=\frac{E\,T}{3}\,,\qquad \Lambda=2\,\mu\,,
\end{equation}
such that
\begin{equation}\label{eq:cabgd2}
    c^{\alpha\beta\gamma\delta}=\frac{2\,E\,T}{3}\mathrm{I}^{\alpha\beta\gamma\delta},\qquad \mathrm{I}^{\alpha\beta\gamma\delta}=\mathrm{I}^{\alpha\beta\gamma\delta}_{\mathrm{dil}}+\mathrm{I}^{\alpha\beta\gamma\delta}_{\mathrm{dev}}\,.
\end{equation}
%
%
%
%
%
%
%-------------------------------------- FE --------------------------------------
\section{FE discretization}\label{s:FE}
This section presents the finite element discretization of the weak form on the basis of isogeometric analysis (IGA) to obtain the FE forces and the FE equilibrium equation following \cite{sauer2014computational} and \cite{duong2017new}. Subsequently, the discretization of the material and its synchronization with the FE analysis mesh is discussed.
\subsection{Surface discretization}
In order to discretize surface $\mathcal{S}$ and to approximately solve Eq.~\eqref{eq:weakform}, NURBS-based shape functions proposed by \citet{hughes2005isogeometric} are used. They take the form
\begin{equation}\label{eq:nurbs}
    N_I(\xi^\alpha)=\frac{w_I\hat{N}^e_I(\xi^\alpha)}{\sum^{n_{e}}_{I=1}w_I\hat{N}^e_I(\xi^\alpha)}\ ,
\end{equation}
where $\{\hat{N}^e_I \}^{n_e}_{I=1}$ are the $n_e$ B-spline basis functions of finite element $\Omega_e$. They are the entries of the matrix
\begin{equation}\label{eq:hatN}
    \hat{\textbf{N}}^e(\xi^\alpha)=\textbf{C}^e_1\,\textbf{B}^e(\xi^1) \otimes \textbf{C}^e_2\, \textbf{B}^e(\xi^2)\ , 
\end{equation}
containing the Bernstein polynomials $\mB^e(\xi^\alpha)$ and the B\'ezier extraction operator $\mC^e_\alpha$ \citep{borden2011isogeometric}. The geometry within an undeformed element $\Omega^e_0$ and its deformed counterpart $\Omega^e$ is then approximated from the position of control points $\textbf{X}^e$ and $\textbf{x}^e$, respectively\,\footnote{Uppercase characters are dedicated to the undeformed configuration, the lowercase characters to the deformed configuration.}, as
\begin{equation}\label{eq:approximation}
    \bX \approx \bX^h = \sum_{I=1}^{n_e} N_I\, \mX_I = \textbf{N}^e\,\textbf{X}^e\ ,\qquad \bx \approx \bx^h = \sum_{I=1}^{n_e}N_I \mx_I=\textbf{N}^e\textbf{x}^e\,,
\end{equation}
where $\textbf{N}^e(\xi^\alpha)=[N_1\textbf{1}, N_2\textbf{1}, ..., N_{n_e}\textbf{1}]$ is an array composed of the NURBS shape functions of Eq.~\eqref{eq:nurbs} and $\textbf{1}$ is the identity matrix in $d$-dimensional space. 
Further, $\mX_I$ is the initial and $\mx_I$ is the current position of the FE node (control point). Likewise, the displacements within element $\Omega^e$ are interpolated as
\begin{equation}\label{eq:uh1}
    \bu \approx \bu^h = \sum_{I=1}^{n_e}N_I\,\textbf{u}_I=\textbf{N}^e\textbf{u}^e\,.
\end{equation}
The covariant tangent vectors of the surface are then determined by
\begin{equation}\label{eq:covariant23}
	\boldsymbol{a}_{\alpha}=\frac{\partial \boldsymbol{x}}{\partial \xi^{\alpha}}\approx \textbf{N}_{,\alpha}^e\,\textbf{x}^e,\qquad \boldsymbol{A}_{\alpha}=\frac{\partial \boldsymbol{X}}{\partial \xi^{\alpha}}\approx \textbf{N}_{,\alpha}^e\,\textbf{X}^e\ ,
\end{equation}
and the variation of $\bx$ and $\ba_\alpha$ are expressed as
\begin{equation}
    \delta\bx \approx \textbf{N}^e\,\delta \textbf{x}^e, \qquad \delta\ba_\alpha \approx \textbf{N}_{,\alpha}^e\,\delta \textbf{x}^e\,. \qquad 
\end{equation}
Based on this, the discretization of all kinematic quantities and their variations can be determined, see \citet{duong2017new}. \par
{\small \textbf{Remark 1:} In classic finite elements, the nodes are lying on the discretized surface $\mathcal{S}^h \approx \sS$. In isogeometric analysis, where spline-based shape functions are used, the control points $\mX_I$ and $\mx_I$, in general, do not lie on the discretized surfaces $\sS_0^h$ and $\sS^h$, respectively, since surface approximation \eqref{eq:approximation} is not interpolating the control points, i.e.
\begin{equation}
    \bX_I := \sum_J N_J(\xi^\alpha_I)\,\mX_J \neq \mX_I 
\end{equation}
due to the property $N_I(\xi^\alpha_J) \neq \delta_{IJ}$. This must be taken into account when displacements resulting from IGA are compared with displacements resulting from experiments.}
\normalsize
%
%%---------------------------------------------------
%
\subsection{Weak form discretization}
The original weak form \eqref{eq:weakform} now yields the discretized version
\begin{equation}\label{eq:weakformFE}
    \sum^{n_{\mathrm{el}}}_{e=1}(G^{e}_{\mathrm{int}}+G^{e}_{\mathrm{ext}})=0\qquad \forall\ \delta \textbf{x} \in \mathcal{V}^h,
\end{equation}
where $n_{\mathrm{el}}$ is the number of finite elements and $G^{e}_{\mathrm{int}}$ and $G^{e}_{\mathrm{ext}}$ are the elemental contributions to the internal and external virtual work, respectively. The former can be written as
\begin{equation}
    G^{e}_{\mathrm{int}}=\delta \textbf{x}_e\,\textbf{f}^{\,e}_{\mathrm{int}} =\delta \textbf{x}_e\left( \textbf{f}^{\,e}_{\mathrm{int}\tau} +\textbf{f}^{\,e}_{\mathrm{int}M}\right).
\end{equation}
Here the internal FE force vectors due to membrane stress $\tau^{\alpha\beta}$ and bending moment $M_0^{\alpha\beta}$ are
\begin{equation}\label{eq:FEvectortau}
    \textbf{f}^{\,e}_{\mathrm{int}\tau}=\int_{\Omega_0^e}\tau^{\alpha\beta}\, \textbf{N}_{,\alpha}^{\mathrm{T}}~\boldsymbol{a}_{\beta}\,\mathrm{d}A\,,
\end{equation}
and
\begin{equation}\label{eq:FEvectorm}
    \textbf{f}^{\,e}_{\mathrm{int}M}=\int_{\Omega_0^e} M^{\alpha\beta}_0\,\textbf{N}^{\mathrm{T}}_{;\alpha\beta}\ \boldsymbol{n}\,\mathrm{d}A\,,
\end{equation}
where $\textbf{N}_{;\alpha\beta}:=\textbf{N}_{,\alpha\beta}-\Gamma_{\alpha\beta}^\gamma\textbf{N}_{,\gamma}$ for $\Gamma^\gamma_{\alpha\beta} := \ba_{\alpha,\beta}\cdot\ba^\gamma$. 
$G^{e}_{\mathrm{ext}}$ follows as
\begin{equation}
    G^{e}_{\mathrm{ext}}=\delta \textbf{x}_e^{\mathrm{T}}\big(\textbf{f}^{\,e}_{\mathrm{ext0}}+\textbf{f}^{\,e}_{\mathrm{ext}p}+\textbf{f}^{\,e}_{\mathrm{ext}t}+\textbf{f}^{\,e}_{\mathrm{ext}m}\big)\,,
\end{equation}
where the external FE force vectors are
\begin{align}
    \textbf{f}^{\,e}_{\mathrm{ext}0}&=\int_{\Omega_0^e} \textbf{N}^{\mathrm{T}} \boldsymbol{f}_{\!0}\ \mathrm{d}A\,,\qquad
    \textbf{f}^{\, e}_{\mathrm{ext}p}=\int_{\Omega_0^e} \textbf{N}^{\mathrm{T}} p\,\boldsymbol{n}\,\mathrm{d}a\,,\\
    \textbf{f}^{\, e}_{\mathrm{ext}t}&=\int_{\partial_t\Omega_0^e} \textbf{N}^{\mathrm{T}} \boldsymbol{t}\,\mathrm{d}s\,,\qquad
    \textbf{f}^{\, e}_{\mathrm{ext}m}=\int_{\partial_m\Omega_0^e} \textbf{N}^{\mathrm{T}}_{,\alpha}\,\nu^{\alpha}\,m_\tau\, \boldsymbol{n}\,\mathrm{d}s\,.
\end{align}
Here $\boldsymbol{f}_{\!0}$ is a constant surface force, $p$ is an external pressure acting always normal to $\mathcal{S}$, $\boldsymbol{t}$ is the effective boundary traction on $\partial_t\Omega^e_0 \subset \partial_t\sS^h$, $\nu^{\alpha}$ is the component of the unit normal to edge $\partial \mathcal{S}$, and $m_\tau$ is the tangential bending moment component on $\partial_m\Omega^e_0 \subset \partial_m\sS^h$. The corresponding tangent matrices can be found in \cite{duong2017new}. \\With the preceding equations, weak form~\eqref{eq:weakformFE} can be rewritten as
\begin{equation}\label{eq:discweak}
\delta \textbf{x}^{\mathrm{T}}\left[\textbf{f}_{\mathrm{int}}-\textbf{f}_{\mathrm{ext}}\right]=0, \quad \forall\ \delta \textbf{x}\in \mathcal{V}^h\,,
\end{equation}
where
\begin{equation}
\textbf{f}_{\mathrm{int}}=\sum_{e=1}^{n_{\mathrm{el}}}\textbf{f}_{\mathrm{int}}^{\,e}\ , \quad \textbf{f}_{\mathrm{ext}}=\sum_{e=1}^{n_{\mathrm{el}}}\textbf{f}_{\mathrm{ext}}^{\,e}
\end{equation}
are obtained from the assembly of the corresponding elemental force vectors and $\sV^h$ is the kinematically admissible set of all nodal variations $\delta\mx$. These are zero for the nodes on the Dirichlet boundary $\partial_u\mathcal{S}^h$. For the remaining nodes, Eq.~\eqref{eq:discweak} implies
\begin{equation}\label{eq:solutionfem}
\textbf{f}(\textbf{u})=\textbf{f}_{\mathrm{int}}-\textbf{f}_{\mathrm{ext}}=\textbf{0}\,,
\end{equation}
which is the discretized equilibrium equation that needs to be solved for the $d\,n_{\mathrm{no}}$ unknown components of the nodal displacement vector
\begin{equation}\label{eq:globalvector}
    \textbf{u}=\begin{bmatrix}\boldsymbol{u}_1\\
\boldsymbol{u}_2\\
\vdots\\
\boldsymbol{u}_{n_{\mathrm{no}}}\\
\end{bmatrix}
\,. 
\end{equation}
{\small\textbf{Remark 2:} In this formulation no mapping of derivatives between master and current configuration is required, also no introduction of a local Cartesian basis is needed.}
%
%%----------------------------------------------------
%
\subsection{Discretization of the material  parameters}
\begin{figure}[!ht]
    \centering
    \includegraphics[width=0.65\textwidth]{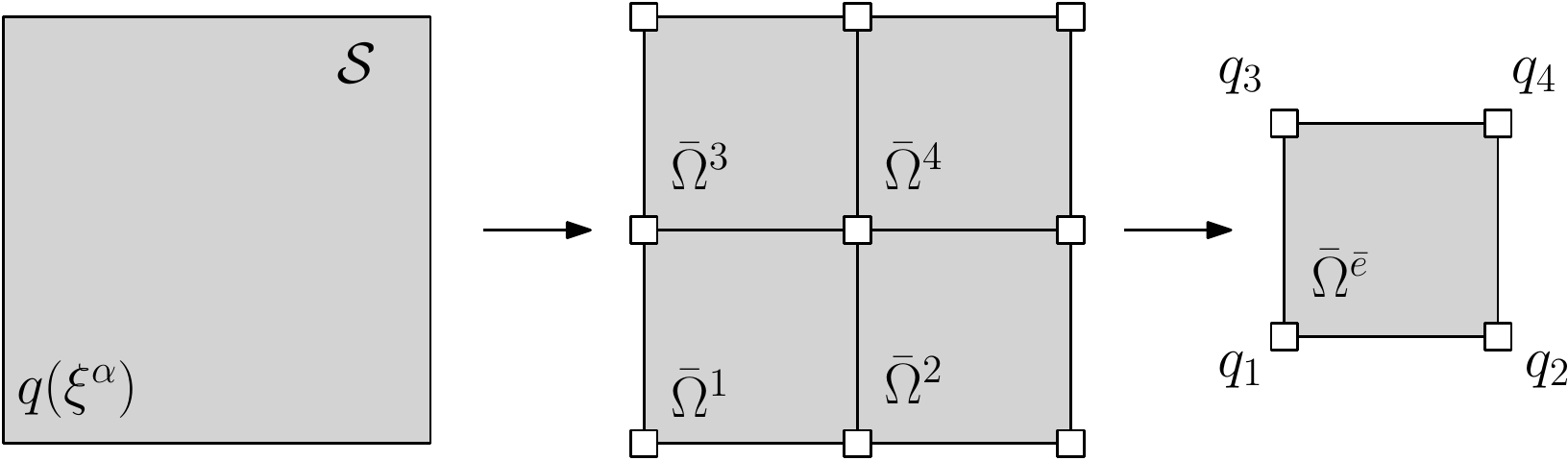}
    \caption{Approximation of the material parameter field $q$ on surface $\sS$ by 4-noded material elements $\bar{\Omega}^{\bar{e}}$.}
    \label{fig:material_el}
\end{figure}

\begin{figure}[!ht]
    \centering
    \includegraphics[width=0.98\textwidth]{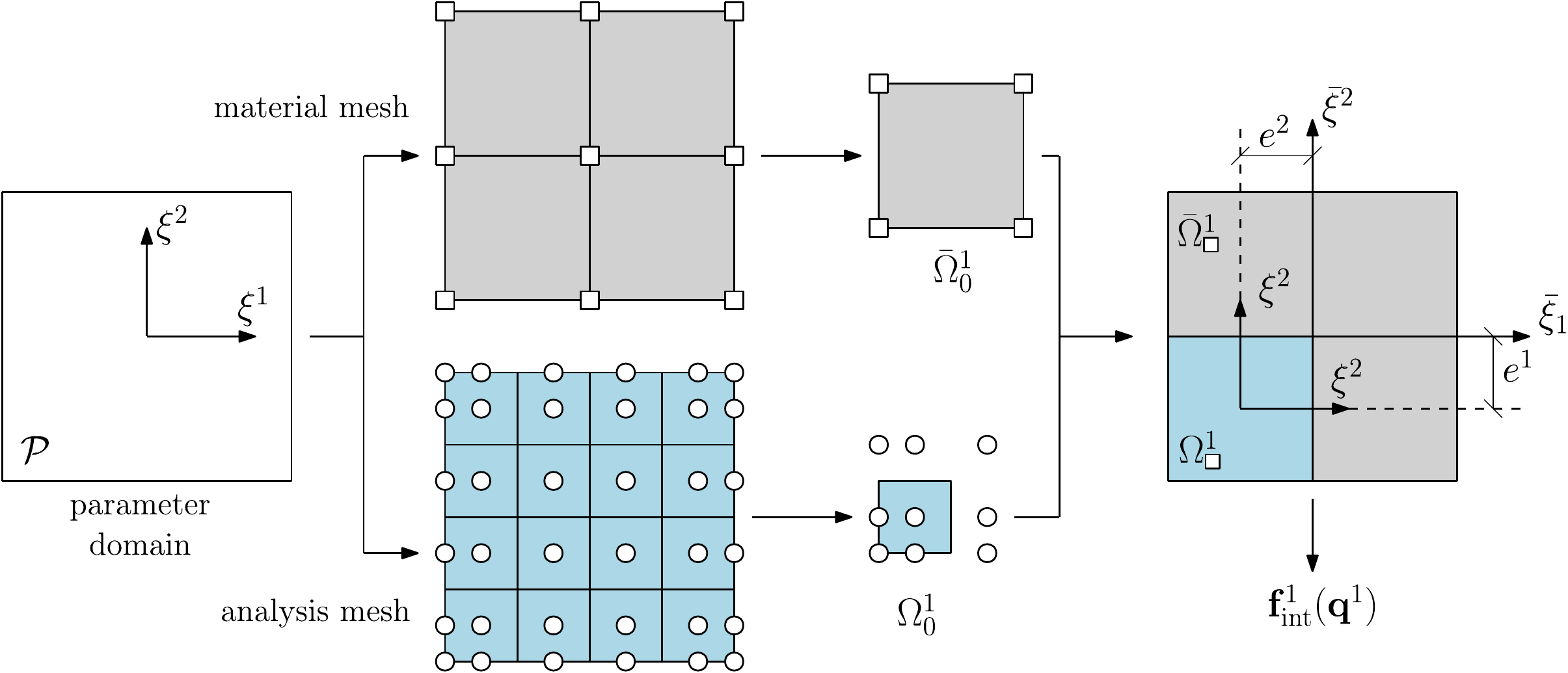}
    \caption{Example of mapping $\xi^\alpha \mapsto \bar{\xi}^\alpha$ for element $\Omega^1_\square\subset\bar\Omega^1_\square$. In this example $n_{\mathrm{el}} = 16$, $\bar n_{\mathrm{el}} = 4,\,n_1 = n_2 = 2$ and $e^1 = e^2 = 1$.}
    \label{fig:conformingmesh}
\end{figure}

The material parameters, here shear modulus $\mu$ and bending modulus $c$, or Young's modulus \textit{E} and thickness \textit{T}, are defined over the surface $\mathcal{S}_0$ as a continuous scalar field $q(\xi^{\alpha})$ that is approximated within the material parameter element $\bar{\Omega}^{\bar{e}}$, by $\bar n_e$ nodal values $q_I$ and interpolation functions $\bar N_I$ as
\begin{equation}\label{eq:interpolationq}
    q=q(\xi^{\alpha})\approx q^h=\sum_{I=1}^{\bar{n}_e} \bar{N}_I(\xi^\alpha)\, q_I=\bar{\textbf{N}}^{\bar{e}} \textbf{q}^{\bar{e}}\,,
\end{equation}
where $\bar{\textbf{N}}^{\bar{e}}:=[\bar{N}_1,\bar{N}_2,...\,,\bar{N}_{\bar{n}_e}]$ and $\textbf{q}^{\bar{e}}:=[q_1,q_2,...\,,q_{\bar{n}_e}]^{\mathrm{T}}$ are the elemental arrays containing all elemental $\bar{N_I}$ and $q_I$. In this work, $q$ is discretized with quadrilateral 4-node elements with bilinear Lagrange interpolation functions $\bar{N}_I$, see Fig.~\ref{fig:material_el}.
In principle, other interpolation functions can be chosen for the material, however we restrict ourselves to bilinear interpolation here, as it provides a good starting point to capture the a priori unknown material distribution. Only if some a priori information about the material is known other possibilities can be justified as better choices. An example are material discontinuities, where constant interpolation might be preferable. Analogously to \eqref{eq:globalvector}, the global vector
\begin{equation}
    \mq = \begin{bmatrix}
    \bq_1\\
    \bq_2\\
    \vdots\\
    \bq_{\bar n_{\mathrm{no}}}
    \end{bmatrix}
\end{equation}
is introduced. It contains the $n_{\mathrm{var}} = \bar d\,\bar n_{\mathrm{no}}$ unknown nodal material parameters to be identified by inverse analysis. Here, each node contains $\bar d = 2$ unknown design variables.\\
Before discussing this, the relation between the two different discretizations for $\muu$ and $\mq$ needs to be addressed, see Fig.~\ref{fig:conformingmesh}. 
Conforming meshes consisting of rectangular elements in the parameter domain $\mathcal{P}(\xi^1,\xi^2)$ are considered here, which preserve the relation $\Omega_\square^e \subset \bar{\Omega}_\square^{\bar{e}}$, where $\Omega_\square^e$ and $\bar{\Omega}_\square^{\bar{e}}$ are the element domains in $\sP$ for the analysis and material mesh respectively. This is also the domain where the numerical integration of the elemental FE force vectors in \eqref{eq:FEvectortau} and \eqref{eq:FEvectorm} is carried out. The tensor product structure of NURBS in \eqref{eq:hatN} automatically leads to rectangular analysis elements on $\sP$. Choosing conforming material elements is then a natural choice. In order to relate the two element domains, analysis element $\Omega_\square^e$, with its NURBS-based shape functions $\textbf{N}(\xi^\alpha)$, is defined on the domain $\xi^\alpha \in [-1,1]$, while the material element $\bar{\Omega}^{\bar{e}}_\square$, with its Lagrange shape functions $\bar{\textbf{N}}(\bar\xi^\alpha)$, is defined on domain $\bar{\xi}^\alpha \in [-1,1]$. 
The mapping between domains $\xi^\alpha \mapsto \bar{\xi}^\alpha$ is obtained by affine linear transformation, i.e.
\begin{equation}\label{eq:mappingL}
    [\bar{\xi}^{\,1},\bar{\xi}^{\,2}]=\left[\frac{1}{n_1}\left(\xi^1 +e^1 \right),\frac{1}{n_2}\left(\xi^2 +e^2 \right)\right]\,,
\end{equation}
such that $\bar N_I = \bar N_I\big(\bar\xi^\alpha(\xi^\beta)\big)$ becomes a function of $\xi^\beta$. 
Here $e^\alpha$ is the offset between the coordinate centers of $\bar{\Omega}_0^{\bar{e}}$ and $\Omega_0^e$ and $n_\alpha$ is the number of $\Omega_0^e \subset \bar{\Omega}_0^{\bar{e}}$ in direction $\alpha$ (see Fig.~\ref{fig:conformingmesh}). 
%
%
%
%
%
%
%-------------------------------------- INVERSE --------------------------------------
\section{Inverse analysis}\label{s:inverse}
In this section the inverse problem is formulated in the context of optimization. A corresponding optimization algorithm and optimality conditions are discussed, followed by an investigation of the different error sources that affect the precision of the inverse problem solution. Finally, the analytical sensitivities for the considered constitutive parameters are derived.
%
%%---------------------------------------------------
%
\subsection{Objective function}
The Finite Element Model Updating method is used to solve the inverse problem. Accordingly, the unknown design vector $\mq$ is obtained from the constrained minimization
\begin{equation}\label{eq:minimization}
\min_{{}\mq} f(\textbf{q})
\end{equation}
subject to the bounds $0 <q_{\min} \leq q_I \leq q_{\max}$ and subject to satisfying the discrete weak form~ \eqref{eq:discweak}. $f$ is a scalar-valued function known as the \textit{objective function}. 
It depends on the $\bar d\,\bar{n}_{\mathrm{no}}$ nodal values of the $\bar d$ discretized constitutive parameters $q(\xi^\alpha)$ according to interpolation \eqref{eq:interpolationq}. 
The function $f$ expresses the discrepancy between the model and the observed experimental behavior. The least squares form
\begin{equation}\label{eq:updatedobjectivefun}
f(\textbf{q})=\frac{\left\Vert\textbf{U}_{\mathrm{exp}}-\textbf{U}_{\mathrm{FE}}(\textbf{q})\right\Vert^2}{2\left\Vert \textbf{U}_{\mathrm{exp}} \right\Vert^2} +\frac{\left\Vert \mR_{\,\mathrm{exp}}-\mR_{\,\mathrm{FE}}(\textbf{q})\right\Vert^2}{2\left\Vert \mR_{\,\mathrm{exp}}\right\Vert^2}
\end{equation}
is considered, where
\begin{equation}
\textbf{U}_{\mathrm{exp}} = \begin{bmatrix} \boldsymbol{u}^{\mathrm{exp}}_1\\[0.5em]
\boldsymbol{u}^{\mathrm{exp}}_2\\
\vdots\\[0.2em]
\boldsymbol{u}^{\mathrm{exp}}_{n_{\mathrm{exp}}}
\end{bmatrix}
\end{equation}
is a vector containing the $n_\mathrm{exp}$ experimentally measured displacements $\boldsymbol{u}_I^{\mathrm{exp}}, I = 1,\, ...\,,n_{\mathrm{exp}},$ at location $\boldsymbol{x}^\mathrm{exp}_I\in\sS$ and
\begin{equation}\label{eq:UFE}
 \textbf{U}_{\rm FE}(\textbf{q}) = \begin{bmatrix} \boldsymbol{u}^h\big(\boldsymbol{x}^{\mathrm{exp}}_1,\textbf{q}\big)\\[0.5em]
 \boldsymbol{u}^h\big(\boldsymbol{x}^{\mathrm{exp}}_2,\textbf{q}\big)\\
 \vdots\\[0.2em]
 \boldsymbol{u}^h\big(\boldsymbol{x}^{\mathrm{exp}}_{n_{\mathrm{exp}}},\textbf{q}\big)\\
 \end{bmatrix}\,,
\end{equation}
is a vector containing the corresponding $n_\mathrm{exp}$ FE results at $\bx_I^\mathrm{exp}$, which according to Eq.~\eqref{eq:uh1} are
\begin{equation}\label{eq:uh}
    \boldsymbol{u}^h\big(\boldsymbol{x}_I^\mathrm{{exp}},\mq\big) = \textbf{N}^e\big(\boldsymbol{x}_I^\mathrm{{exp}}\big)\, \textbf{u}^e\big(\mq\big)\,.
\end{equation}
Thus \eqref{eq:minimization} determines the material distribution $\mq$ that minimizes the difference between $n_\mathrm{exp}$ measured displacements $\bu_I^\mathrm{exp}$ and their numerically calculated counterparts. The $n_\mathrm{exp}$ measurements can come from a single experiment or from multiple experiments that are for example conducted at $n_\mathrm{ll}$ different load levels and concatenated in the global vector $\mU$. This is useful for capturing the nonlinear response of $\mU_\mathrm{FE}$ on $\mq$ for increasing loads, as is seen in the examples of Sec.~\ref{s:examples}. Additionally, to ensure that system \eqref{eq:minimization} is uniquely  determinable for pure Dirichlet problems, the reaction forces on various boundaries are included in objective \eqref{eq:updatedobjectivefun} via vector $\mR$. Without support reactions, each material parameter is only determinable up to a constant for pure Dirichlet problems. 
%
%%---------------------------------------------------
%
\subsection{Optimization algorithm}
\begin{figure}[!ht]
    \centering
    \includegraphics[width=0.92\textwidth]{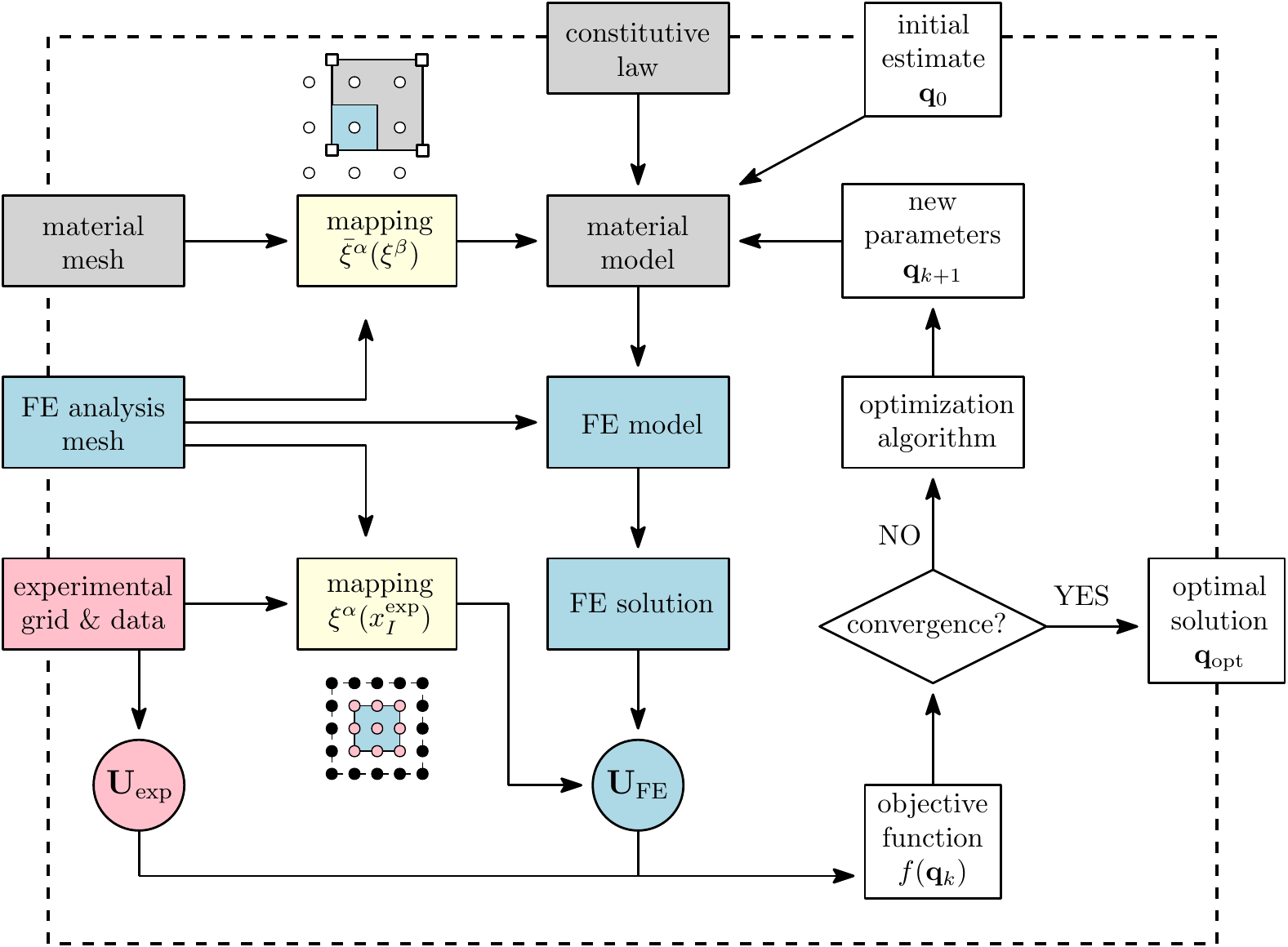}
    \caption{Flowchart of the material identification algorithm: Based on a chosen FE analysis mesh and chosen material mesh, the algorithm determines the optimal solution for the material parameters $\mq$ for given initial guess $\mq_0$, given experimental data and given constitutive law.}
    \label{fig:theflowchart}
\end{figure}

In order to satisfy Eq.~\eqref{eq:minimization}, a trust-region method is employed \citep{conn2000trust}. Trust region methods are iterative methods that construct an approximation of the function $f(\textbf{q})$ in the neighborhood (trust region) $\sN$ of the current iterate $\textbf{q}_k$. One of the advantages of trust-region methods over line search methods, is that non-convex approximate models can be used, which makes this class of iterative methods reliable, robust and applicable to ill-conditioned problems \citep{yuan2000review}. Minimization over $\sN$  (the trust-region subproblem) results in solution $\textbf{s}_k$, called the trial step. The current design is then updated by $\textbf{q}_k+\textbf{s}_k$ if $f(\textbf{q}_k+\textbf{s}_k)<f(\textbf{q}_k)$. Otherwise, it remains unchanged, $\sN$ is shrunk and the computation of $\textbf{s}_k$ is repeated. The algorithm keeps updating $\textbf{q}_k$ until certain conditions are satisfied. The iteration is terminated when the two stopping criteria
\begin{align}
\lvert f(\textbf{q}_{k+1})-f(\textbf{q}_k)\rvert\leqslant\epsilon\label{eq:criteria1}\\\quad \Vert \textbf{q}_{k+1}-\textbf{q}_k\Vert\leqslant\epsilon\label{eq:criteria2}\,,
\end{align}
are satisfied. Here $\epsilon$ is a small positive tolerance chosen on the order of machine precision ($\epsilon\approx2.2\cdot10^{-16}$). Together, both conditions ensure stopping the procedure after both $f$ and $\mq$ have converged. The \texttt{lsqnonlin} solver from the MATLAB Optimization Toolbox\texttrademark~\citep{MATLAB2018} is employed, which takes advantage of the Trust-region Interior Reflective (TIR) approach proposed by \cite{coleman1996interior} and allows adding analytical Jacobians (see Appendix~\ref{s:appendix1}). \\
The flowchart of the inverse identification algorithm is shown in Fig.~\ref{fig:theflowchart}.
Given a constitutive law, the initial estimate $\mq_0$, the FE analysis mesh and material mesh are provided first. They are needed to define the material model, determine the mapping between material and FE mesh and determine the FE solution. Next, the experimental data (experimental grid and measurements $\textbf{U}_\mathrm{exp}$) are provided. The mapping between experimental grid and FE mesh is determined, and used to compute the displacements $\mU_{\mathrm{FE}}$ at locations $\bx_I^{\mathrm{exp}}$. Given $\textbf{U}_\mathrm{exp}$ and $\textbf{U}_\mathrm{FE}$, the objective function $f(\mq)$ is evaluated. As long as the convergence criteria \eqref{eq:criteria1} \& \eqref{eq:criteria2} are not met, the optimization algorithm proceeds to find a new estimate and repeats the process.
%
%
%
%%---------------------------------------------------
%
\subsection{Error sources}
The inverse identification process is based on three discretized fields, see Fig.~\ref{fig:mesh_sync}:
\begin{figure}[!ht]
    \centering
    \includegraphics[width=0.7\textwidth]{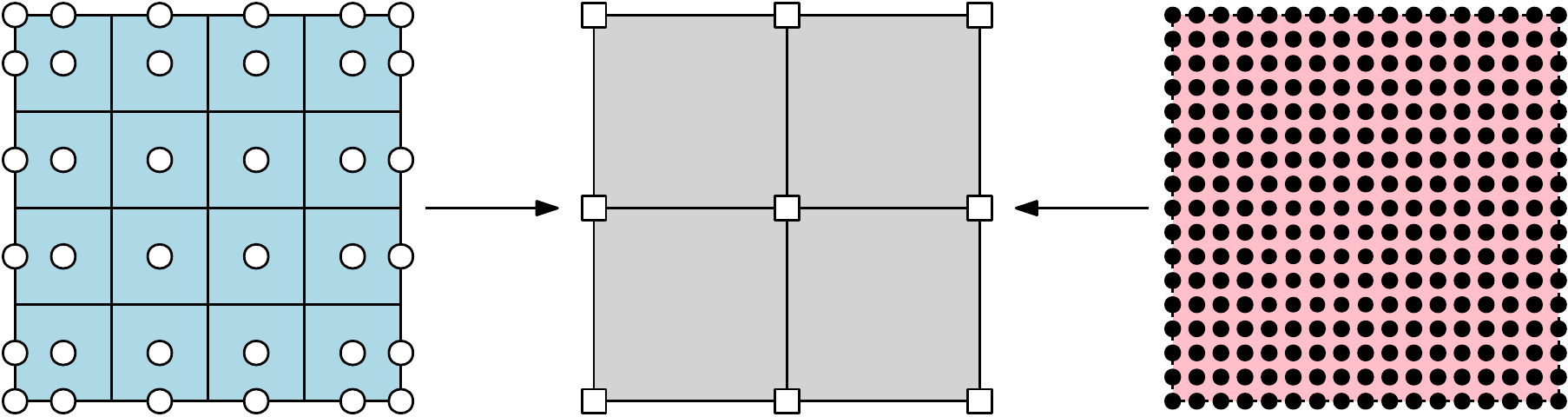}
    \caption{Inverse analysis based on three differently discretized fields: The central objective is to reconstruct the unknown parameters of the material mesh (\textit{center}). This requires a sufficiently dense FE analysis mesh (\textit{left}) and experimental grid (\textit{right}).}
    \label{fig:mesh_sync}
\end{figure}
First, the FE discretization of the displacement field of the forward problem \eqref{eq:uh1}. It determines the FE accuracy and computational cost  of the forward problem. Second, the material parameter discretization \eqref{eq:interpolationq}. It captures the material distribution and establishes the size and computational cost of the inverse problem (characterized by $n_{\mathrm{var}}$). Third, the chosen experimental displacement grid. It determines the amount of available data points used in the inverse analysis, which also contributes to the computational cost of the inverse problem. There are three corresponding error sources:
\begin{itemize}
    \item finite element approximation error, resulting from the difference between the finite element solution for the displacement and the (unknown) exact displacement field,
    \item material approximation error, resulting from the difference between the material interpolation and the (unknown) exact material distribution,
    \item experimental measurement error (i.e.~noise), resulting from the difference between the measured data and the (unknown) exact specimen behavior. 
\end{itemize}
The following numerical examples demonstrate that the proposed inverse algorithm converges w.r.t. all three error sources. For this, exact (or at least highly accurate) solutions are required. They are determined here by (1) conducting a convergence study of the forward FE problem, thus determining suitable FE meshes with acceptable error levels, (2) providing known material distributions, and (3) manufacturing ``experimental" results based on highly accurate FE results and then mimicking the effect of measurement error through the subsequent application of random noise. The FE mesh for these synthetic results is always chosen much denser than the discretization used for the subsequent inverse analysis to avoid the analysis bias commonly referred to as \textit{inverse crimes} \citep{colton1998inverse,invcrime}. 
Further, by examining the sensitivities, preliminary conclusions on the nature of the inverse problem can be drawn.
%
%%---------------------------------------------------
%
\subsection{Analytical sensitivities}
Gradient-based optimization algorithms typically require the gradient $\textbf{g}(\textbf{q})$ and Hessian $\textbf{H}(\textbf{q})$ of the objective function. Often, those are approximated through computationally expensive finite differences. On the other hand, the available constitutive formulations and FE force vectors given in Sec.~\ref{s:shelltheory} \& \ref{s:FE} allow for the derivation and implementation of the analytical gradient and Hessian. In order to provide the analytical gradient, differentiation of the internal FE force vectors in Eqs.~\eqref{eq:FEvectortau} and \eqref{eq:FEvectorm} w.r.t.~the elemental material parameter vector $\textbf{q}^{{\bar{e}}}$ is needed. Denoting the elemental sensitivity matrix
\begin{equation}\label{eq:sensitivitymatrixS}
 \textbf{S}^{e\bar{e}}:=\frac{\partial \textbf{f}^{\,e}_{\mathrm{int}}}{\partial \textbf{q}^{\bar{e}}}\,,
\end{equation}
and applying \eqref{eq:interpolationq} to $\mu$ and $c$, the change of the internal force vector due to material changes becomes
\begin{equation}\label{eq:deltaf}
    \Delta\textbf{f}^{\,e}_{\mathrm{int}}=\frac{\partial \textbf{f}^{\,e}_{\mathrm{int}\tau}}{\partial \boldsymbol{\mu}^{\bar{e}}} \Delta\boldsymbol{\mu}^{\bar{e}} +\frac{\partial \textbf{f}^{\,e}_{\mathrm{int}M}}{\partial \textbf{c}^{\bar{e}}} \Delta\textbf{c}^{\bar{e}} =\textbf{S}^{e\bar{e}}_{\mu}\Delta\boldsymbol{\mu}^{\bar{e}} + \textbf{S}^{e\bar{e}}_{c}\Delta\textbf{c}^{\bar{e}}\ ,
\end{equation}
where
\begin{equation}\label{eq:sensitivitymu}
    \textbf{S}^{e\bar{e}}_{\mu}:=\int_{\Omega_0^e}{\mN^e}^\mathrm{T}_{\!\!\!,\alpha} \boldsymbol{a}_\beta \left(A^{\alpha \beta}-\frac{a^{\alpha\beta}}{J^2}\right) \bar{\textbf{N}}^{\bar e}\,\mathrm{d}A
\end{equation}
and
\begin{equation}\label{eq:sensitivityc}
    \textbf{S}^{e\bar{e}}_{c}:=\int_{\Omega_0^e}J\, b^{\alpha\beta}\,{\mN^e}^\mathrm{T}
_{\!\!\!;\alpha\beta}\,\boldsymbol{n}\,\bar{\mN}^{\bar{e}}\,\mathrm{d}A\,
\end{equation}
are the Neo-Hookean membrane and the Canham bending sensitivity, respectively, that follow directly from \eqref{eq:FEvectortau}, \eqref{eq:FEvectorm}, \eqref{eq:tauneohooke}, \eqref{eq:momentcanham}, \eqref{eq:interpolationq} and \eqref{eq:sensitivitymatrixS}.
For the Koiter model, \eqref{eq:interpolationq} is applied to $E$ and $T$. \eqref{eq:deltaf} then becomes
\begin{equation}
    \Delta\textbf{f}^{\,e}_{\mathrm{int}}=\frac{\partial \textbf{f}^{\,e}_{\mathrm{int}}}{\partial \textbf{E}^{\bar{e}}} \Delta\textbf{E}^{\bar{e}} +\frac{\partial \textbf{f}^{\,e}_{\mathrm{int}}}{\partial \textbf{T}^{\bar{e}}} \Delta\textbf{T}^{\bar{e}} =\textbf{S}^{e\bar{e}}_{E}\Delta\textbf{E}^{\bar{e}} + \textbf{S}^{e\bar{e}}_{T}\Delta\textbf{T}^{\bar{e}}\ ,
\end{equation}
where
\begin{equation}\label{eq:sensitivityE} 
    \textbf{S}^{e\bar{e}}_{E}:=\int_{\Omega_0^e}\bigg(\frac{2\,T}{3}{\mN}^\mathrm{T}_{,\alpha}\,\mathrm{I}^{\alpha\beta\gamma\delta} \,\eps_{\gamma\delta}\,\boldsymbol{a}_\beta+\frac{T^3}{18}\,{\mN}^\mathrm{T}_{;\alpha\beta}\,\mathrm{I}^{\alpha\beta\gamma\delta}\,\kappa_{\gamma\delta}\,\boldsymbol{n}\bigg){\bar{\textbf{N}}^{\bar{e}}}\,\mathrm{d}A\,,
\end{equation}
\begin{equation}\label{eq:sensitivityT}
    \textbf{S}^{e\bar{e}}_{T}:=\int_{\Omega_0^e}\bigg(\frac{2E}{3}{\mN}^\mathrm{T}_{,\alpha}\,\mathrm{I}^{\alpha\beta\gamma\delta} \,\eps_{\gamma\delta}\,\boldsymbol{a}_\beta+\frac{E\,T^2}{6}\,{\mN}^\mathrm{T}_{;\alpha\beta}\,\mathrm{I}^{\alpha\beta\gamma\delta}\,\kappa_{\gamma\delta}\,\boldsymbol{n}\bigg){\bar{\textbf{N}}^{\bar{e}}}\,\mathrm{d}A\,.
\end{equation}
are the sensitivities w.r.t.~Young's modulus $E$ and shell thickness $T$, that follow from \eqref{eq:FEvectortau}, \eqref{eq:FEvectorm}, \eqref{eq:tauKoiter}, \eqref{eq:MKoiter}, \eqref{eq:cabgd2}, \eqref{eq:interpolationq} and \eqref{eq:sensitivitymatrixS}.\\
All $\textbf{S}^{e\bar{e}}_\bullet$ are of size\footnote{number of degrees of freedom for finite element $e$ $\times$ number of material parameters per material element $\bar{e}$.} $27\times4$ and require numerical integration over element $\Omega^e_0$ and subsequent assembly for all $e=1,...\,,n_{\mathrm{el}}$ and $\bar{e}=1,...\,,\bar{n}_e$. This results in the global sensitivity matrix $\textbf{S}$ with dimension $d\,n_{\mathrm{no}} \times \bar d\,\bar{n}_{\mathrm{no}}$. 
If $\textbf{f}^{\,e}_{\mathrm{int}}$ is linear in \textbf{q}, the sensitivities are constant (in \textbf{q}) and the global force vector is simply given by
\begin{equation}\label{eq:globalsensitivity}
    \textbf{f}_{\mathrm{int}}=\textbf{S}\,\textbf{q}\,,
\end{equation}
which is the case for \eqref{eq:sensitivitymu} and \eqref{eq:sensitivityc}, but not \eqref{eq:sensitivityE} and \eqref{eq:sensitivityT}. 
As seen in \eqref{eq:sensitivitymu} and \eqref{eq:sensitivityc}, the membrane and bending sensitivities depend differently on the deformation. Since the deformation varies in space and time, the two sensitivities can thus be expected to play different roles in space and time\footnote{or computational pseudo-time marking load stepping.}. This is seen in the example of Sec.~\ref{sec:inf}. \\ 
Given $\mS$, the discretization of the gradient and Hessian follow as outlined in Appendix \ref{s:appendix1}.
%
%
%
%
%
%
%-------------------------------------- EXAMPLES --------------------------------------
\section{Numerical examples}\label{s:examples}
In this section four different identification examples are examined: uniaxial tension in Sec.~\ref{sec:uni} -- a pure membrane problem that according to \eqref{eq:tauneohooke} only depends on the unknown shear modulus~$\mu$, pure bending in Sec.~\ref{sec:purebending}, which according to \eqref{eq:momentcanham} only depends on the unknown bending stiffness~$c$, inflation in Sec.~\ref{sec:inf} -- a coupled problem involving unknowns $c$ and $\mu$, and abdominal wall pressurization -- a coupled problem involving unknowns $E$ and $T$. 
Quasi experimental deformations are used in all cases. They are generated by accurately solving the forward problem defined by the given analytical reference distribution $q(\boldsymbol{\bX})$.\footnote{$q$ can be equivalently expressed as a function of $\xi^\alpha$, $\bx$ or  $\bX$, since a 1-to-1 mapping between the three configurations is assumed here.} In order to mimic measurement uncertainties, random noise is added to the deformation obtained from the reference parameters such that the measured displacement is
\begin{equation}\label{eq:noise}
    u^{\mathrm{exp}}_{Ii} = u^h_i\big(\boldsymbol{x}^{\mathrm{exp}}_I)(1+\gamma_{Ii})\,, 
\end{equation}
where $i = 1,2,3$ are the Cartesian components, $u^h_i$ is the solution for a very fine FE mesh and each $\gamma_{Ii}$ is a random number picked uniformly from the range $[-1,1]$ and multiplied by the considered noise level, which is up to 4\% in all examples.
%
%
%
%
%
%
%-------------------------------------- UNI-AXIAL --------------------------------------
\subsection{Uniaxial tension}\label{sec:uni}
\begin{figure}[ht]
\begin{center} \unitlength1cm
\begin{picture}(0,5.4)
\put(-7.8,-.2){\includegraphics[width=65mm]{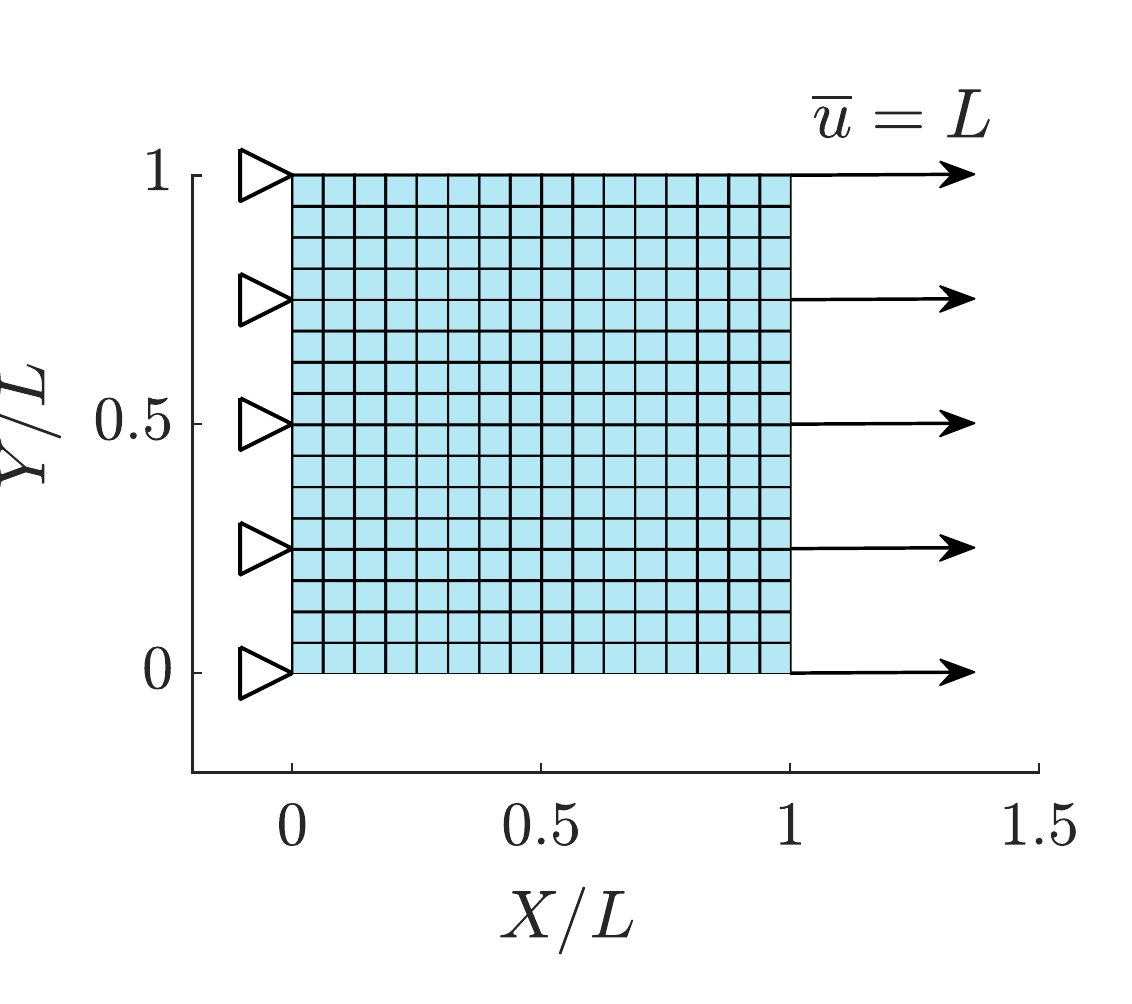}}
\put(-1,0){\includegraphics[width=83mm]{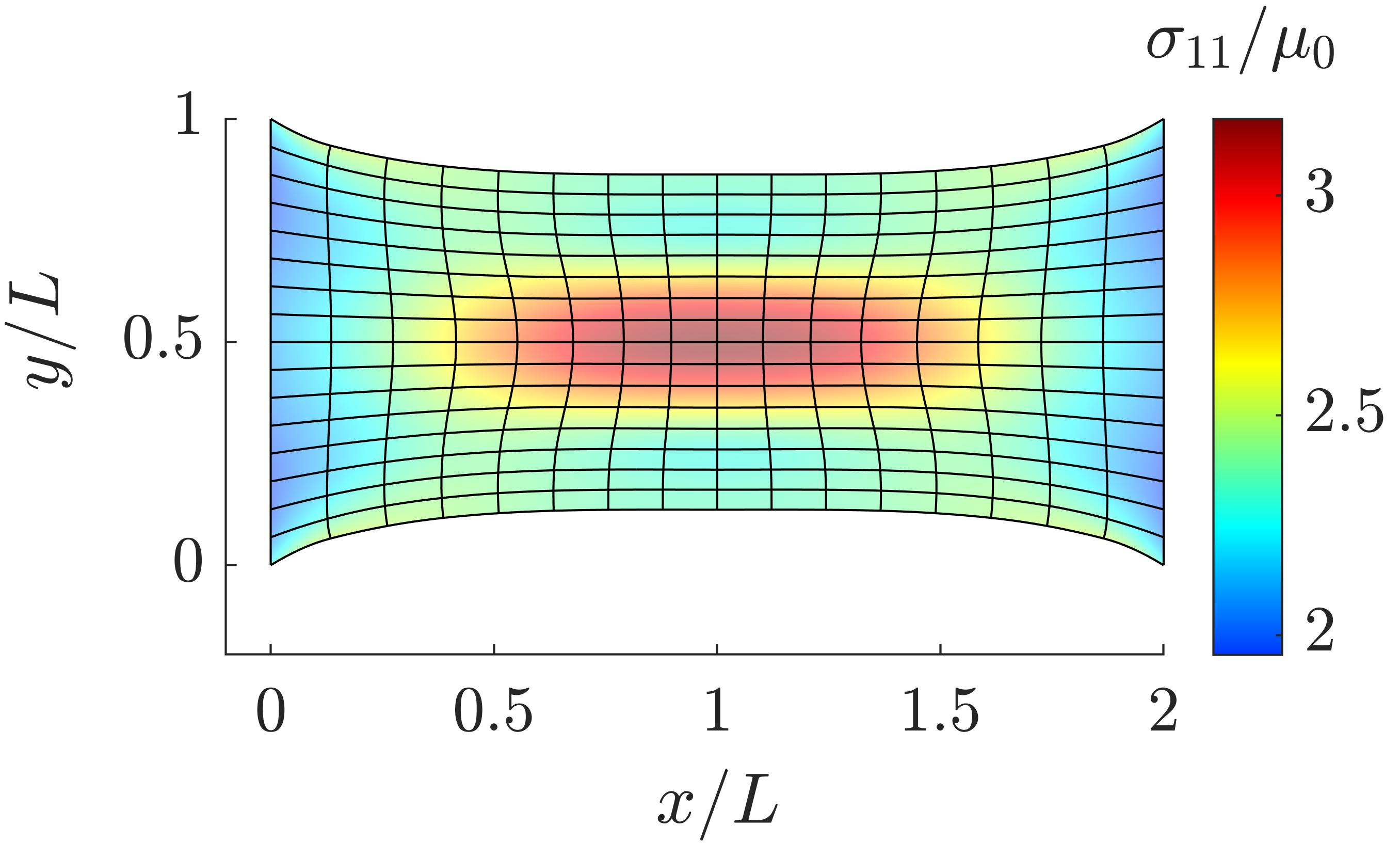}}
\put(-7.4,0){\small{a.}}
\put(0,0){\small{b.}}
\end{picture}
\caption{Uniaxial tension: a.~undeformed configuration; b.~deformed configuration colored by the membrane stress $\sigma_{11}=\be_1\cdot\bsig\be_1$, ranging between $1.96\mu_0$ and $3.17\mu_0$.}
\label{fig:uniaxialBC}
\end{center}
\end{figure}

The first example considers uniaxial tension, which induces pure membrane deformations without any bending. It thus allows to study material reconstruction for a single ($\bar d = 1$) unknown field -- in this example the shear modulus $\mu = \mu(\bX)$. To induce uniaxial tension, a flat sheet with dimension $L_x \times L_y = L \times L$ is fixed in all directions on the left edge ($X=0$), fixed in the $Z$-~direction on the entire surface, and stretched by the prescribed displacement $\overline{\boldsymbol{u}}=L\be_1$ applied to the opposite edge ($X=L$), see Fig.~\ref{fig:uniaxialBC}.
\par
\begin{figure}[!ht]
\begin{center} \unitlength1cm
\begin{picture}(0,6)
\put(-8,0){\includegraphics[width=85mm]{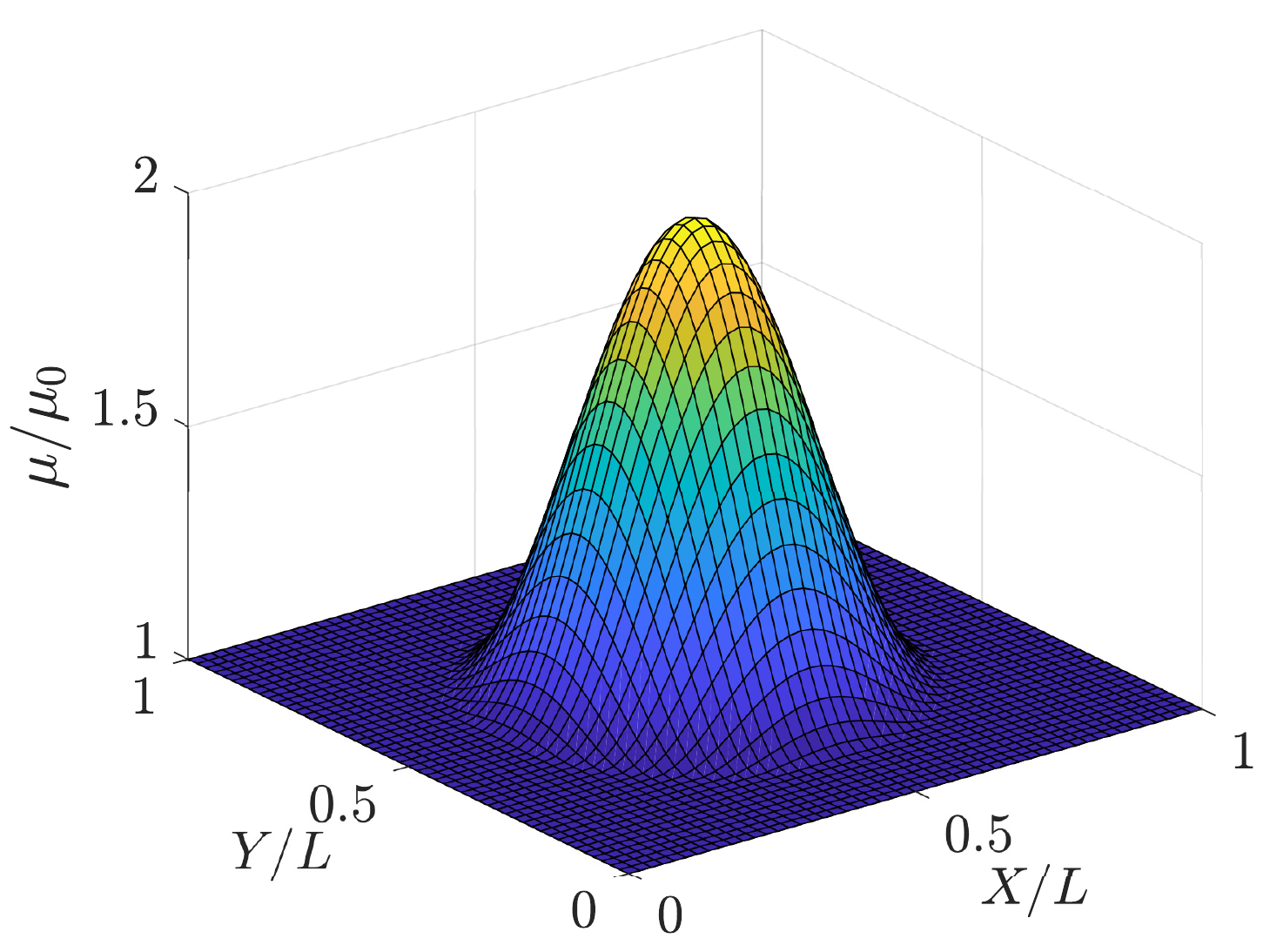}}
\put(1.5,0.5){\includegraphics[width=55mm]{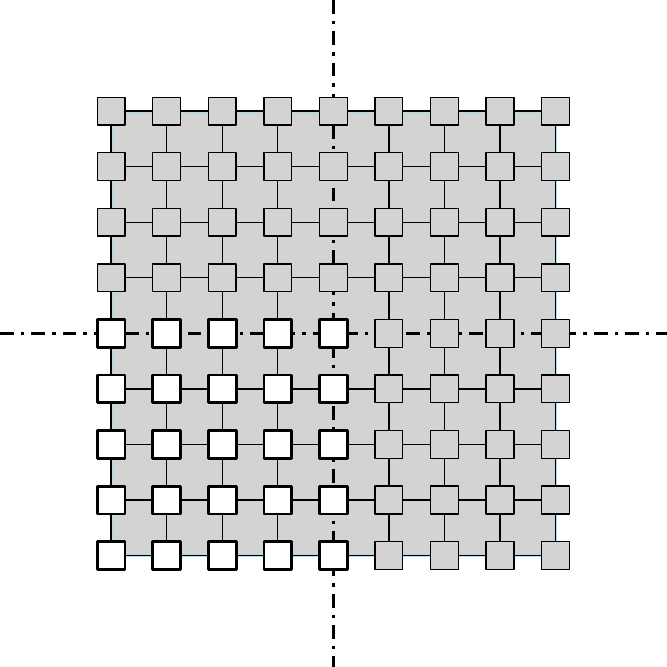}}
\put(-7.2,0){\small{a.}}
\put(1.8,0){\small{b.}}
\end{picture}
\caption{Uniaxial tension: a.~reference  shear modulus distribution $\mu(X,Y)$; b.~material mesh $\bar{n}_{\mathrm{el}}=8\times8$, reduced to $n_{\mathrm{var}}=25$ unknowns due to symmetry.}
\label{fig:uniaxialother}
\end{center}
\end{figure}

\begin{figure}[!ht]
\begin{center} \unitlength1cm
\begin{picture}(0,6.5)
\put(-7.7,0){\includegraphics[width=70mm]{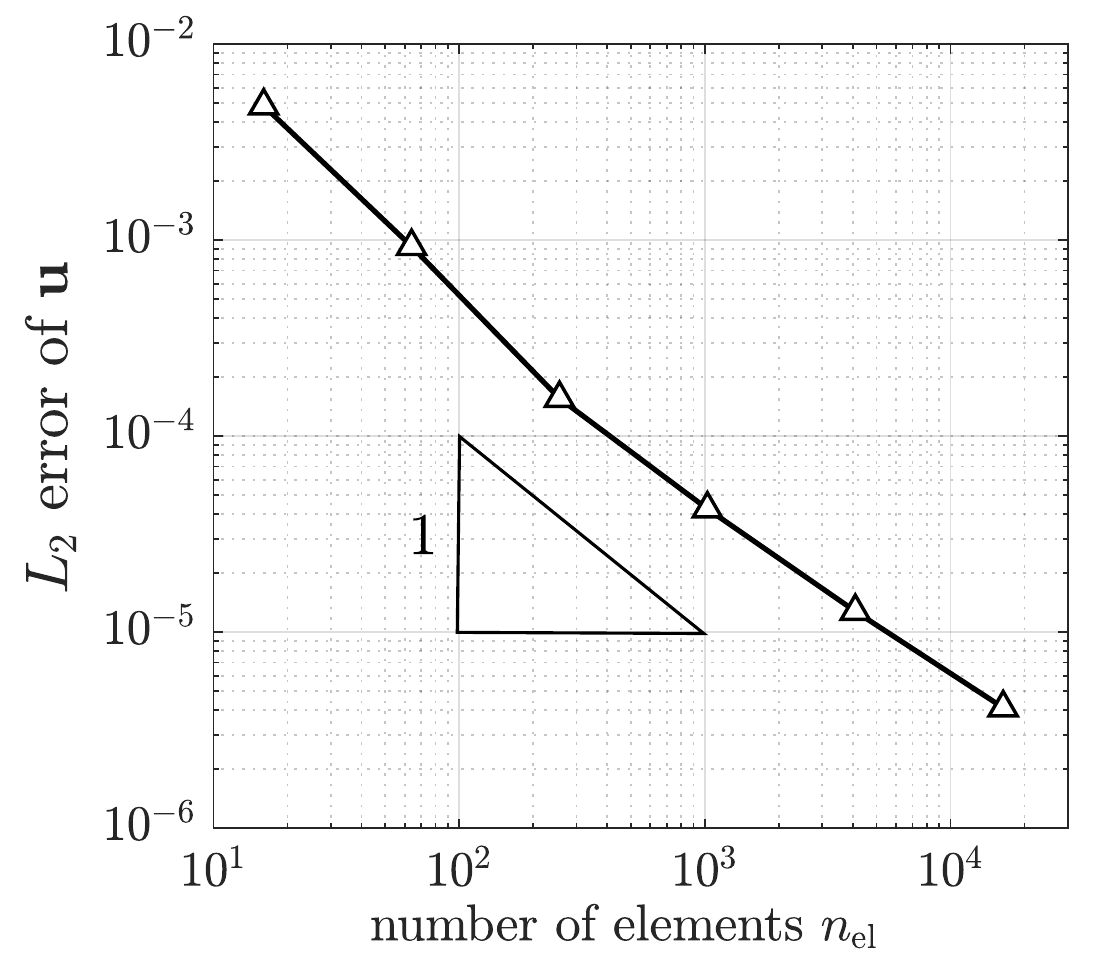}}
\put(.3,0){\includegraphics[width=70mm]{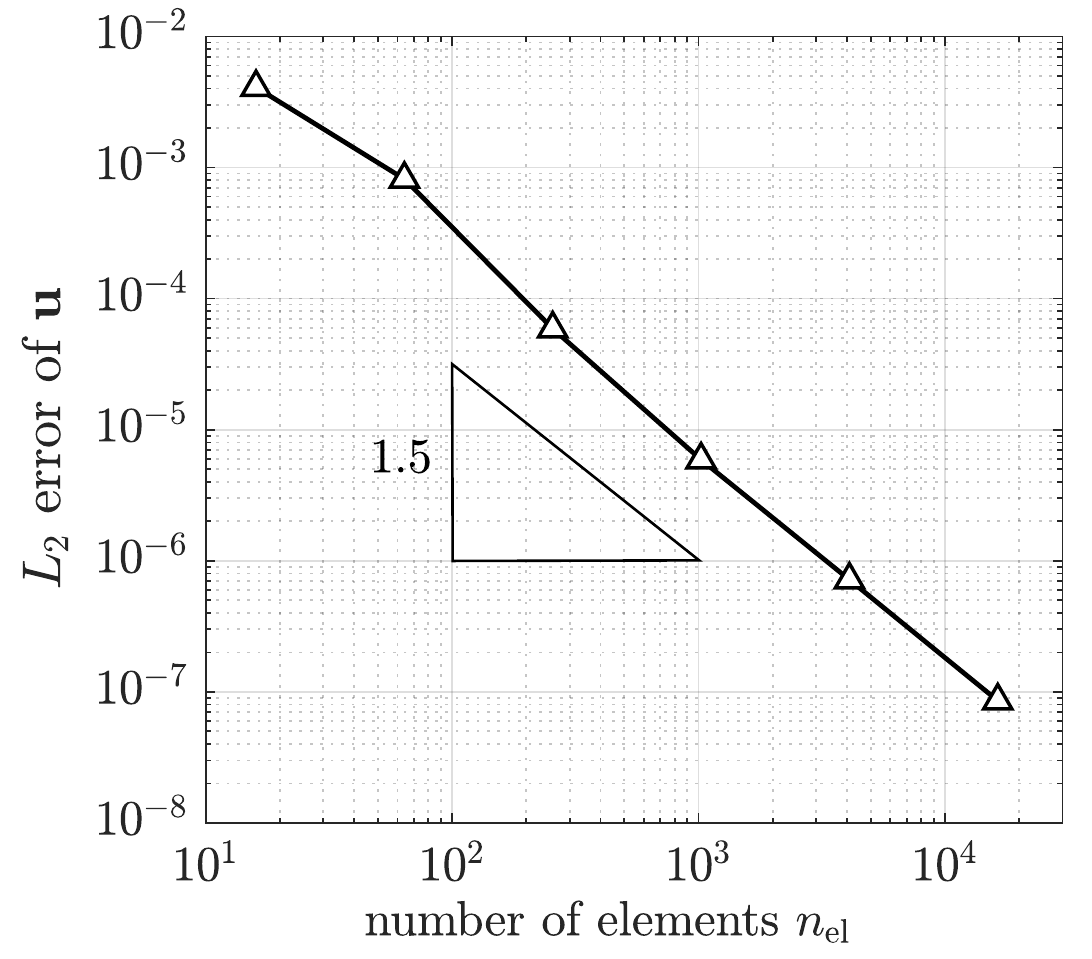}}
\put(-7.4,0){\small{a.}}
\put(.5,0){\small{b.}}
\end{picture}
\caption{Uniaxial tension: a.~FE convergence of the discrete $L_2$ error $\Vert\textbf{u}_{\mathrm{exact}}-\textbf{u}_{\mathrm{FE}}\Vert/\Vert\textbf{u}_{\mathrm{exact}}\Vert$, where $\textbf{u}_{\mathrm{exact}}$ is the FE solution for $n_{\mathrm{el}}=512\times512$ elements; b.~FE convergence for the case without corner singularities.}
\label{fig:uniaxial2333}
\end{center}
\end{figure}

 The incompressible Neo-Hooke model \eqref{eq:tauneohooke} is used with the heterogeneous reference shear modulus distribution
\begin{equation}\label{eq:uniaxial_distr}
   \mu_{\mathrm{ref}}(X,Y) = \left\{ \begin{array}{lcl}
\mu_0 & \mbox{for}
& R \geq R_0\,, \\ \mu_0 + \displaystyle\frac{\Delta\mu_1}{2} \cdot \bigg(1 + \cos\bigg(\displaystyle\pi\frac{R}{R_0}\bigg)\bigg) & \mbox{for} & R < R_0\,,
\end{array}\right.
\end{equation}
where $\Delta\mu_1 = \mu_0, R^2 := X^2 + Y^2$ and $R_0=0.35L$, see Fig.~\ref{fig:uniaxialother}a. 
$L$ and $\mu_0$ are used for normalization and do not need to be specified. Based on the separate convergence study shown in Fig.~\ref{fig:uniaxial2333}a, $n_{\mathrm{el}}=16\times16$ FE are chosen for most of the following cases, since the FE analysis error is below $2.0\cdot10^{-4}$. As seen, the convergence rate is only linear for this example, even though quadratic NURBS are used. This is due to the four inherent corner singularities. If corner singularities are avoided, e.g.~by fixing the Y-direction at the top and bottom edges ($Y=0,Y=L$) the ideal convergence rate $O\big(n_\mathrm{el}^{-1.5}\big) = O(h^3)$ is obtained \citep{strangfix1973}, see Fig.~\ref{fig:uniaxial2333}b.
\par
\begin{figure}[!ht]
\begin{center} \unitlength1cm
\begin{picture}(0,6)
\put(-8,-.2){\includegraphics[width=80mm]{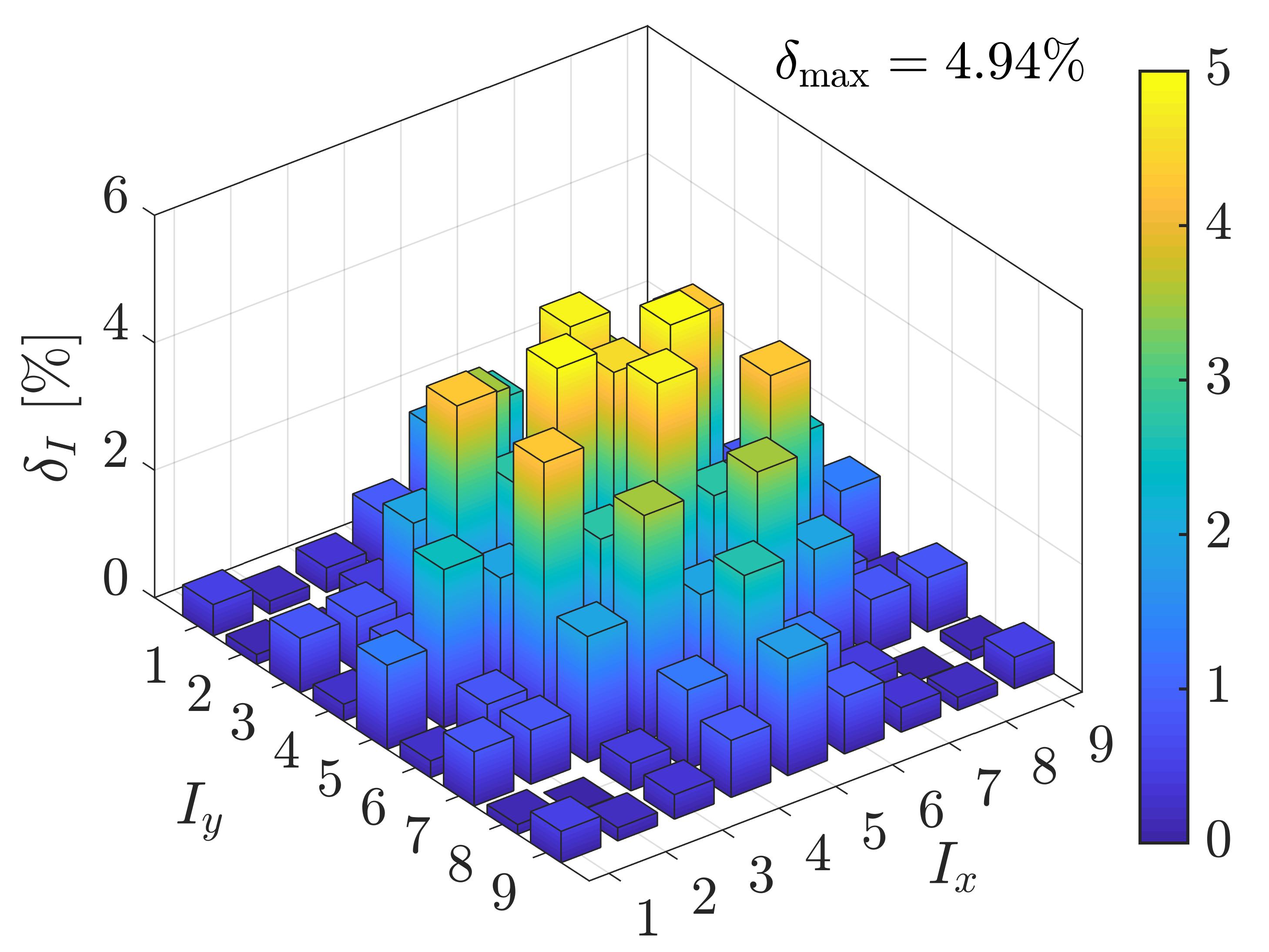}}
\put(0,-.2){\includegraphics[width=80mm]{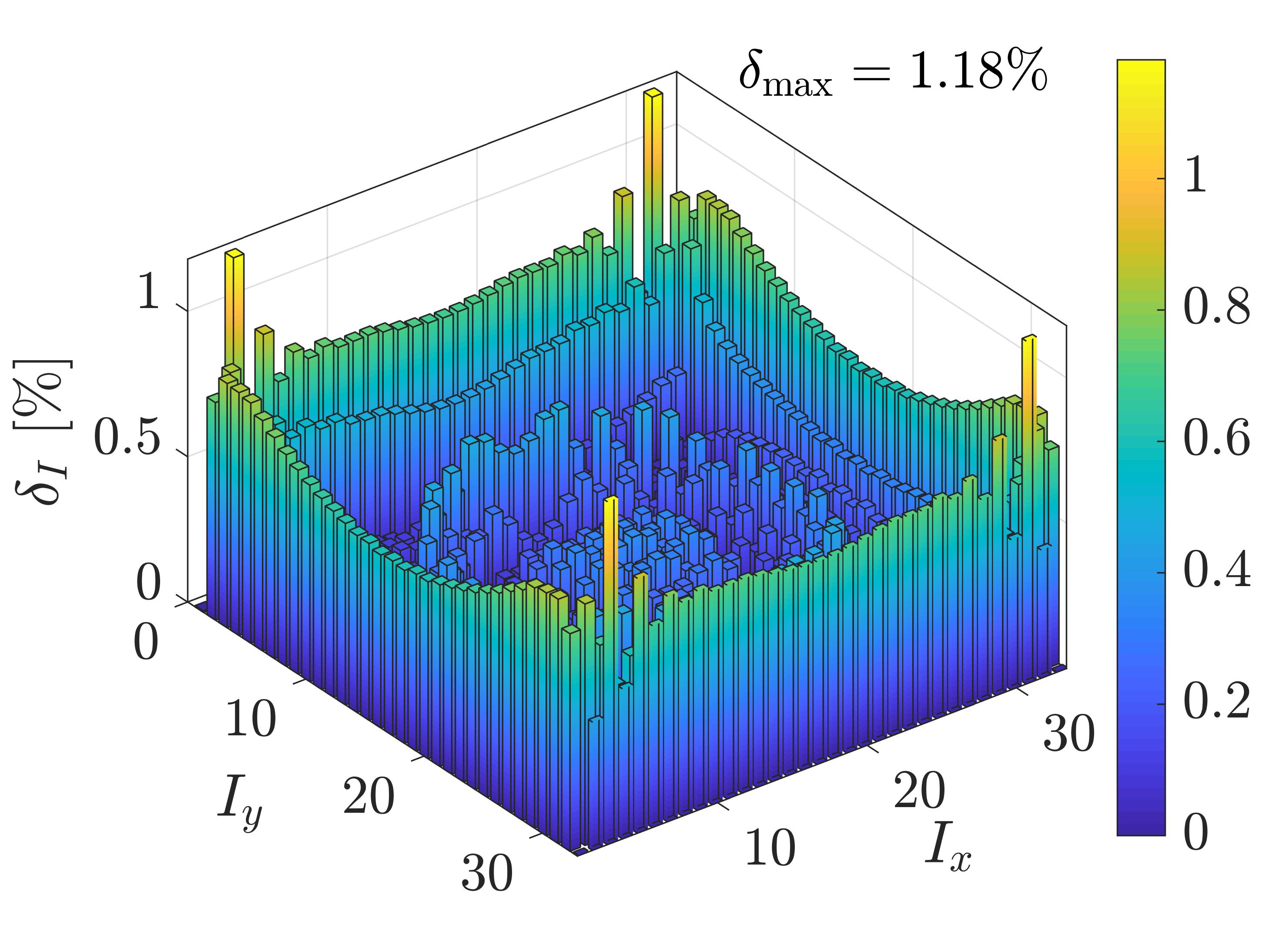}}
\put(-7.4,0){\small{a.}}
\put(1,0){\small{b.}}
\end{picture}
\caption{Uniaxial tension: Error between identified and reference $\mu_I$ values for: a.~Case 1.2 with $n_{\mathrm{var}}=81$ unknowns; b.~Case 1.4 with $n_{\mathrm{var}}=1089$ unknowns.}
\label{fig:uniaxialresults}
\end{center}
\end{figure}

 The material reconstruction mesh varies between $\bar{n}_{\mathrm{el}}=8\times8$ ($n_{\mathrm{var}}=\bar{n}_{\mathrm{no}}=9\times9$) and $\bar{n}_{\mathrm{el}}=32\times32$ ($n_{\mathrm{var}}=\bar{n}_{\mathrm{no}}=33\times33$). Exploiting symmetry, only 1/4 of the material elements can be used in this particular example to speed-up computations (see Fig.~\ref{fig:uniaxialother}b). However, it should be noted that in general the material symmetry is not known a priori, and hence should not be used. The net reaction force $R_x$ is included in the objection function in order to ensure that the problem is well-posed. Objective minimization is done with the chosen lower and upper bounds $\mu_{\mathrm{min}} = 0.1\,\mu_0$ and $\mu_{\mathrm{max}} = 5.0\,\mu_0$. The initial estimate is a vector of random numbers from the range $[\mu_{\mathrm{min}},\mu_{\mathrm{max}}]$. The reconstruction results for different noise levels are presented in Tab.~\ref{tab:1} in terms of the maximum, $\delta_{\mathrm{max}}$, and average, $\delta_{\mathrm{ave}}$, of the relative error
\begin{equation}
    \delta_I = \bigg| \frac{q_{I,\mathrm{ref}}-q_{I,\mathrm{opt}}}{q_{I,\mathrm{ref}}}\bigg|\,, \qquad I = 1,...\,,\bar{n}_{\mathrm{no}}
\end{equation}
between reference and estimated parameters. The error distribution for selected cases is shown in Fig.~\ref{fig:uniaxialresults}. As long as there is no noise, the error distribution is symmetric, even if symmetry is not exploited computationally, as is the case in Fig.~\ref{fig:uniaxialresults}.\par
\begin{figure}[!ht]
\begin{center} \unitlength1cm
\begin{picture}(0,5.7)
\put(-7.8,-.1){\includegraphics[width=78mm]{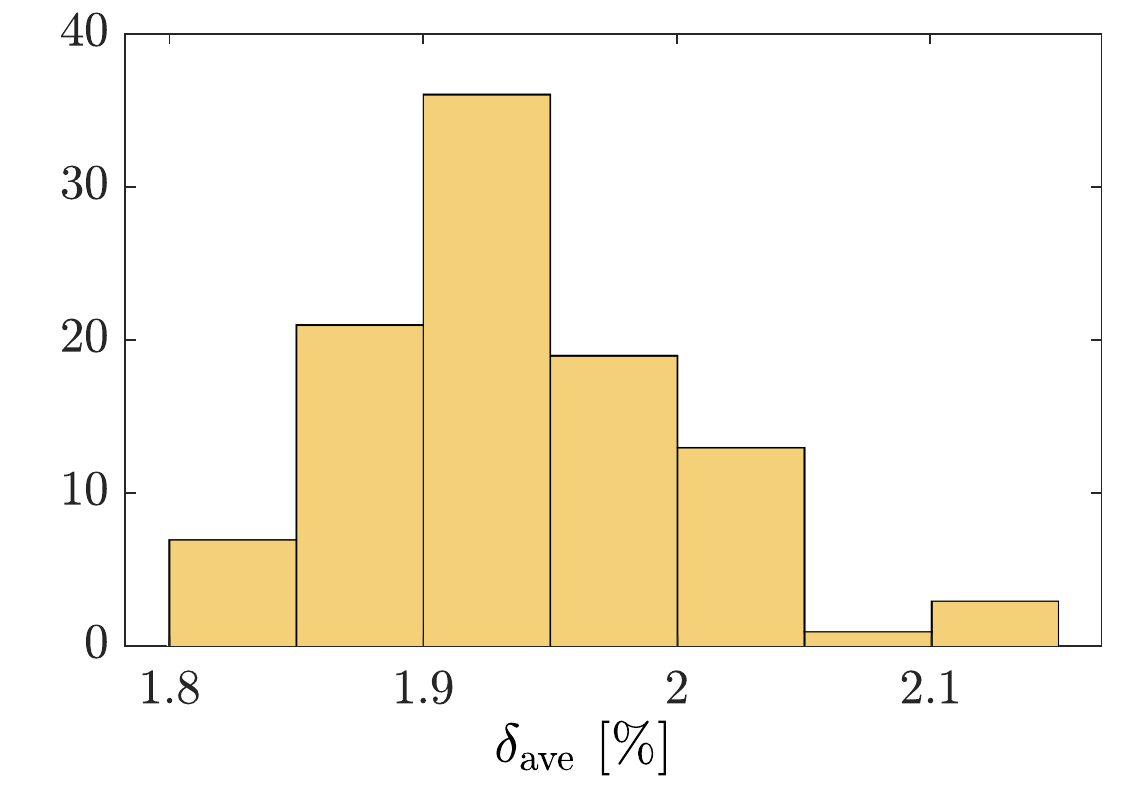}}
\put(.2,.1){\includegraphics[width=75mm]{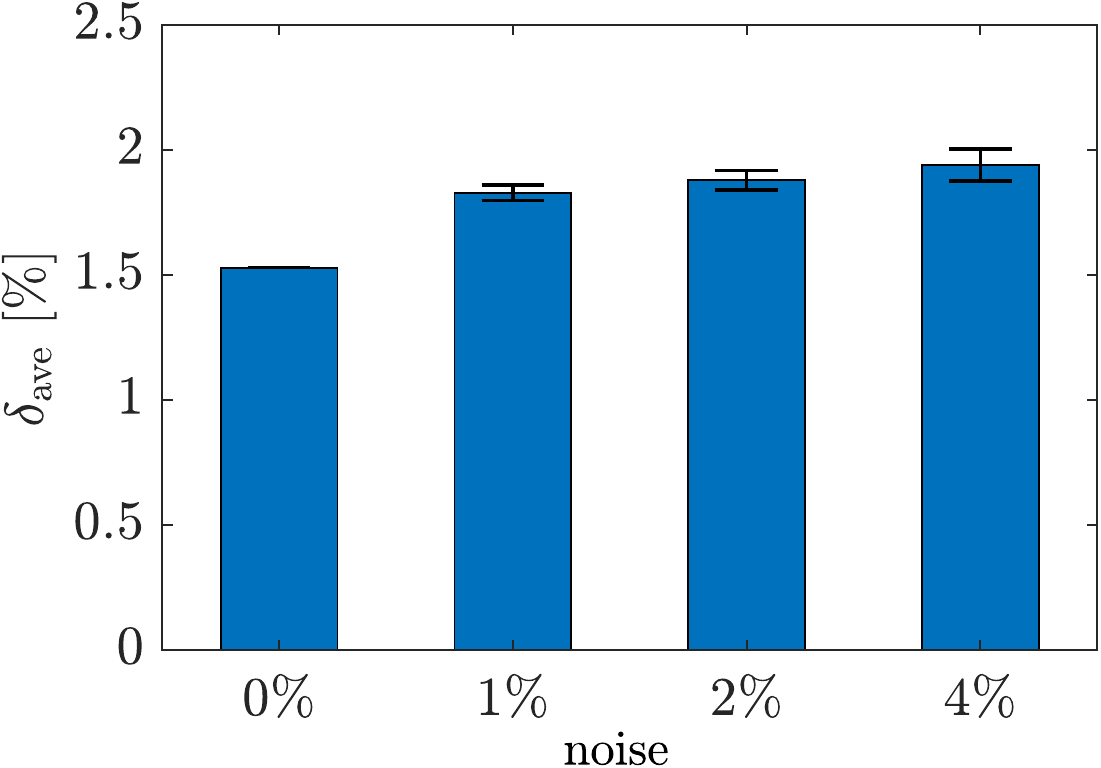}}
\put(-7.4,0){\small{a.}}
\put(.8,0){\small{b.}}
\end{picture}
\caption{Uniaxial tension: a.~histogram of $\delta_{\mathrm{ave}}$ for 100 runs of Case 1.\textcolor{col2}{11}; b.~error $\delta_{\mathrm{ave}}$ with its standard deviation for 0\%  (Case 1.2), 1\%  (Case 1.\textcolor{col2}{6}), 2\%  (Case 1.\textcolor{col2}{9}) and 4\% (Case 1.\textcolor{col2}{11}) noise.}
\label{fig:uniaxial_statistics}
\end{center}
\end{figure}

\begin{table}[!ht]
\centering
\begin{tabulary}{\textwidth}{CCCCCCCCC}
\toprule
Case & FE & mat. & \textcolor{col2}{mat.} & exp. & load & noise & $\delta_\mathrm{{max}}$  & $\delta_\mathrm{{ave}}$ \\
 & $n_{\mathrm{el}}$ & \textcolor{col2}{$\bar{n}_{\mathrm{el}}$} & \textcolor{col2}{$n_{\mathrm{var}}$} & $n_{\mathrm{exp}}\textcolor{col2}{/\mathrm{n_{ll}}}$ & $\mathrm{n_{ll}}$ & $[\%]$ & $[\%]$ & $[\%]$  \\
\midrule
1.1 & $16\times16$ & $4\times4$ & \textcolor{col2}{25} & \textcolor{col2}{$130^2$} & 1 & 0 &24.45  & 3.72   \\
\rowcolor{lightgray} 1.2 & $16\times16$ & $8\times8$ & \textcolor{col2}{81} & \textcolor{col2}{$130^2$} & 1 & 0 & 4.94  & 1.53  \\
1.3 & $16\times16$ & $16\times16$ & \textcolor{col2}{289} & \textcolor{col2}{$130^2$} & 1 & 0 &  1.81  & 0.54 \\
1.4 & $32\times32$ & $32\times32$ & \textcolor{col2}{1089} & \textcolor{col2}{$130^2$} & 1 & 0 & 1.18  & 0.24 \\
\midrule
1.5 & $16\times16$ & $8\times8$ & \textcolor{col2}{25} & \textcolor{col2}{$130^2$} & 1 & 1 & $5.64\pm0.49$ & $1.99\pm0.17$  \\
\rowcolor{lightgray}\textcolor{col2}{1.6} & \textcolor{col2}{$16\times16$} & \textcolor{col2}{$8\times8$} & \textcolor{col2}{25} & \textcolor{col2}{$514^2$} & \textcolor{col2}{1} & \textcolor{col2}{1} & \textcolor{col2}{$5.01\pm0.12$} & \textcolor{col2}{$1.83\pm0.031$}  \\
\textcolor{col2}{1.7} & $16\times16$ & $8\times8$ & \textcolor{col2}{25} & \textcolor{col2}{$130^2$} & 1 & 2 & $6.34\,\pm\,0.85$ & $2.29\,\pm\,0.23$ \\
\textcolor{col2}{1.8} & $16\times16$ & $8\times8$ & \textcolor{col2}{25} & \textcolor{col2}{$130^2$} & 2 & 2 & $6.22\,\pm\,0.81$ & $2.12\,\pm\,0.16$  \\
\rowcolor{lightgray} \textcolor{col2}{1.9} & $16\times16$ & $8\times8$ & \textcolor{col2}{25} & \textcolor{col2}{$514^2$} & 2 & 2 & $5.08\,\pm\,0.14$ & $1.88\,\pm\,0.039$ \\
\midrule
\textcolor{col2}{1.10} & $16\times16$ & $8\times8$ & \textcolor{col2}{25} & \textcolor{col2}{$514^2$} & 2 & 4 & $5.29\,\pm\,0.26$ & $1.93\,\pm\,0.072$   \\
\rowcolor{lightgray} \textcolor{col2}{1.11} & $16\times16$ & $8\times8$ & \textcolor{col2}{25} & \textcolor{col2}{$514^2$} & 4 & 4 & $5.05\pm0.22$ & $1.94\pm0.064$ \\
\textcolor{col2}{1.12} & $16\times16$ & $16\times16$ & \textcolor{col2}{289} & \textcolor{col2}{$514^2$} & 4 & 4 & \textcolor{col2}{$16.21\,\pm\,3.01$} & \textcolor{col2}{$3.53\,\pm\,0.54$}  \\
\bottomrule
\end{tabulary}
\caption{Uniaxial tension: Studied inverse analysis cases and their errors $\delta_{\mathrm{max}}$ and $\delta_{\mathrm{ave}}$ for different FE meshes, material meshes, number of design variables, experimental grid resolutions, load levels and noise levels. 
At least 25 repetitions were used for the statistical analysis of Cases 1.5--1.12. 
The highlighted four cases are compared in Fig.~\ref{fig:uniaxial_statistics}b. 
Cases 1.5--1.11 use material mesh symmetry resulting in lower $n_{\mathrm{var}}$.}
\label{tab:1}
\end{table}
\normalsize
As Tab.~\ref{tab:1} shows, for Cases 1.1--1.4 with 0$\%$ noise, the reconstruction error decreases when refining FE and material meshes -- ultimately down to $\delta_{\mathrm{ave}} = 0.24\%$ for Case 1.4 (also shown in Fig.~\ref{fig:uniaxialresults}b). The results are for $\Delta\mu_1 = \mu_0$. Other $\Delta\mu_1$ values, including $\Delta\mu_1<0$, give similar results. Case 1.2 has been repeated for various initial estimates resulting all in the same error distribution (Fig.~\ref{fig:uniaxialresults}a). Hence, no sensitivity w.r.t.~the initial estimate is observed, and the constant initial estimate $\mu=\mu_0$ is considered in all further cases.\\
Cases 1.5--1.\textcolor{col2}{12} examine the reconstruction for experimental data with the addition of 1--4\% noise according to \eqref{eq:noise}. The statistical effect of random noise is taken into account by repeating each case \textcolor{col2}{25} times. Here, $\delta_{\mathrm{max}}$ and $\delta_{\mathrm{ave}}$ are obtained in each run, and then the mean and standard deviation of all $\delta_{\mathrm{max}}$ and $\delta_{\mathrm{ave}}$ are calculated, see Tab.~\ref{tab:1}. \\
The reconstruction algorithm is expected to overcome noise by incorporating a sufficient number of measurements into the objection function, either by refining the experimental grid or by increasing the number of considered load levels. This is confirmed by Cases 1.5--1.\textcolor{col2}{12}. Eventually, a similar error is achieved for $4\%$ noise (Case 1.\textcolor{col2}{11}) as for zero noise (Case 1.2) when $514\times514$ experimental points and \textcolor{col2}{four} load levels (2\textcolor{col2}{5}\%, 50\%, 75\% and 100\% of the load) are used. For Case 1.\textcolor{col2}{11} errors $\delta_\mathrm{max}$ and $\delta_{\mathrm{ave}}$ were found at $5.05\,\pm\,0.22\%$ (mean $\,\pm\,$ standard deviation) and $1.94\,\pm\,0.064\%$, respectively. The histogram of the average error $\delta_{\mathrm{ave}}$ calculated from 100 runs for Case 1.\textcolor{col2}{11} is shown in Fig.~\ref{fig:uniaxial_statistics}a. As seen, the histogram is close to a normal distribution.
\\
The mean error and its standard deviation for Cases 1.2, 1.6, 1.9 and 1.11 (0\%, 1\%, 2\% and 4\% noise, respectively) are compared in Fig.~\ref{fig:uniaxial_statistics}b. No significant difference in the mean of error $\delta_{\mathrm{ave}}$ is observed for 1\%--4\% noise. This indicates that, as long as a sufficiently large experimental dataset is used, the mean error is insensitive to noise. \\Although the available amount of data makes the problem determined even for a larger number of unknowns, further material discretization with a high noise level can lead to oscillations in solution $\mq$ due to \textit{overfitting}, as the comparison between Case 1.11 and Case 1.12 shows. In the latter case the error is much larger, even though the material mesh has been refined. Filtering techniques, known from topology and shape optimization \citep{sigmund1998numerical,bletzinger2014consistent}, can be applied to deal with the overfitting phenomenon. In this sense, a lower number of material unknowns acts like an inherent filter. This shows that, in order to bring the mean reconstruction error below $2\%$, at least $n_{\mathrm{el}} = 16\times16$ analysis elements, $\bar n_{\mathrm{el}} = 8\times8$ material reconstruction elements, $n_{\mathrm{exp}}/n_{\mathrm{ll}} = 514\times514$ and $n_{\mathrm{ll}} = 2$ need to be used in the presence of 4\% noise. \\ 
In all cases, the solution of \eqref{eq:minimization} was found in 9--13 iterations.
The Jacobian based on the analytical sensitivities (see App.~\ref{s:appendix1}) provides a tremendous speed-up over numerical (finite-difference based) Jacobians in case of a large number of unknowns: for Case 1.12 the speed-up is about 32.8.\footnote{For forward finite differences, the cost of one iteration is $n_{\mathrm{var}}$ forward problem evaluations, while for the analytic Jacobian only one forward evaluation per iteration is needed. However, building the Jacobian from \eqref{eq:dUFEdq} and \eqref{eq:dUFE} involves solving \eqref{eq:dudq} -- a system of $n_{\mathrm{var}}$ equations.}
%
%
%
%
%
%
%-------------------------------------- BENDING1 --------------------------------------
\subsection{Pure bending}\label{sec:purebending}
\begin{figure}[!ht]
\begin{center} \unitlength1cm
\begin{picture}(0,12)
\put(-8.2,6.3){\includegraphics[width=85mm]{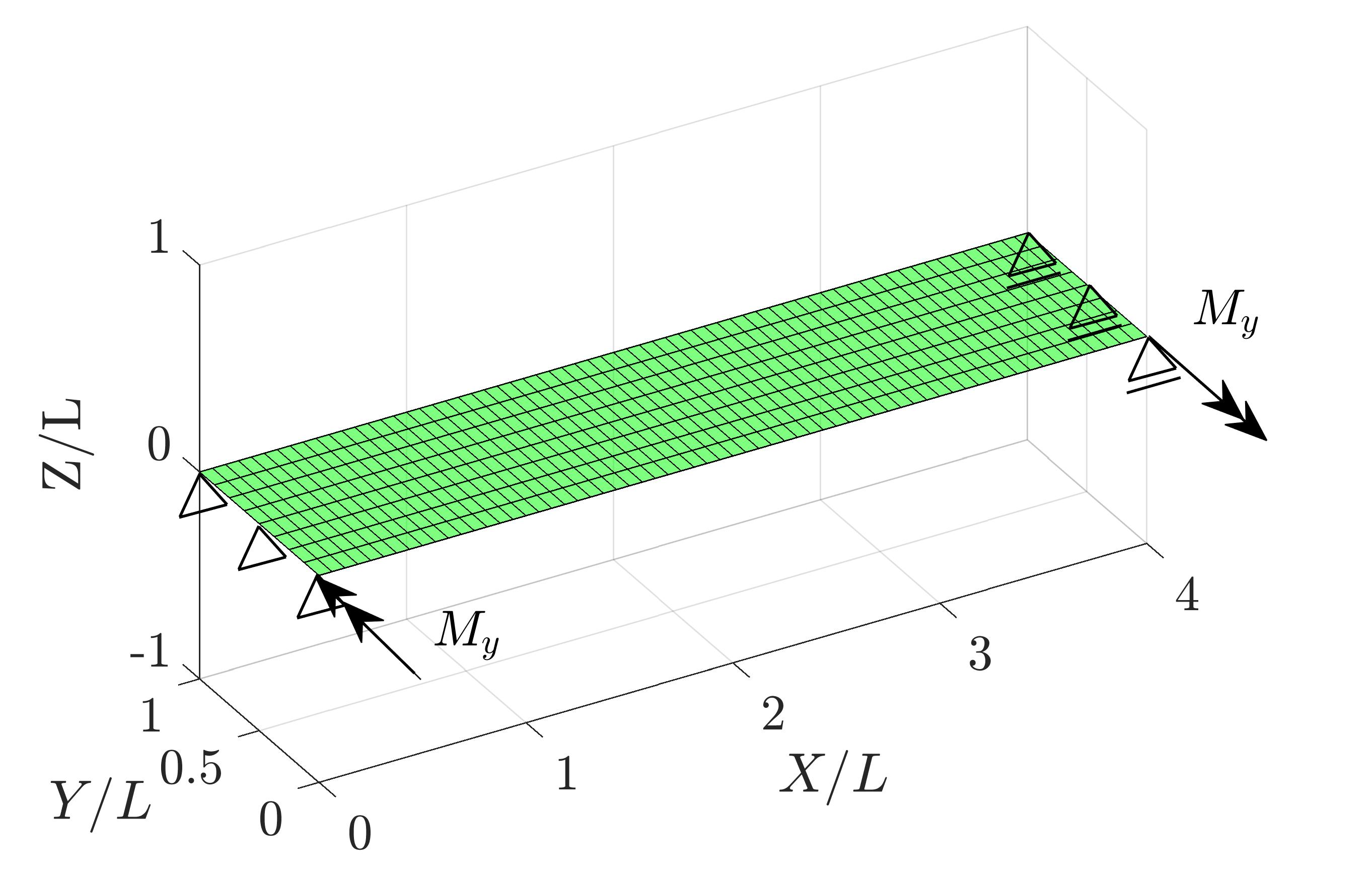}}
\put(.1,6){\includegraphics[width=80mm]{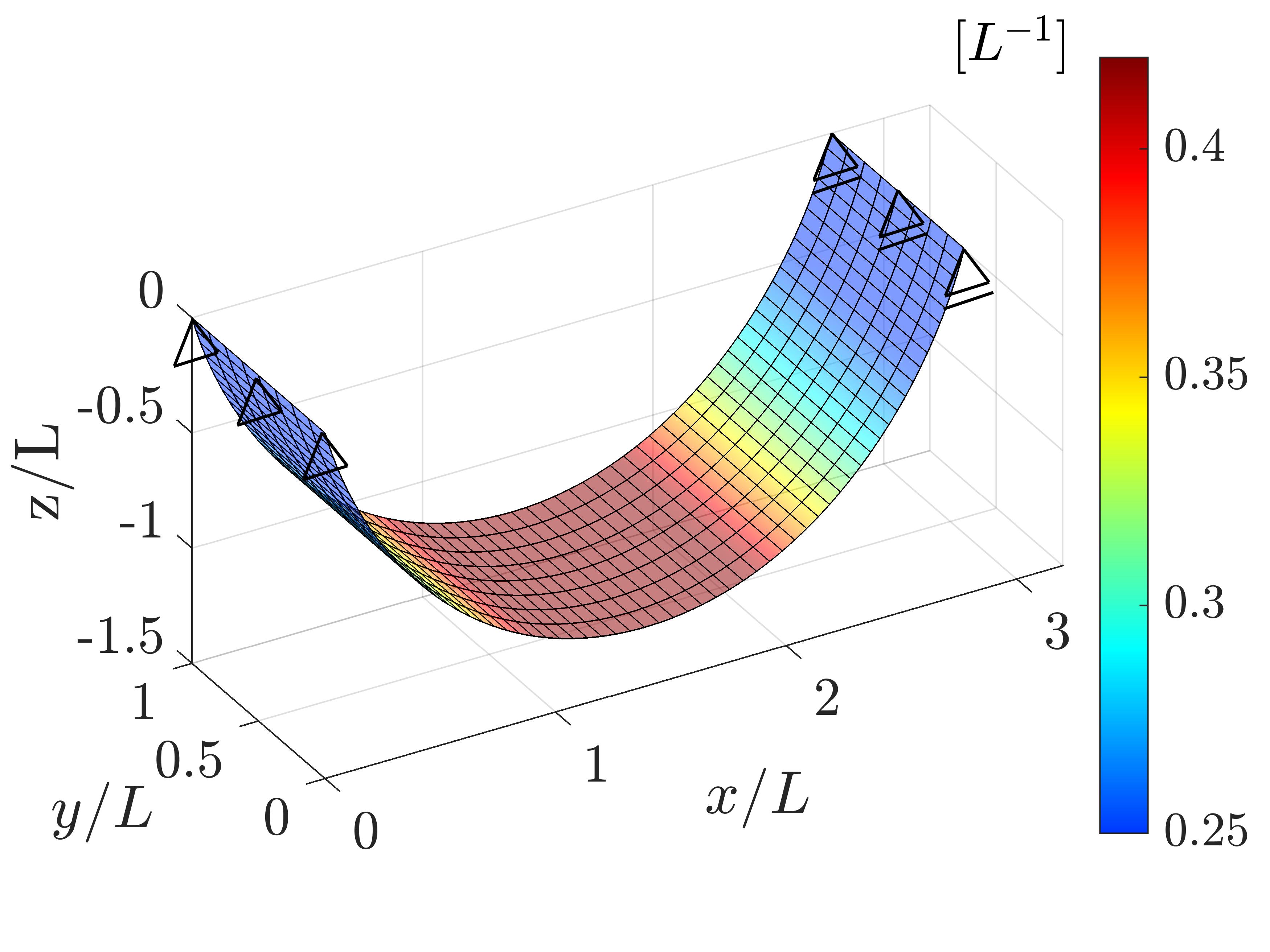}}
\put(-7.4,0.6){\includegraphics[width=73mm]{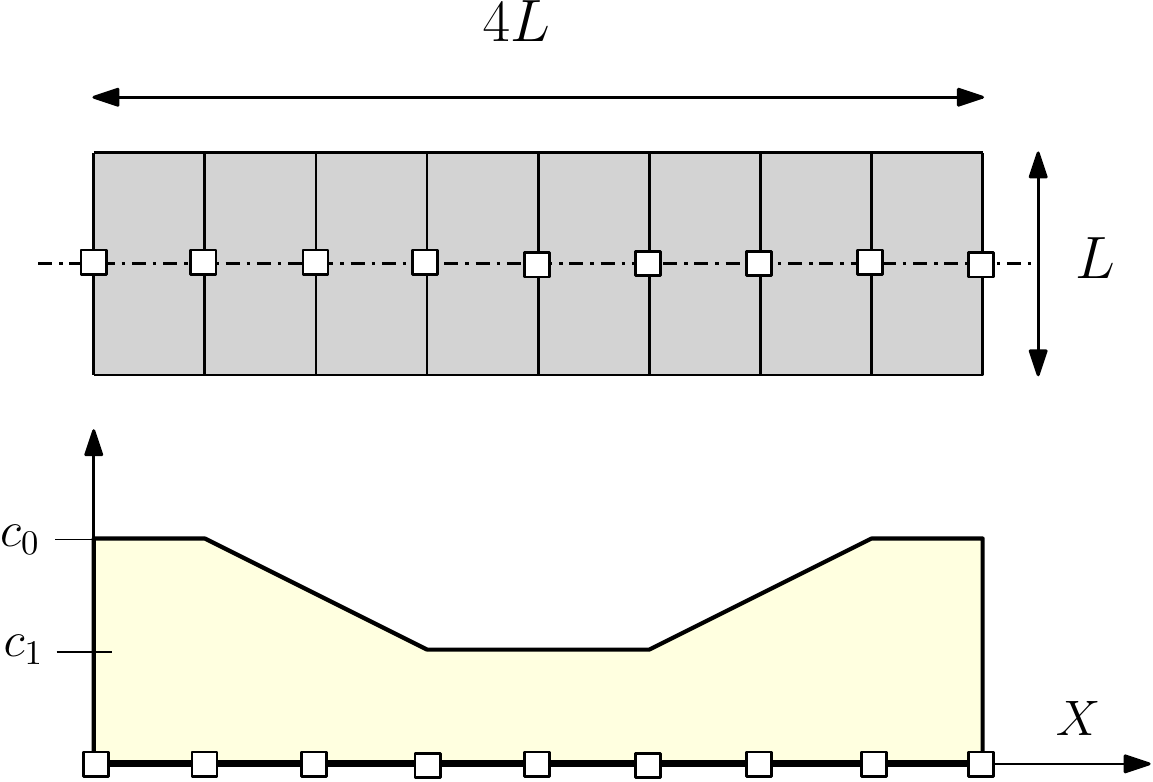}}
\put(.3,0){\includegraphics[width=75mm]{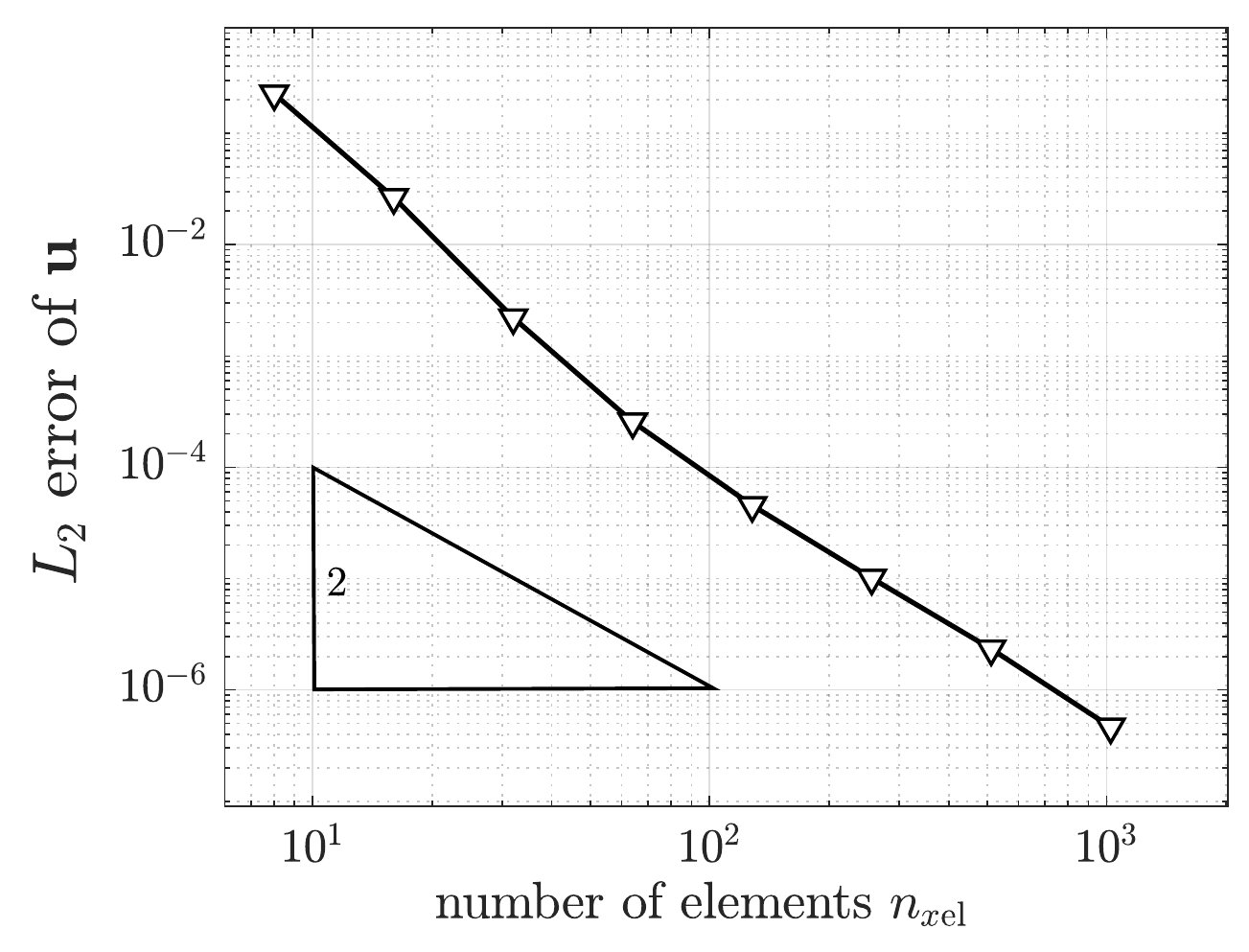}}
\put(-7.5,6){\small{a.}}
\put(0,6){\small{b.}}
\put(-7.5,0){\small{c.}}
\put(1.1,0){\small{d.}}
\end{picture}
\caption{Pure bending (gradual distribution): a.~undeformed configuration and boundary conditions; b.~deformed configuration colored by the mean curvature, ranging between $0.25/L$ and $0.42/L$; c.~material mesh with reference bending stiffness distribution; d.~FE convergence of the discrete $L_2$ error $\Vert\textbf{u}_{\mathrm{exact}}-\textbf{u}_{\mathrm{FE}}\Vert/\Vert\textbf{u}_{\mathrm{exact}}\Vert$ w.r.t.~the refined `exact' FE solution for $n_{x\mathrm{el}} = 4096$ elements.}
\label{fig:bend}
\end{center}
\end{figure}
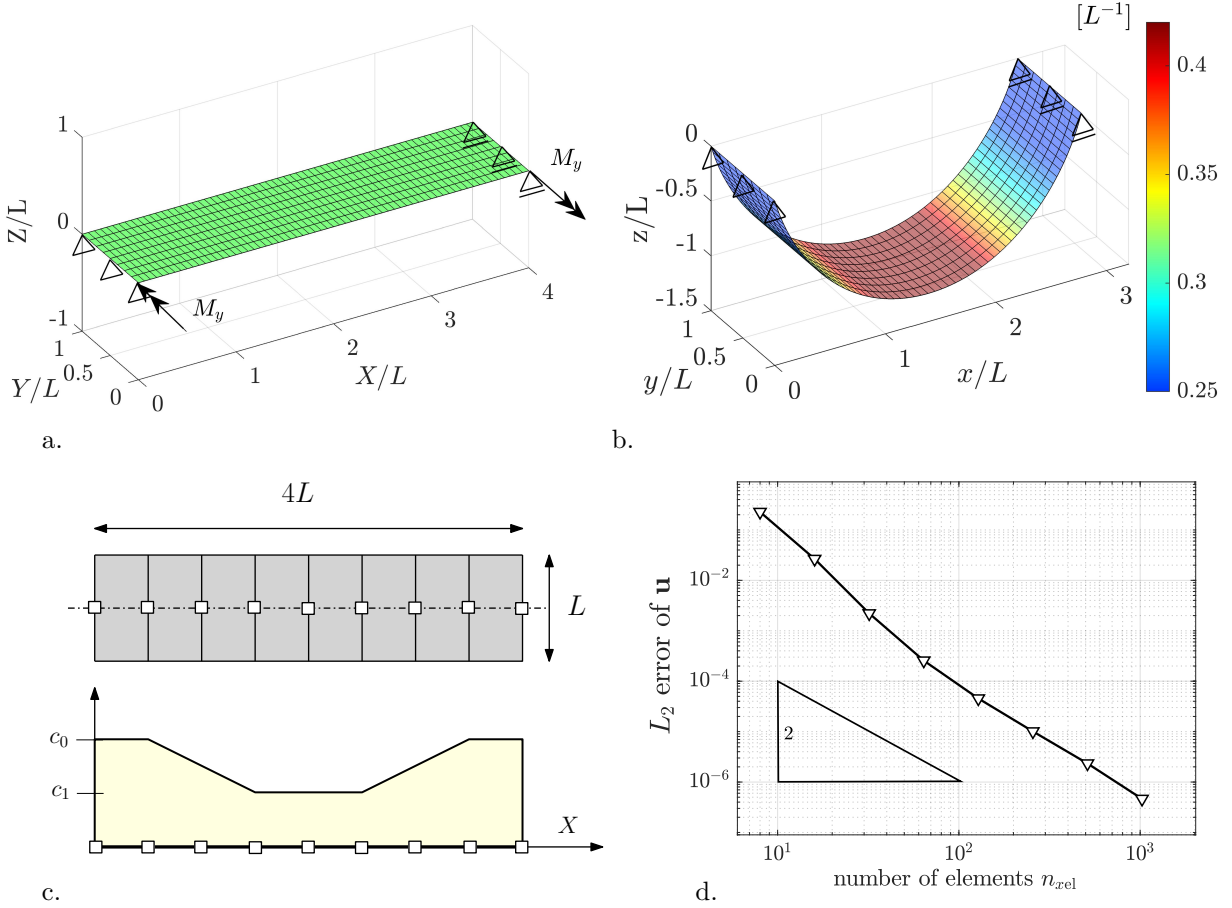

The second example, illustrated in Fig.~\ref{fig:bend}a, considers pure unaxial bending of a thin strip, which allows to study the isolated material reconstruction of bending stiffness $c = c(X)$. 
Two different distributions are examined -- a gradual variation in Sec.~\ref{s:bend_d1} and a discontinuous variation in Sec.~\ref{s:bend_d2}. 
The strip has dimension $L_x\times L_y=4L\times L$. Edge $X=0$ is fully fixed, while edge $X=4L$ is only fixed in $Z$ direction (slider support). Both edges are subjected to the distributed bending moment $M_y = 0.5 \cdot 10^{-4}$ $FL/L$, where $F$ and $L$ are force and length scales that remain unspecified. Since geometry, load and material only vary along $X$, but not along $Y$, the mesh refinement along $X$ is the only relevant discretization parameter in this example.
\subsubsection{Gradual material distribution}\label{s:bend_d1}
\begin{figure}[!ht]
    \centering
    \includegraphics[width=0.5\linewidth]{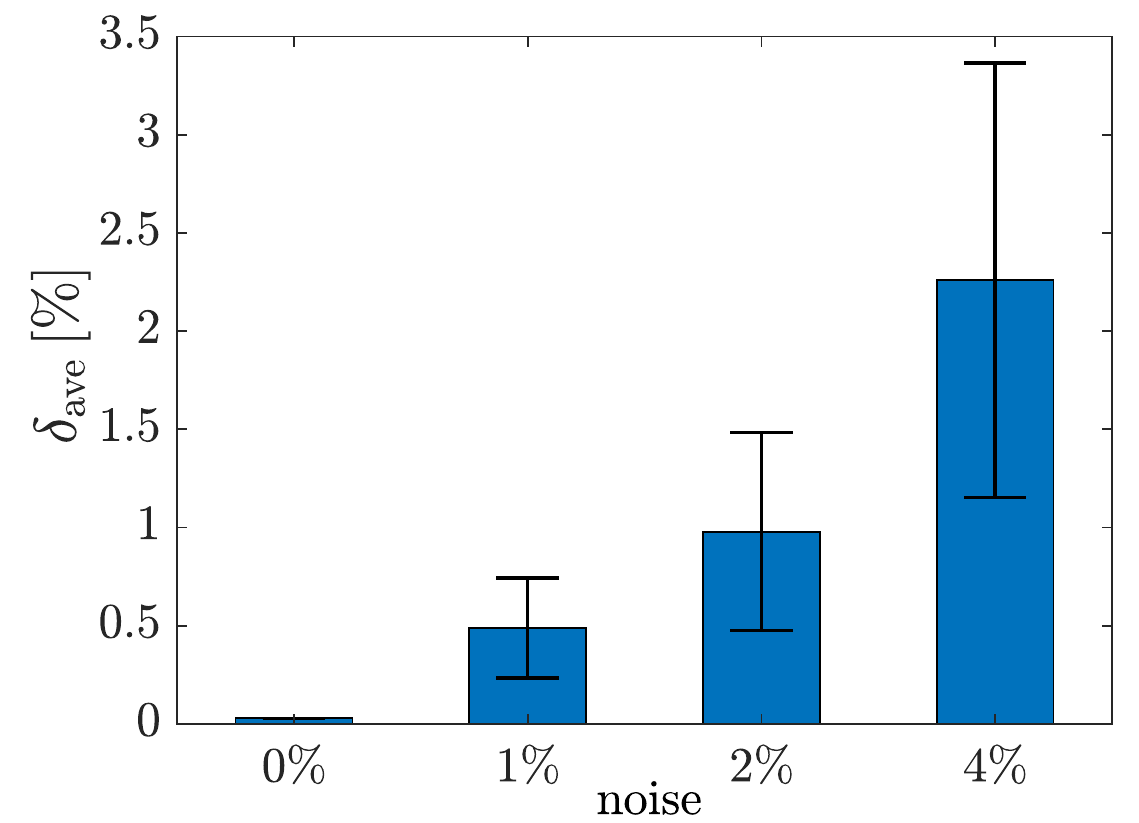}
    \caption{Pure bending (gradual distribution): Mean error and standard deviation of $\delta_{\mathrm{ave}}$ for noise levels 0\%--4\%.}
    \label{fig:bend_res}
\end{figure}

First, the chosen reference bending stiffness is gradually varying in $X$- direction by the piecewise linear function
\begin{equation}\label{eq:benddistr}
   c(X) = \left\{ \begin{array}{rcl}
c_0 & \mbox{for} & X \leq 0.5L\,  \vee X \geq 3.5L, \\
(c_1-c_0)\cdot(X/L-0.5) + c_0 & \mbox{for} & 0.5L<X<1.5L,\\
c_1 & \mbox{for} & 1.5L \leq X \leq 2.5L,\\
(c_0-c_1) \cdot(X/L-2.5) + c_1 & \mbox{for} & 2.5L < X < 3.5L,\\
\end{array}\right.
\end{equation}
with $c_0 = 1.0\cdot 10^{-3} FL$ and $c_1 = 0.6\,c_0 $, shown in Fig.~\ref{fig:bend}c. This material distribution is captured exactly by $\bar n_{x\mathrm{el}} = 8$ material elements ($n_{\mathrm{var}} = \bar n_{x\mathrm{no}} = 9$ material nodes), which will be used for all cases studied here. The shear modulus, which does not affect pure bending, but is required for the well-posedness of the FE model, is chosen as $\mu = F/L$ and considered known ($\bar d = 1$). This example is not a pure Dirichlet problem, as a bending moment is applied, and so reactions forces are not needed in $f$. Based on the separate convergence study shown in Fig.~\ref{fig:bend}d, $n_{x\mathrm{el}} =64$ FE (along the $x$- direction) are chosen as the analysis mesh for the subsequent inverse analysis. 
Experiment-like reference results are generated from four load levels (at 25, 50, 75, 100 [\%] load) using a fine analysis mesh ($n_{x\mathrm{el}}=4096$). Objective minimization is conducted with the bounds $c_{\mathrm{min}} = 0.4\,c_0$ and $c_{\mathrm{max}} = 5.0\,c_0$. The initial estimate for the minimization is a vector of random numbers from the range $[c_{\mathrm{min}},c_{\mathrm{max}}]$.\\
The results for different noise levels are presented in Tab.~\ref{tab:2} and Fig.~\ref{fig:bend_res} using the same error definition as in Sec.~\ref{sec:uni}. Every case was repeated 25 times to show the influence of the random noise distribution.
\begin{table}[!ht]
\centering
\begin{tabular}{ccccccccc}
\toprule
Case & FE & mat. & \textcolor{col2}{mat.} & exp. & \textcolor{col2}{load} & noise  & $\delta_\mathrm{{max}}$  & $\delta_{\mathrm{ave}}$ \\
 & $n_{x\mathrm{el}}$ & $\bar{n}_{x\mathrm{el}}$ & \textcolor{col2}{$n_{\mathrm{var}}$} & $n_{\mathrm{exp}}/n_{\mathrm{ll}}$ & $n_{\mathrm{ll}}$ & $[\%]$  & $[\%]$ & $[\%]$ \\
\midrule
2.1 & $64$ & $8$ & \textcolor{col2}{9} & $4098$ & 1 & 0 &  0.076  & 0.029 \\
2.2 & $64$ & $8$ & \textcolor{col2}{9} & $4098$ & 4 & 1   & $1.31\pm0.68$ & $0.49\pm0.25$ \\
2.3 & $64$ & $8$ & \textcolor{col2}{9} & $4098$ & 4 & 2   & $ 2.62\pm1.36$ & $0.98\pm0.51$ \\
2.4 & $64$ & $8$ & \textcolor{col2}{9} & $4098$ & 4 & 4   & $5.51\pm2.52$ & $2.26\pm1.11$ \\
\bottomrule
\end{tabular}
\caption{Pure bending (gradual distribution): Studied inverse analysis cases with their FE mesh, material mesh, design variables, experimental grid resolution, load level, noise levels, and resulting errors $\delta_{\mathrm{max}}$ and $\delta_{\mathrm{ave}}$. 
25 repetitions were used for the statistical analysis of Cases 2.2--2.4. 
Fig.~\ref{fig:bend_res} shows a graphical representation of $\delta_{\mathrm{ave}}$ for the four cases.}
\label{tab:2}
\end{table}
Since material distribution \eqref{eq:benddistr} is captured exactly by the material mesh and the combined reconstruction error hence becomes negligible ($\delta_{\mathrm{ave}}=0.03\%$ for 0\% noise in Case 2.1), the isolated influence of noise on the material reconstruction can be studied in this example. As Tab.~\ref{tab:2} and Fig.~\ref{fig:bend_res} show, a proportional increase of $\delta_{\mathrm{max}}$, $\delta_{\mathrm{ave}}$ and its standard deviation is observed. In contrast to uniaxial tension, the relation between $\delta_{\mathrm{ave}}$ and the noise level is nearly linear. It is explainable by the fact that noise applied to both the $x$- and $z$- components, as is considered here, also induces membrane deformations, which cannot be eliminated by changing the bending stiffness, i.e.~in-plane membrane deformations are insensitive w.r.t.~the bending stiffness, as Eq.~\eqref{eq:sensitivitymu} also shows. This illustrates that noise cannot be contained if it affects material parameters that are not part of $\mq$. \\The solution of \eqref{eq:minimization} was found in 20--24 iterations for all cases.
%
%
%
%
%
%
%-------------------------------------- BENDING2 --------------------------------------
\subsubsection{Discontinuous material distribution}\label{s:bend_d2}
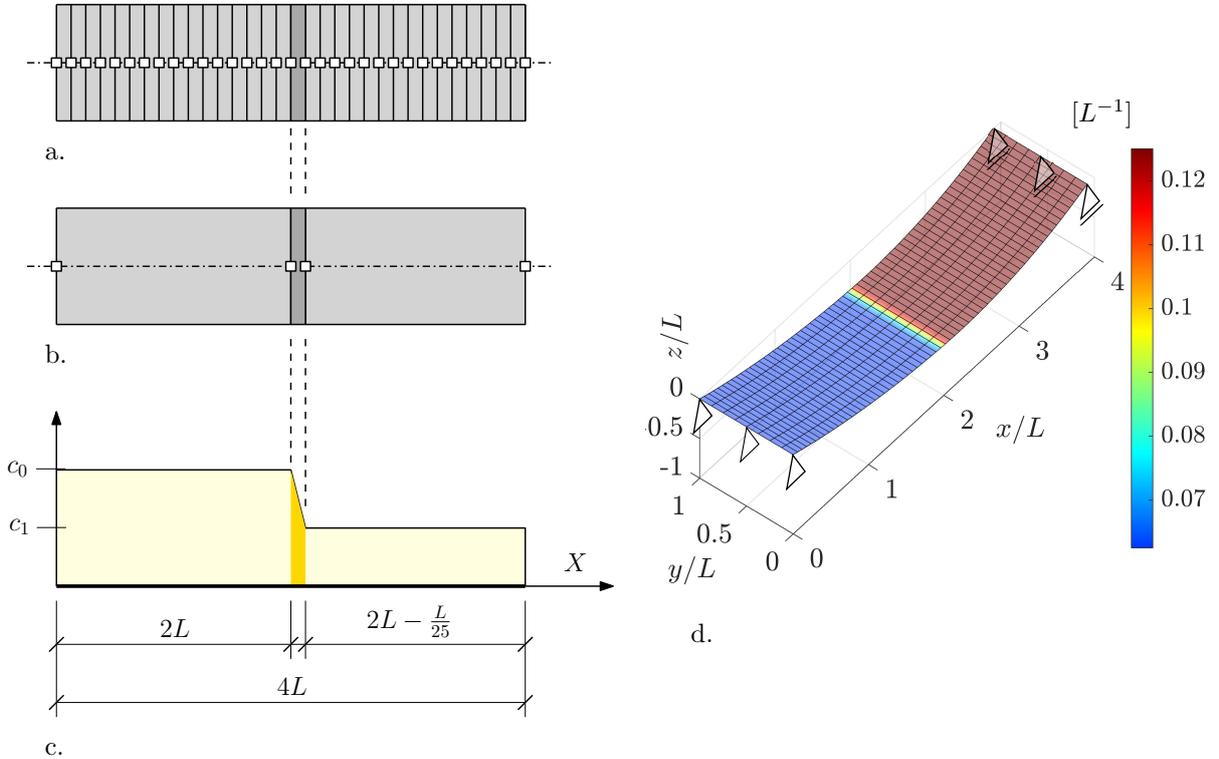
\begin{figure}[!ht]
\begin{center} \unitlength1cm
\begin{picture}(0,10.5)
\put(-8,.5){\includegraphics[width=80mm]{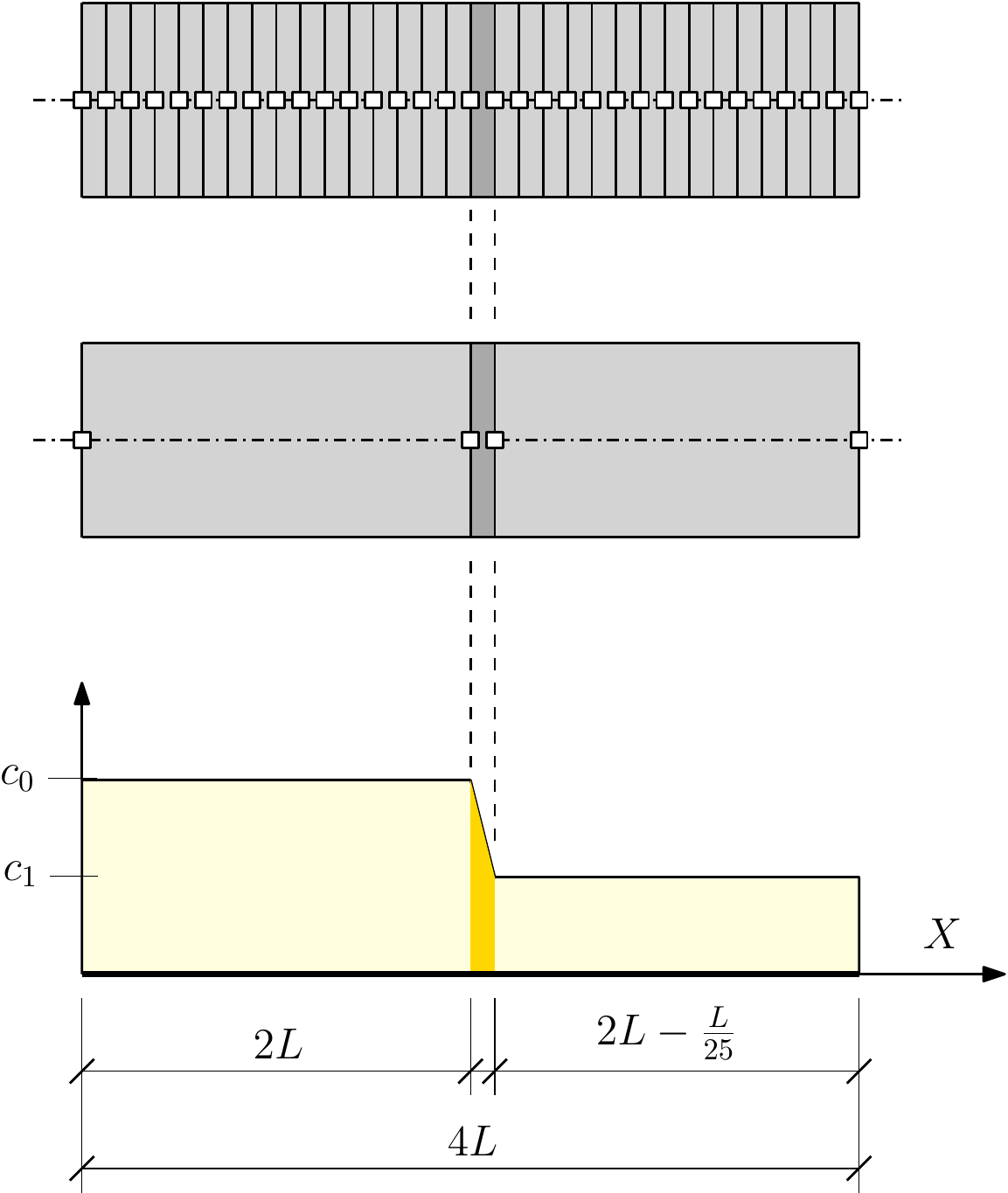}}
\put(0.4,2){\includegraphics[width=74mm]{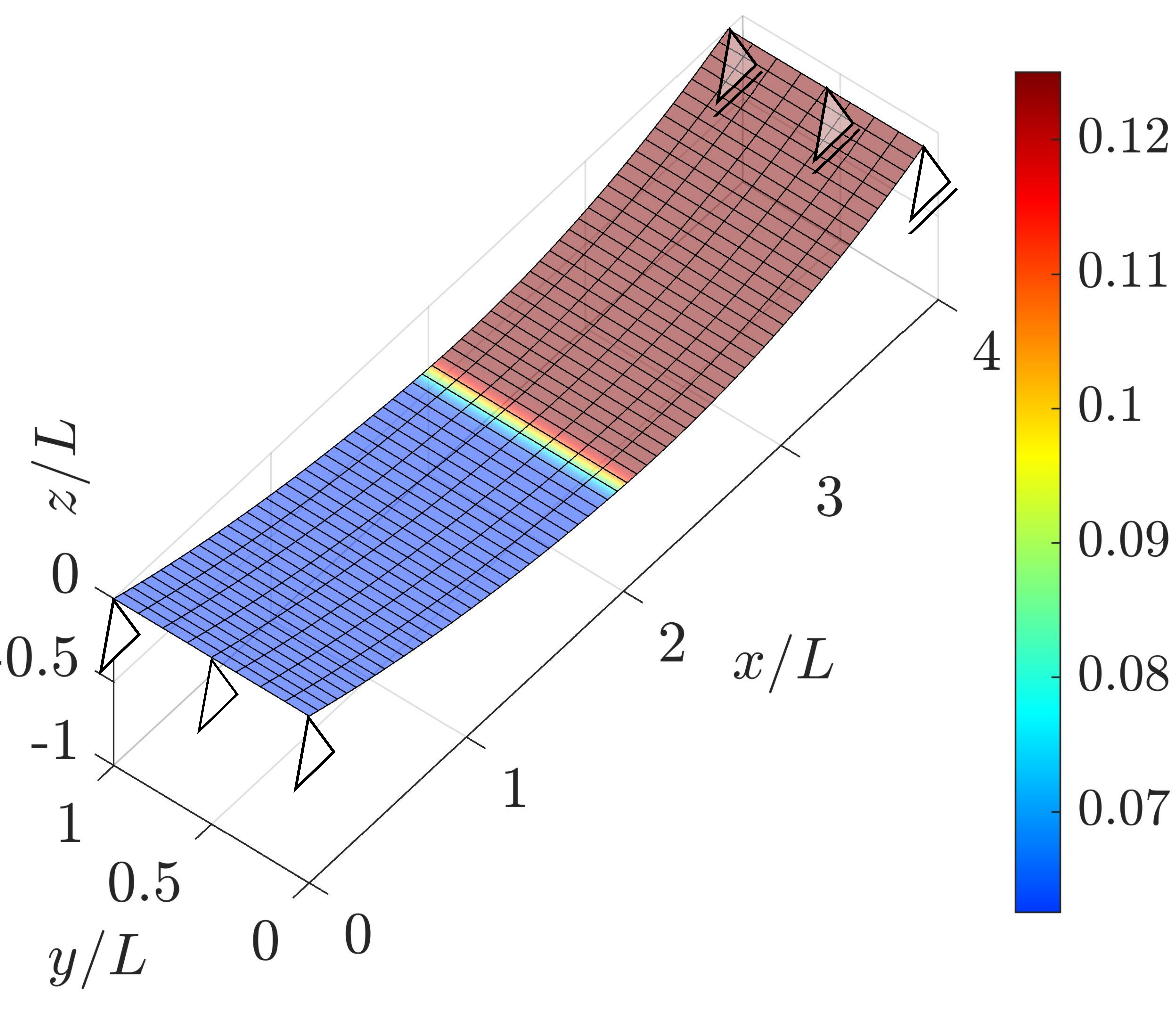}}
\put(-7.5,7.9){\small{a.}}
\put(-7.5,5.2){\small{b.}}
\put(-7.5,0){\small{c.}}
\put(1,1.5){\small{d.}}
\put(6,8.4){\small{$[L^{-1}]$}}
\end{picture}
\caption{Pure bending (discontinuous distribution): uniform (a.) and adapted (b.) material mesh with reference bending stiffness distribution (c.); d.~deformed configuration colored by the mean curvature, ranging between $0.062/L$ and $0.125/L$.}
\label{fig:bend2distribution}
\end{center}
\end{figure}

\begin{figure}[!ht]
\begin{center} \unitlength1cm
\begin{picture}(0,13.4)
\put(-7.9,8.9){\includegraphics[width=77mm]{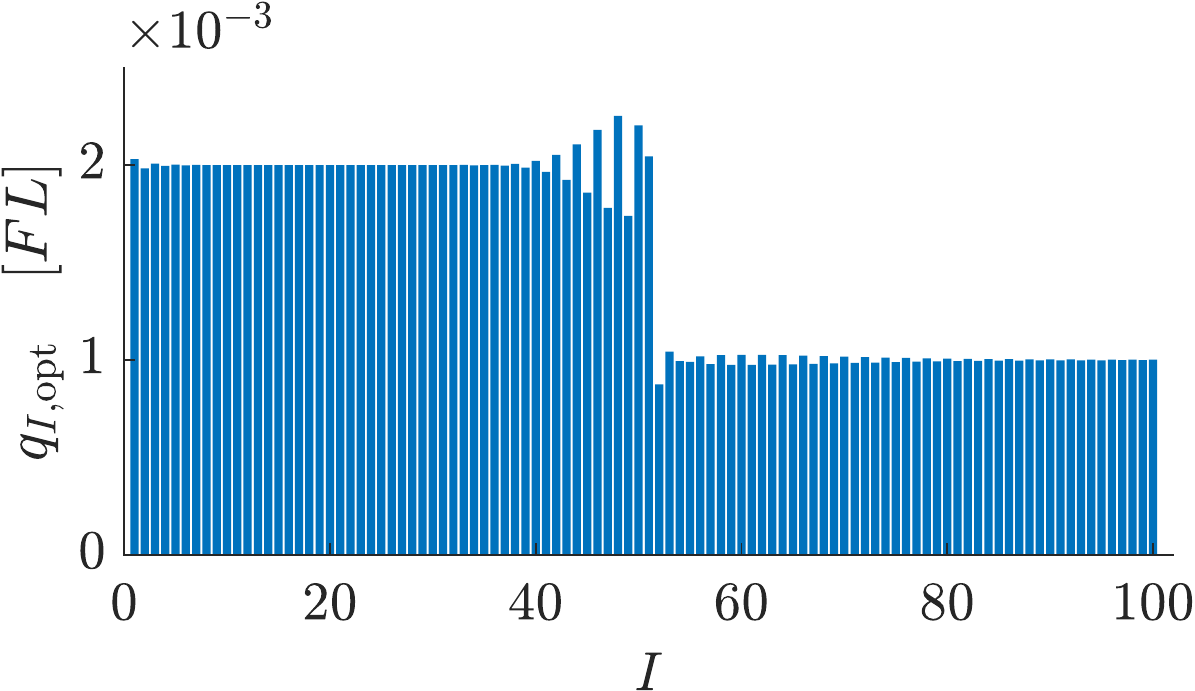}}
\put(0,7){\includegraphics[width=80mm]{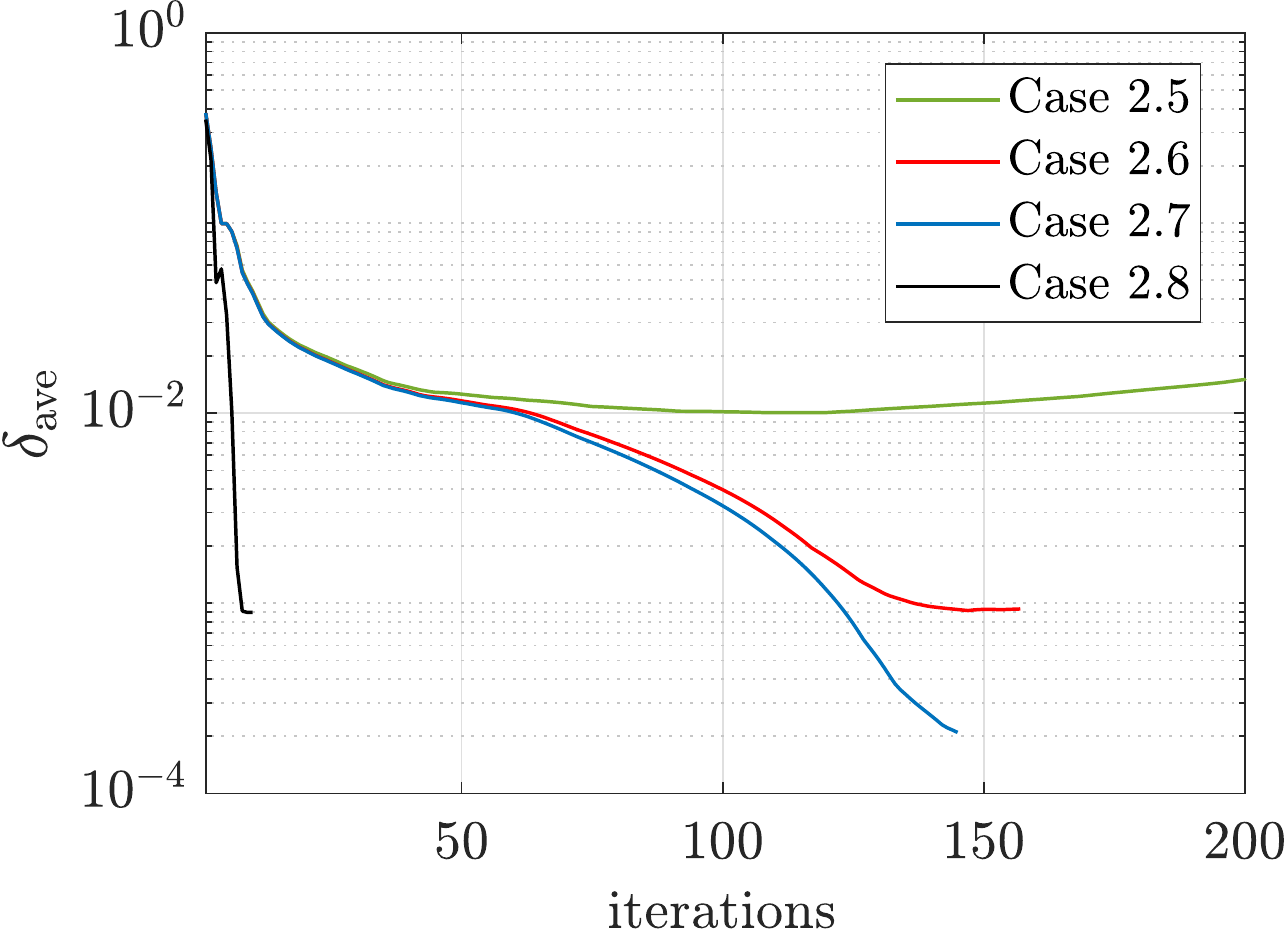}}
\put(-7.9,4.5){\includegraphics[width=77mm]{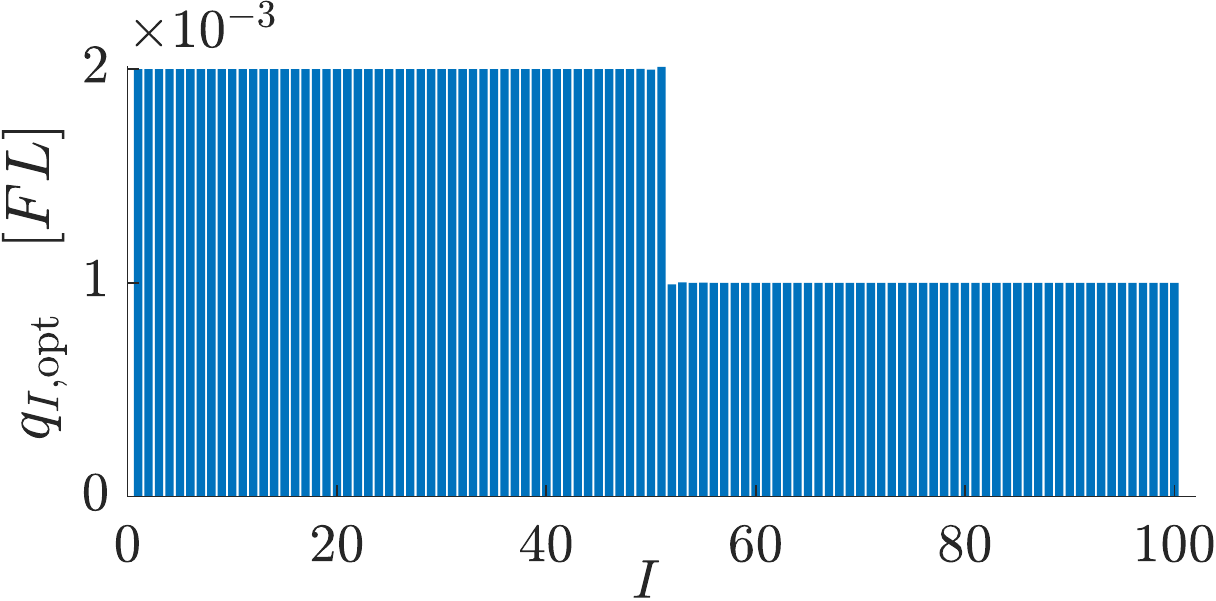}}
\put(-7.9,.2){\includegraphics[width=65mm]{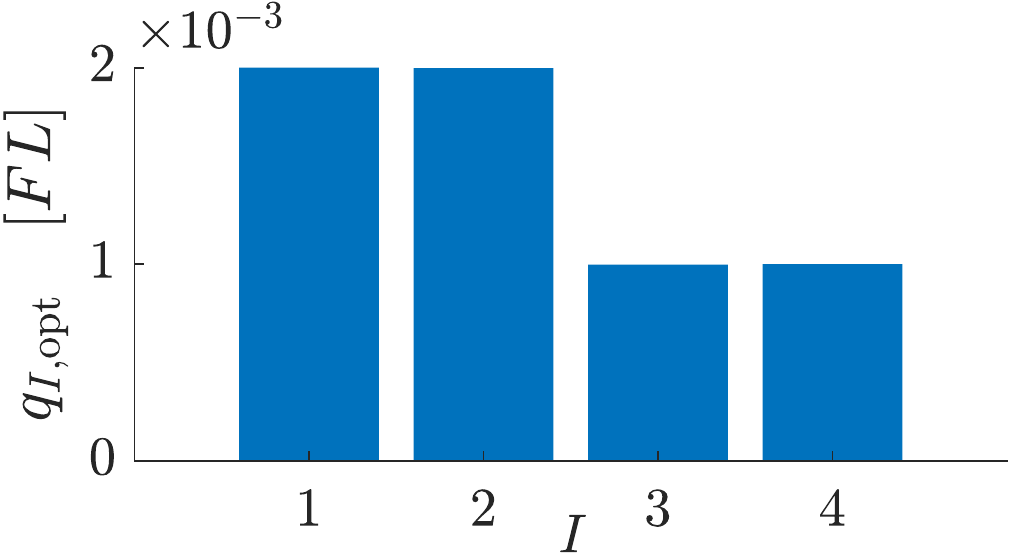}}
\put(0,.3){\includegraphics[width=80mm]{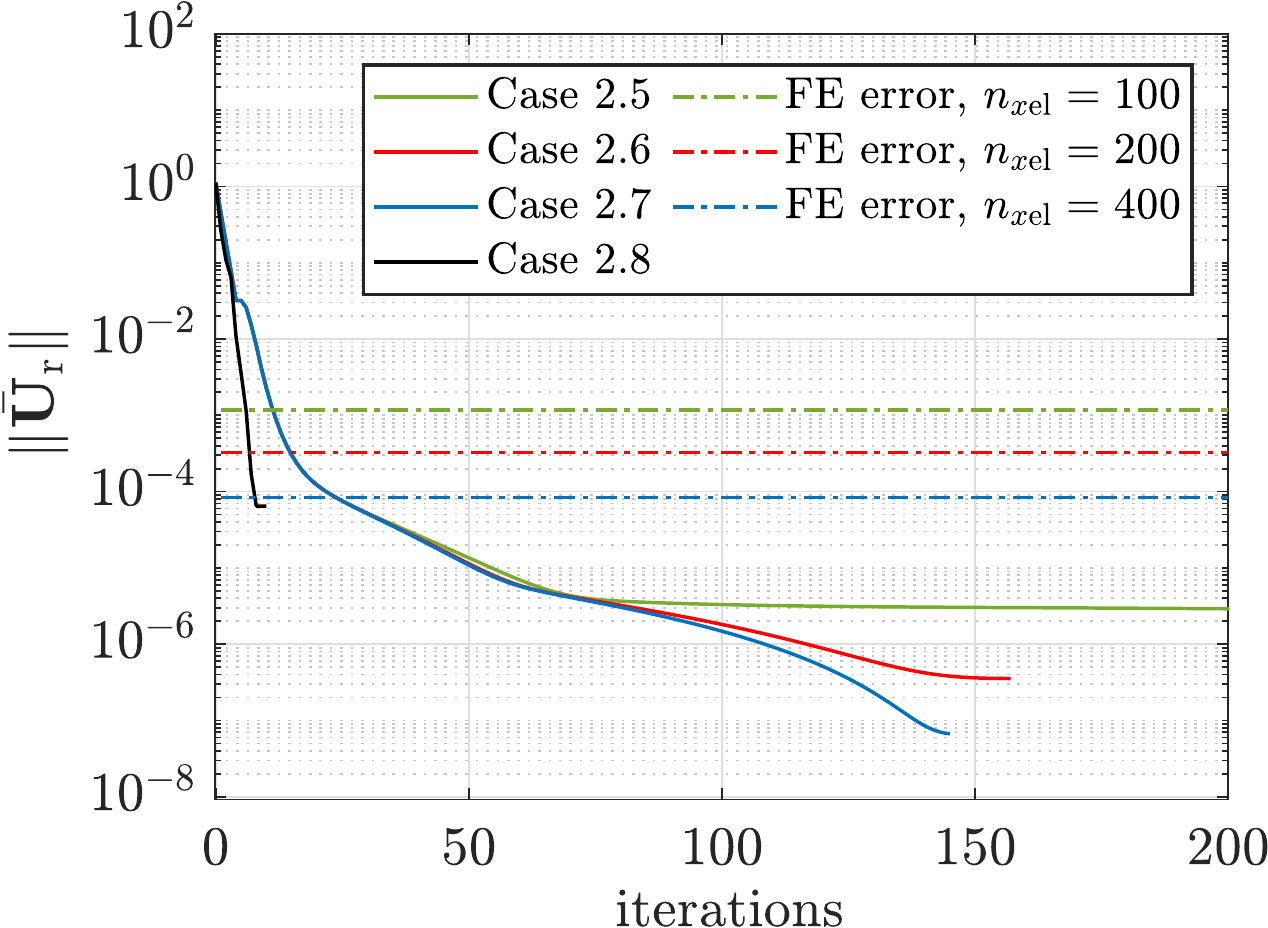}}
\put(-7.5,8.8){\small{a.}}
\put(-7.5,4.2){\small{b.}}
\put(.5,6.8){\small{d.}}
\put(-7.5,0){\small{c.}}
\put(.7,0){\small{e.}}
\end{picture}
\caption{\textcolor{col1}{Pure bending (discontinuous distribution): a.~Case 2.5 (uniform mesh); b.~Case 2.7 (uniform mesh); c.~Case 2.8 (adapted mesh); d.~evolution of reconstruction error $\delta_{\mathrm{ave}}$; e.~objective evolution.}}
\label{fig:basd}
\end{center}
\end{figure}
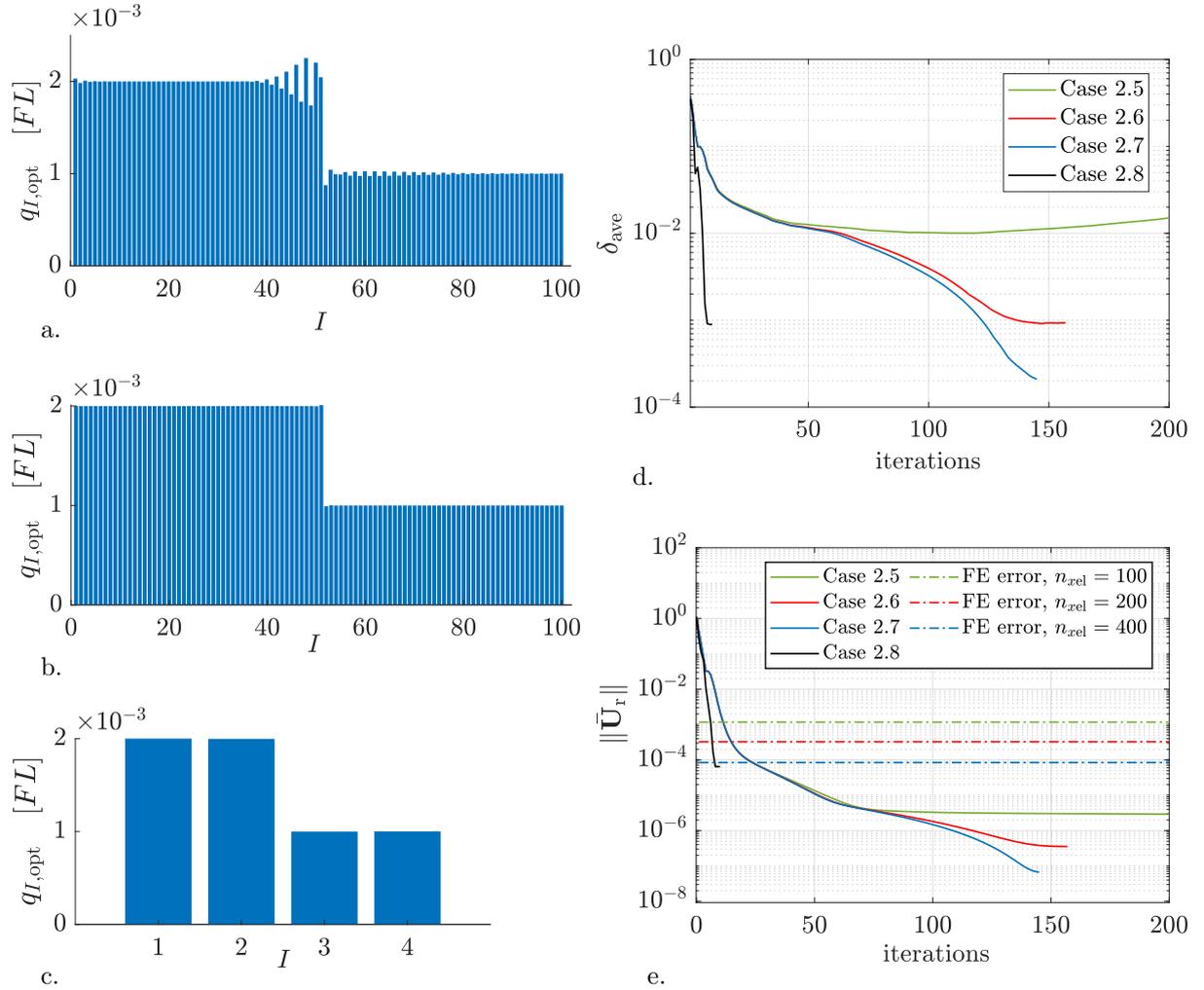

Second, the more challenging material distribution
\begin{equation}\label{eq:distrlenght}
   c(X) = \left\{ \begin{array}{rcl}
c_0 & \mbox{for} & X \leq 2L,\\
25(c_1-c_0)\cdot(X/L-2)+c_0 & \mbox{for} & 2L<X<2.04L,\\
c_1 & \mbox{for} & 2.04L \leq X \leq 4L,\\
\end{array}\right.
\end{equation}
is considered with $c_0 = 2.0\cdot 10^{-3} FL$ and $c_1 = c_0/2$, see Fig.~\ref{fig:bend2distribution}c. This distribution has a sharp jump characterized by the relative length scale $1/100$ (w.r.t.~the strip length $4L$) and results in the discontinuous curvature shown in Fig.~\ref{fig:bend2distribution}d. Shear stiffness $\mu$ is again kept constant at $\mu = F/L$.
Experiment-like reference results are generated from four load levels (at $25,50,75,100$ [\%] load) using a fine mesh ($n_{x\mathrm{el}}=4096$). Noise is not applied in order to isolate the influence of the discontinuity. A uniform material mesh with $\bar{n}_{x\mathrm{el}}=100$ elements of length $0.04L$ is used (Fig.~\ref{fig:bend2distribution}a), as it captures the material distribution exactly, and hence allows to also eliminate the influence of material mesh errors. Tab.~\ref{tab:bend_newde}
\begin{table}[!ht]
\centering
{\color{col1}\begin{tabular}{ccccccccc}
\toprule
Case & FE & mat. & mat. & exp. & load & noise  & $\delta_\mathrm{{max}}$  & $\delta_{\mathrm{ave}}$ \\
 & $n_{x\mathrm{el}}$ & $\bar{n}_{x\mathrm{el}}$ & $n_{\mathrm{var}}$ & $n_{\mathrm{exp}}/n_{\mathrm{ll}}$ & $n_{\mathrm{ll}}$ & $[\%]$  & $[\%]$ & $[\%]$ \\
\midrule
2.5 & $100$ & $100$ & 101 & $4098$ & 4 & 0 &  $13.03$  & $1.52$ \\
2.6 & $200$ & $100$ & 101 & $4098$ & 4 & 0   & $3.48$ & $0.093$ \\
2.7 & $400$ & $100$ & 101 & $4098$ & 4 & 0   & $0.76$ & $0.021$ \\
2.8 & $100$ & $3$ & 4 & $4098$ & 4 & 0   & $0.12$ & $0.089$ \\
\bottomrule
\end{tabular}}
\caption{Pure bending (discontinuous distribution): Studied inverse analysis cases with their FE mesh, material mesh, design variables, experimental grid resolution, load levels, noise level, and resulting errors $\delta_{\mathrm{max}}$ and $\delta_{\mathrm{ave}}$.}
\label{tab:bend_newde}
\end{table}
shows the results of the inverse analysis with the uniform material mesh for three FE analysis meshes (Cases 2.5--2.7). As seen, there are large errors in $\textbf{q}$ even when 100 FE are used (Case 2.5). These errors are coming from the discontinuity as Fig.~\ref{fig:basd}a
shows. Only for 400 FE accurate results for $\textbf{q}$ are obtained (Case 2.7), see Fig.~\ref{fig:basd}b.\\The evolution of $\delta_{\mathrm{ave}}$ is shown in Fig.~\ref{fig:basd}d. The elimination of the error in $\textbf{q}$ is only possible for a high number of FE and inverse iteration steps. Thus, the proposed method is able to reconstruct discontinuous material distributions, but at the cost of efficiency. Dense FE meshes are needed, since otherwise the FE error in $\textbf{U}$ overshadows the differences in $\textbf{U}$ coming from $\textbf{q}$, making an accurate reconstruction of $\textbf{q}$ impossible.\par 
The material distribution in \eqref{eq:distrlenght} suggests an obvious alternative to uniform material meshes: Suppose an adapted material mesh is available, such as the 3-element mesh shown in Fig.~\ref{fig:bend2distribution}b. In this case (denoted 2.8 in Tab.~\ref{tab:bend_newde}), the material can be accurately reconstructed within few iterations even for a comparably coarse analysis mesh, as Fig.~\ref{fig:basd}c--e show. Interestingly, the displacement residual $\Vert\bar\mU_\mathrm{r}\Vert$ is not the lowest for Case 2.8, even though the material error $\delta_{\mathrm{ave}}$ is, as Fig.~\ref{fig:basd}d \& e show. This illustrates the problem of overfitting: Decreasing $\Vert\bar\mU_\mathrm{r}\Vert$ beyond the FE error tends to wrongly fit the material parameters, which is especially problematic for a high number of design variables.\\
These results demonstrate that the proposed method, in conjunction with adapted mesh refinement, has the potential to speed-up simulations tremendously, while avoiding overfitting at the same time.
%
%
%
%
%
%
%-------------------------------------- INFLATION --------------------------------------
\subsection{Sheet inflation}\label{sec:inf}
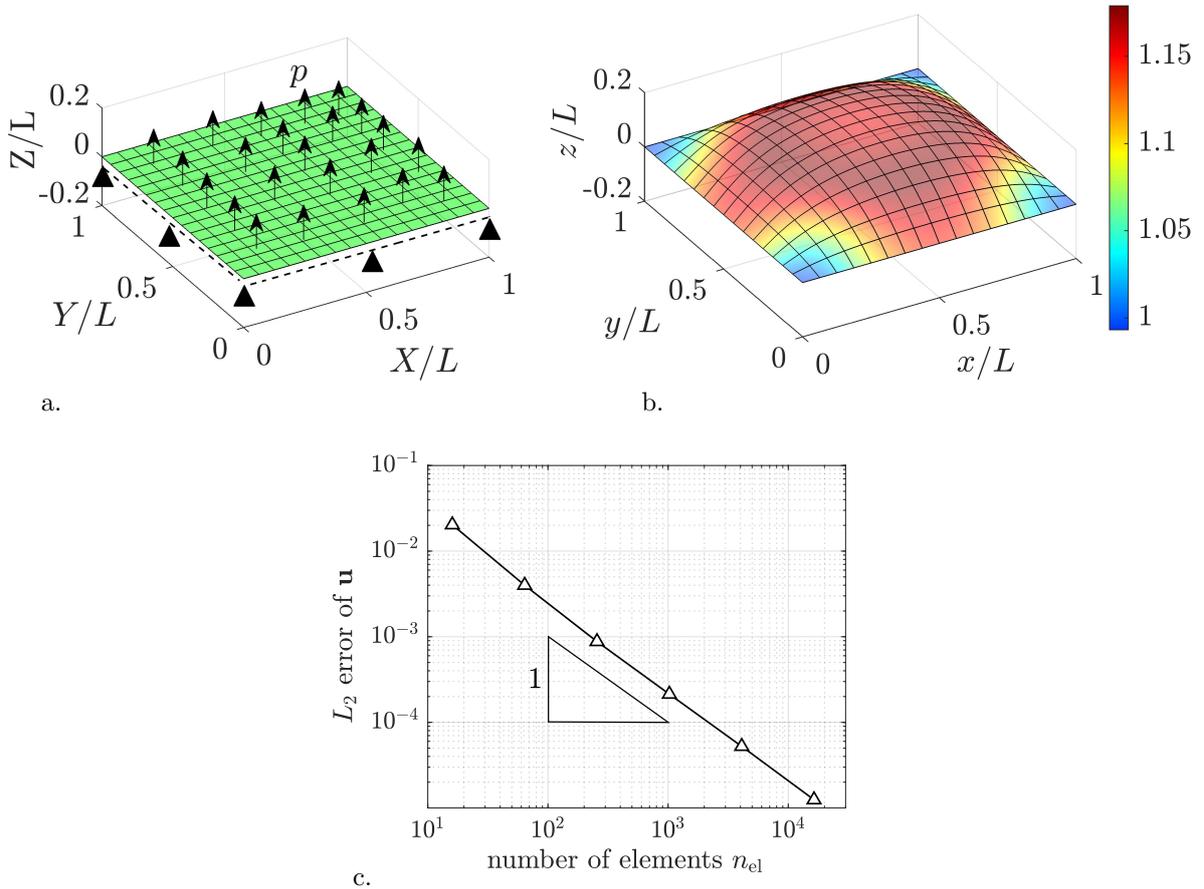
\begin{figure}[!ht]
\begin{center} \unitlength1cm
\begin{picture}(0,12)
\put(-8,5.8){\includegraphics[width=75mm]{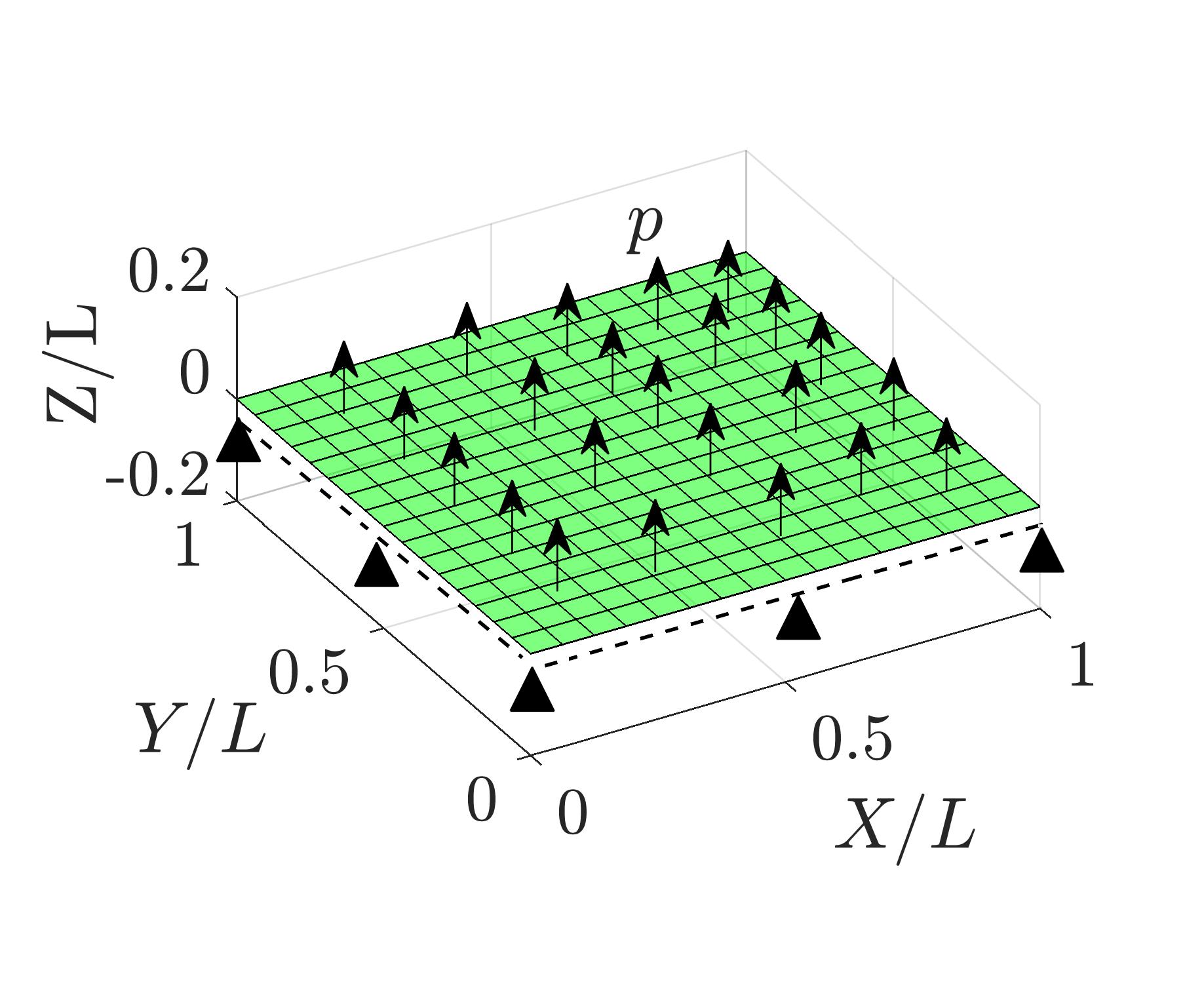}}
\put(-.6,6.6){\includegraphics[width=85mm]{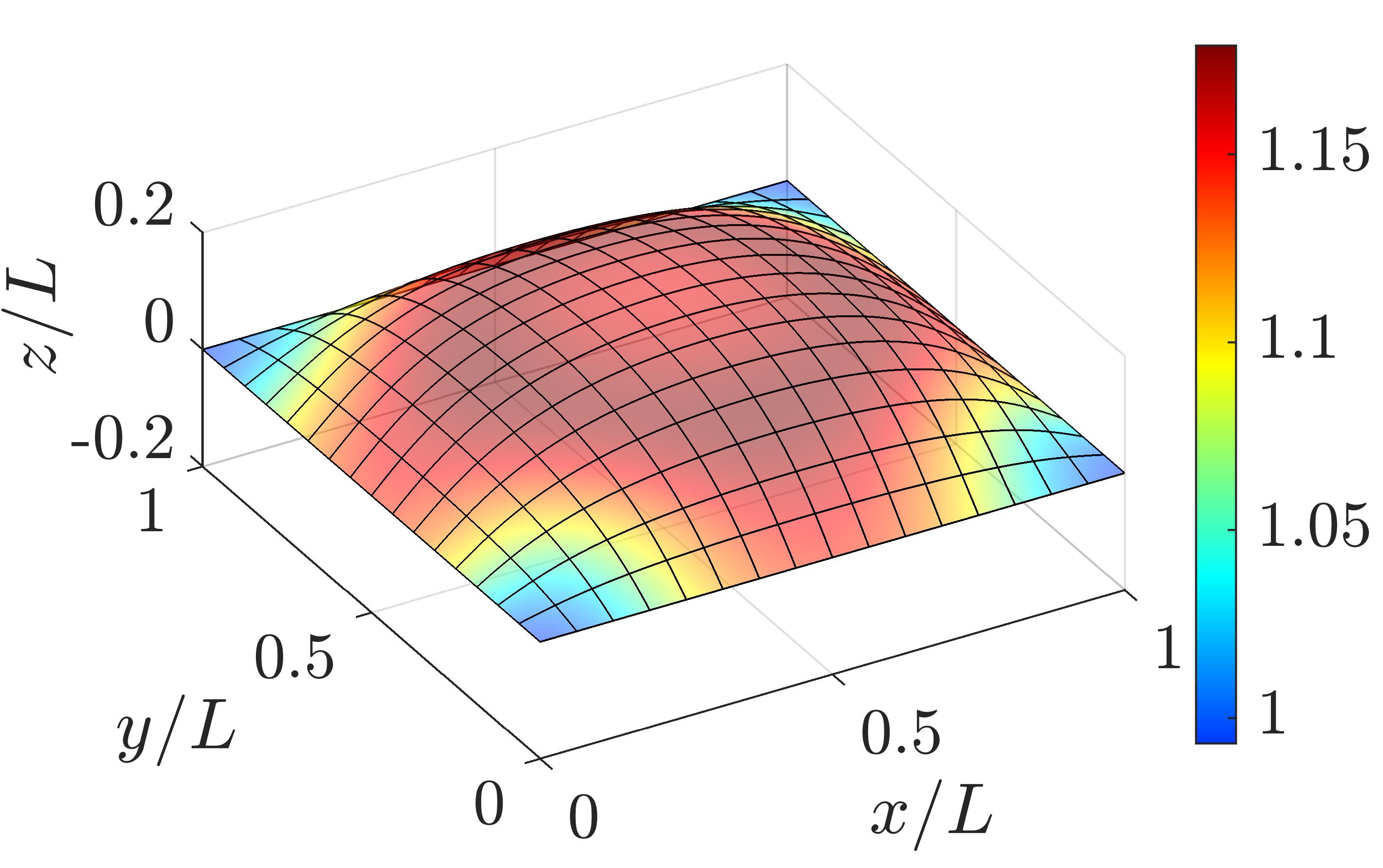}}
\put(-3.6,.1){\includegraphics[width=70mm]{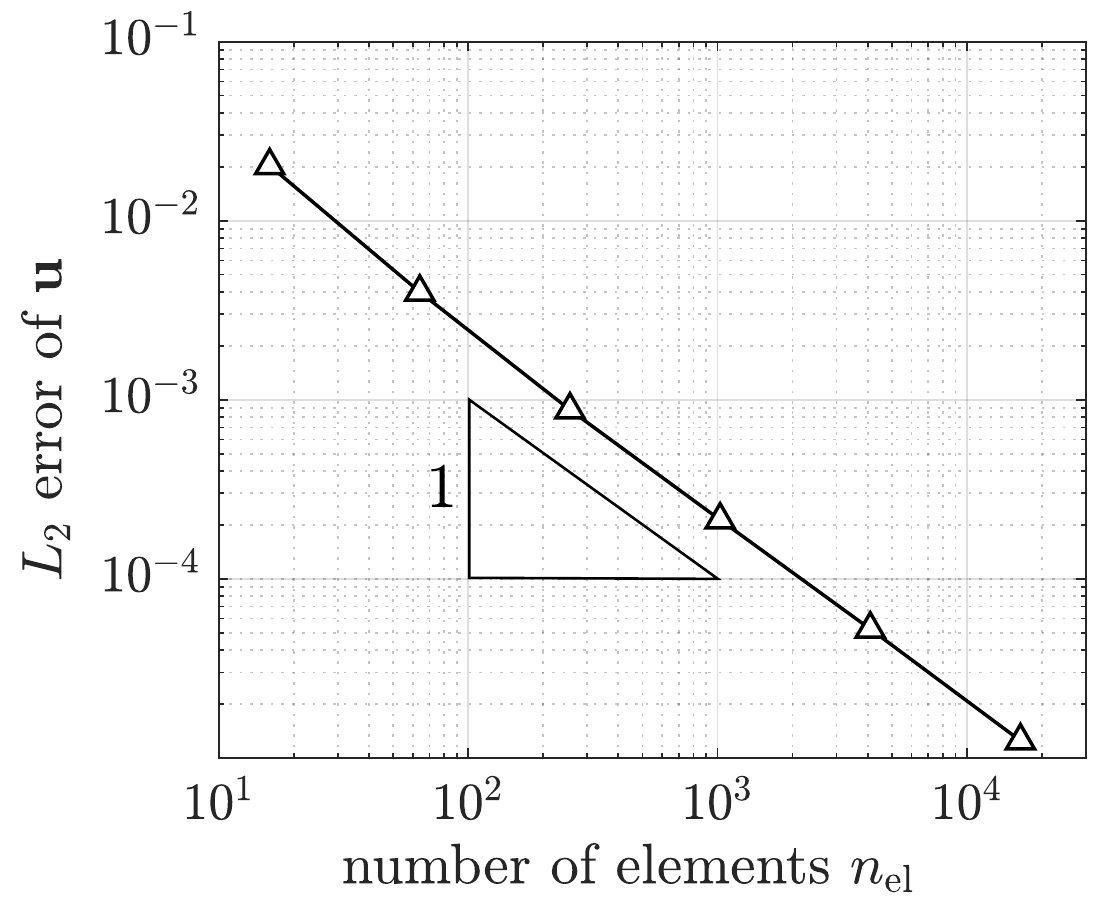}}
\put(-7.3,6.3){\small{a.}}
\put(.6,6.3){\small{b.}}
\put(-3.2,0){\small{c.}}
\end{picture}
\caption{Sheet inflation: a.~undeformed configuration and boundary conditions; b.~deformed configuration colored by the surface stretch  $J$, ranging between 1.0 and 1.18; c.~convergence of the discrete $L_2$ error $\Vert\textbf{u}_{\mathrm{exact}}-\textbf{u}_{\mathrm{FE}}\Vert/\Vert\textbf{u}_{\mathrm{exact}}\Vert$, where $\textbf{u}_{\mathrm{exact}}$ is the FE solution for $n_{\mathrm{el}}=512\times512$ elements.}
\label{fig:inf}
\end{center}
\end{figure}

\begin{figure}[!ht]
\begin{center} \unitlength1cm
\begin{picture}(0,6)
\put(-8,-.1){\includegraphics[width=75mm]{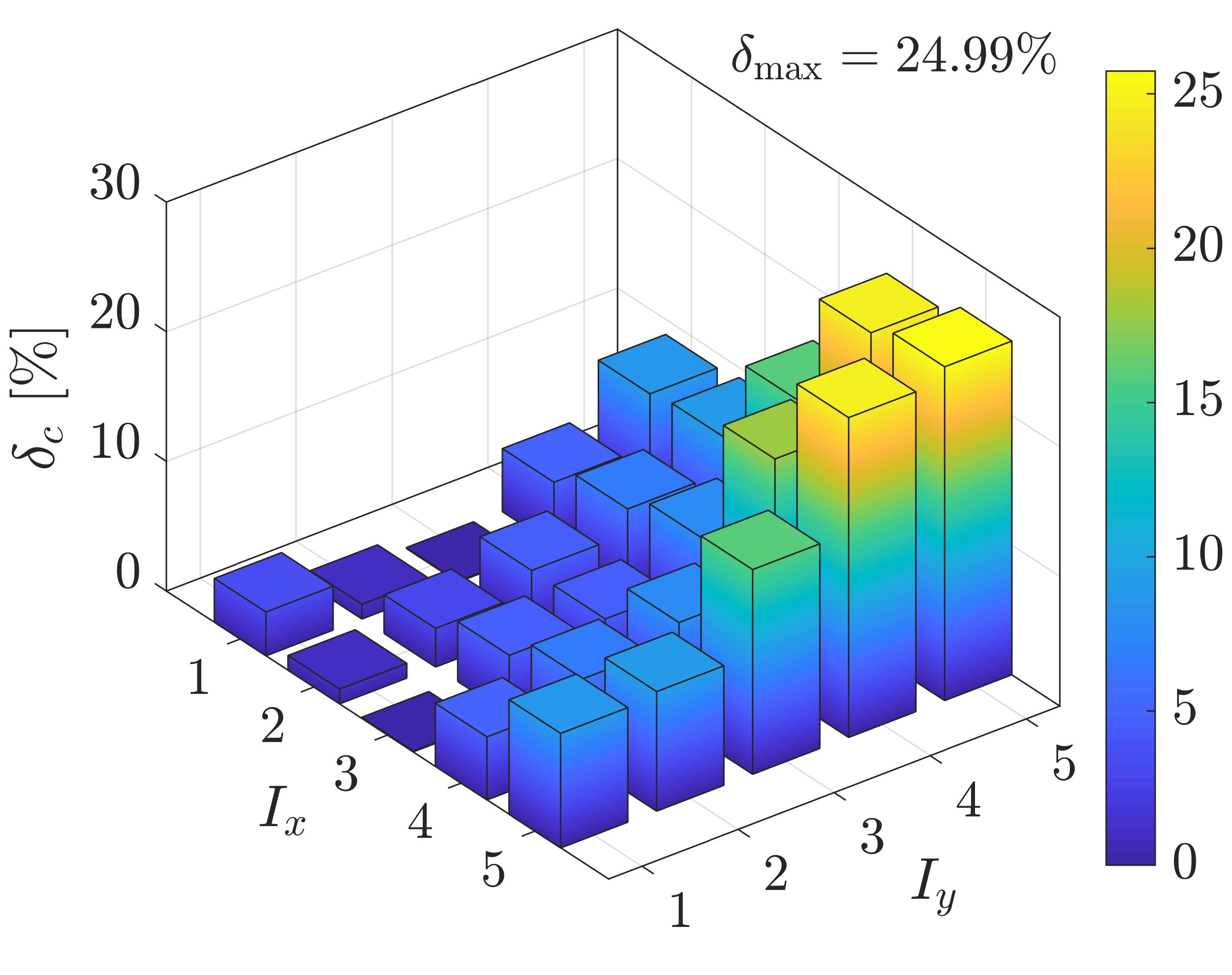}}
\put(0,0){\includegraphics[width=75mm]{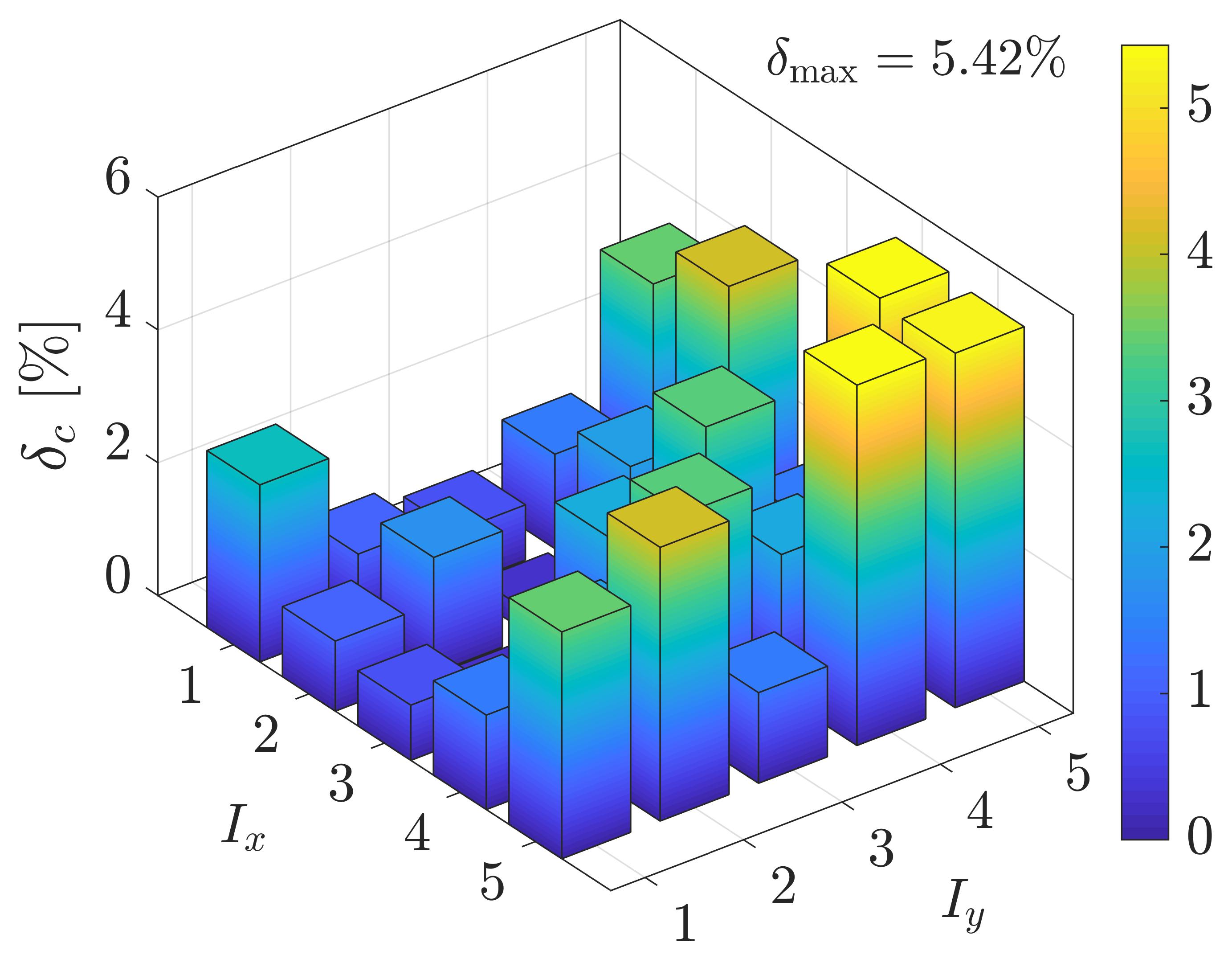}}
\put(-7.3,0){\small{a.}}
\put(.8,0){\small{b.}}
\end{picture}
\caption{Sheet inflation: Comparison of nodal error $\delta_I$ of the bending stiffness~$c$ for: a.~Case 3.3, b.~Case 3.4.}
\label{fig:infcomp1}
\end{center}
\end{figure}

\begin{figure}[!ht]
\begin{center} \unitlength1cm
\begin{picture}(0,5.3)
\put(-8,0){\includegraphics[width=72mm]{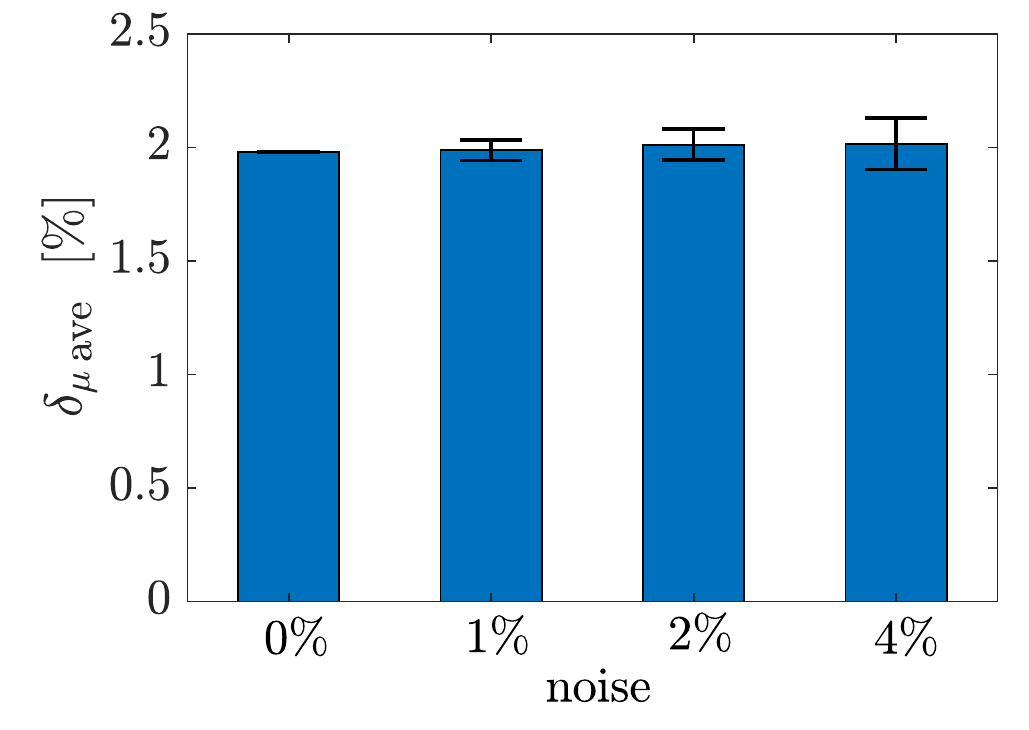}}
\put(0,0){\includegraphics[width=72mm]{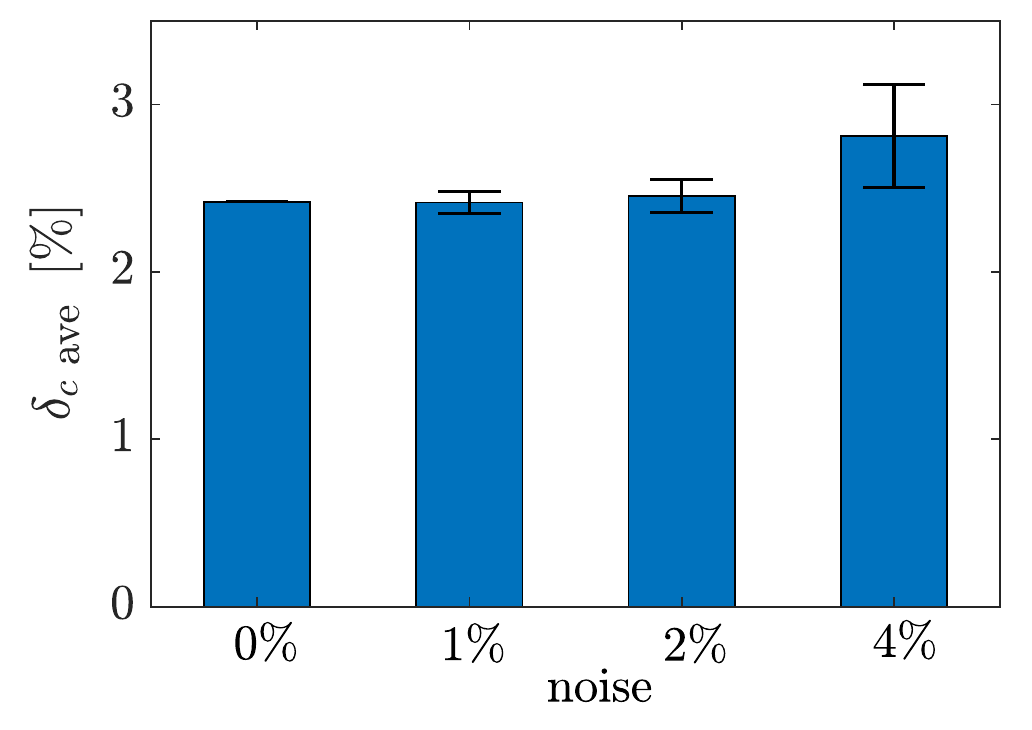}}
\put(-7.3,0){\small{a.}}
\put(.4,0){\small{b.}}
\end{picture}
\caption{Sheet inflation: Comparison of the mean and standard deviation of error $\delta_{\mathrm{ave}}$ for 0\% (Case 3.4), 1\% (Case 3.5), 2\% (Case 3.6) and 4\% (Case 3.7) noise for: a.~shear modulus~$\mu$, b.~bending stiffness~$c$.}
\label{fig:inf_stat}
\end{center}
\end{figure}

The third example studies the inflation of a square sheet, which induces coupled biaxial membrane and bending deformations and thus tests the capability to simultaneously reconstruct the $\bar d = 2$ unknown fields $\mu(\bX)$ and $c(\bX)$. The initially flat sheet with dimension $L_x\times L_y=L\times L$ is pinned on all boundaries and exposed to the uniform pressure $p=0.5\,FL^2$ prescribed over its entire surface (Fig.~\ref{fig:inf}a,~\ref{fig:inf}b). 
The material is defined with heterogeneous bending and membrane stiffness according to distribution \eqref{eq:uniaxial_distr} (see Fig.~\ref{fig:uniaxialother}a) with $\mu_0=FL\,,\Delta\mu_1=\mu_0$, $c_0=0.001FL$ and $\Delta c_1=c_0$. $8\times8$ material elements are chosen to reconstruct the material distribution based on the results from the uniaxial tension example, where the same distribution was used. This example is not a pure Dirichlet problem, and so no reaction force are required in $f$. Due to symmetry, only $\approx$1/4 of the material nodes ($\bar{n}_{\mathrm{no}}=25$) are treated as design variables (see Fig.~\ref{fig:uniaxialother}b). Objective minimization is performed with the lower bounds $c_{\mathrm{min}} = 0.4\,c_0,\,\mu_{\mathrm{min}} = 0.1\,\mu_0$ and the upper bounds $c_{\mathrm{max}} = 5.0\,c_0\,,\mu_{\mathrm{max}} = 5.0\,\mu_0$. The initial estimate for the reconstruction  algorithm is a vector of random numbers from the range $[c_{\mathrm{min}},c_{\mathrm{max}}]$ and $[\mu_{\mathrm{min}},\mu_{\mathrm{max}}]$. The results of the identification are collected in Tab.~\ref{tab:3} and illustrated for selected cases in Figs.~\ref{fig:infcomp1} and \ref{fig:inf_stat}.
\begin{table}[!ht]
\centering
\begin{tabulary}{\textwidth}{cccccccccc}
\toprule
Case & FE & \textcolor{col2}{mat.} & \textcolor{col2}{mat.} & $q(\boldsymbol{X})$  & \textcolor{col2}{exp.} & load & noise & $\delta_\mathrm{{max}}$  & $\delta_{\mathrm{ave}}$ \\
& $n_{\mathrm{el}}$ & $\bar{n}_{\mathrm{el}}$ & \textcolor{col2}{$n_{\mathrm{var}}$}  &  & $n_{\mathrm{exp}}/n_{\mathrm{ll}}$ & $\mathrm{n_{ll}}$ & $[\%]$ & $[\%]$ & $[\%]$\\
\midrule
3.1 & $16\times16$ & $8\times8$ & \textcolor{col2}{25} & $\mu$ & $130^2$ & 1 & 0 & 4.99  & 1.53\\
3.2 & $16\times16$ & $8\times8$ & \textcolor{col2}{25} & $c$   & $130^2$ & 1 & 0 & 4.75  & 1.96\\
\midrule
3.3 & $32\times32$ & $8\times8$ & \textcolor{col2}{50} & $\mu$ & $130^2$ & 1 & 0 & 4.19  & 1.86\\
& & & & $c$ & & & & 24.99 & 8.58\\
\midrule
\rowcolor{lightgray}3.4 & $32\times32$ & $8\times8$ & \textcolor{col2}{50} & $\mu$ & $130^2$ & 4 & 0 & 5.04 & 1.98\\
& & & & $c$ & & & & $5.42$ & $2.42$\\
\midrule
\rowcolor{lightgray}3.5 & $32\times32$ & $8\times8$ & \textcolor{col2}{50} & $\mu$ & $514^2$ & 4 & 1 & $5.08\pm0.036$ & $1.99\pm0.045$\\
& & & & $c$ & & & & $5.97\pm0.33$  & $2.41\pm0.065$\\
\midrule
\rowcolor{lightgray}3.6 & $32\times32$ & $8\times8$ & \textcolor{col2}{50} & $\mu$ & $514^2$ & 4 & 2 & $5.28\pm0.63$  & $2.01\pm0.068$\\
& & & & $c$ & & & & $6.75\pm0.71$  & $2.45\pm 0.098$\\
\midrule
\rowcolor{lightgray}3.7 & $32\times32$ & $8\times8$ & \textcolor{col2}{50} & $\mu$ & $514^2$ & 4 & 4 & $5.71\pm1.36$  & $2.02\pm0.11$\\
& & & & $c$ & & & & $8.99\pm1.65$  & $2.81\pm0.31$\\
\bottomrule
\end{tabulary}
\caption{Sheet inflation: Studied inverse analysis cases with their FE mesh, material mesh, design variables, experimental grid resolution, load levels, noise level, and resulting errors $\delta_\mathrm{max}$ and $\delta_{\mathrm{ave}}$. 
25 repetitions were used for the statistical analysis of Cases 3.5--3.7.
The four highlighted cases are compared in Fig.~\ref{fig:inf_stat}.}
\label{tab:3}
\end{table}

In Cases 3.1 and 3.2 only one parameter is identified, while the remaining parameter is assumed given by distribution Eq.~\eqref{eq:uniaxial_distr}. The maximum and average errors are similar to Case 1.2 in Tab.~\ref{tab:1}, which shows a predictable performance of the optimization when the two material parameters are uncoupled. Problems with coupled membrane and bending behavior exhibit additional challenges compared to uncoupled problems such as those in Sec.~\ref{sec:uni} \& \ref{sec:purebending}. One, are the different units of membrane stiffness $\mu$ and bending stiffness $c$. Working with normalized quantities, as is done here, alleviates this problem. Another challenge are the different sensitivities of parameters $\mu$ and $c$. As a remedy (not considered here) different weights for parameters $\mu$ and $c$ could be considered in the objective function \eqref{eq:minimization}. However, this can still not be expected to fully solve the problem, as the two sensitivities dependent differently on location and deformation. The bending sensitivity, for instance, can be expected to play a larger role at the boundary than in the center of the sheet. As is known from nonlinear plate and shell theory, the early load stages engage bending more than membrane deformations, i.e.~the bending sensitivity \eqref{eq:sensitivityc} can be expected to be more dominant at small loads, while the membrane sensitivity \eqref{eq:sensitivitymu} can be expected to be more influential at high loads. Therefore the bending stiffness cannot be expected to be reconstructed very well, if only the last load step is taken into account. This can be seen by comparing Cases 3.3 and 3.4 as is done in Fig.~\ref{fig:infcomp1}. As seen, using four load steps (at $25,50,75,100$ [\%] load) reduces the error in $\delta_c$ significantly.\\
The parameters $n_\mathrm{el}$, $\bar n_\mathrm{el}$ and $n_\mathrm{ll}$ of case 3.4 are further used in Cases 3.5--3.7 to study the influence of noise. Each case is repeated 25 times and the influence of the noise on the error $\delta_{\mathrm{ave}}$ for both parameters is shown in Fig.~\ref{fig:inf_stat}. Ultimately, for Case 3.7 (4\% noise), errors $\delta_\mathrm{max}$ and $\delta_{\mathrm{ave}}$ were found at $8.99\pm1.65\%$ (mean $\,\pm\,$ standard deviation) and $2.81\pm0.31\%$, respectively, for bending stiffness $c$, and $5.71\pm1.36\%$ and $2.02\pm0.11\%$ for shear modulus $\mu$. The comparison with Case 3.4, where no noise was applied, shows good performance of the algorithm in the presence of increasing noise. The results indicate that the mean error (but not its variation) is independent of the noise, as long as the experimental dataset is sufficiently large, which does seem to be the case for $c$ in Case 3.7. This behavior can be expected for noise that is symmetrically distributed around zero, as is considered here. If the noise is distributed differently, it can be expected to affect the mean error also.\\
Throughout all cases, the inverse solution was found after 9--11 iterations.
%
%
%
%
%
%
%
%
%-------------------------------------- ABDOMINAL --------------------------------------
\subsection{Abdominal wall under pressure loading}\label{sec:abdominalwall}
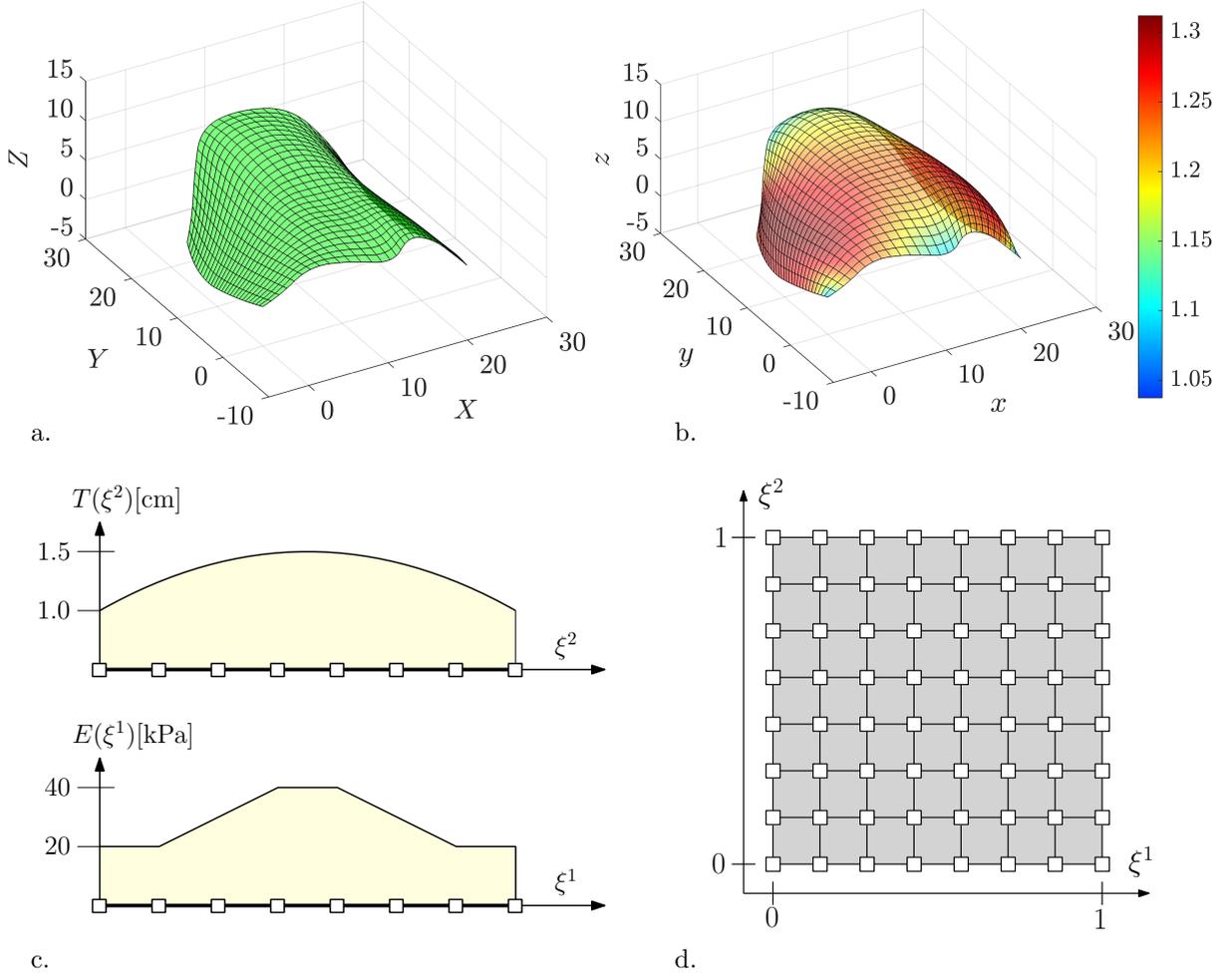
\begin{figure}[!ht]
\begin{center} \unitlength1cm
\begin{picture}(0,13.2)
\put(-7.9,6.9){\includegraphics[width=79mm]{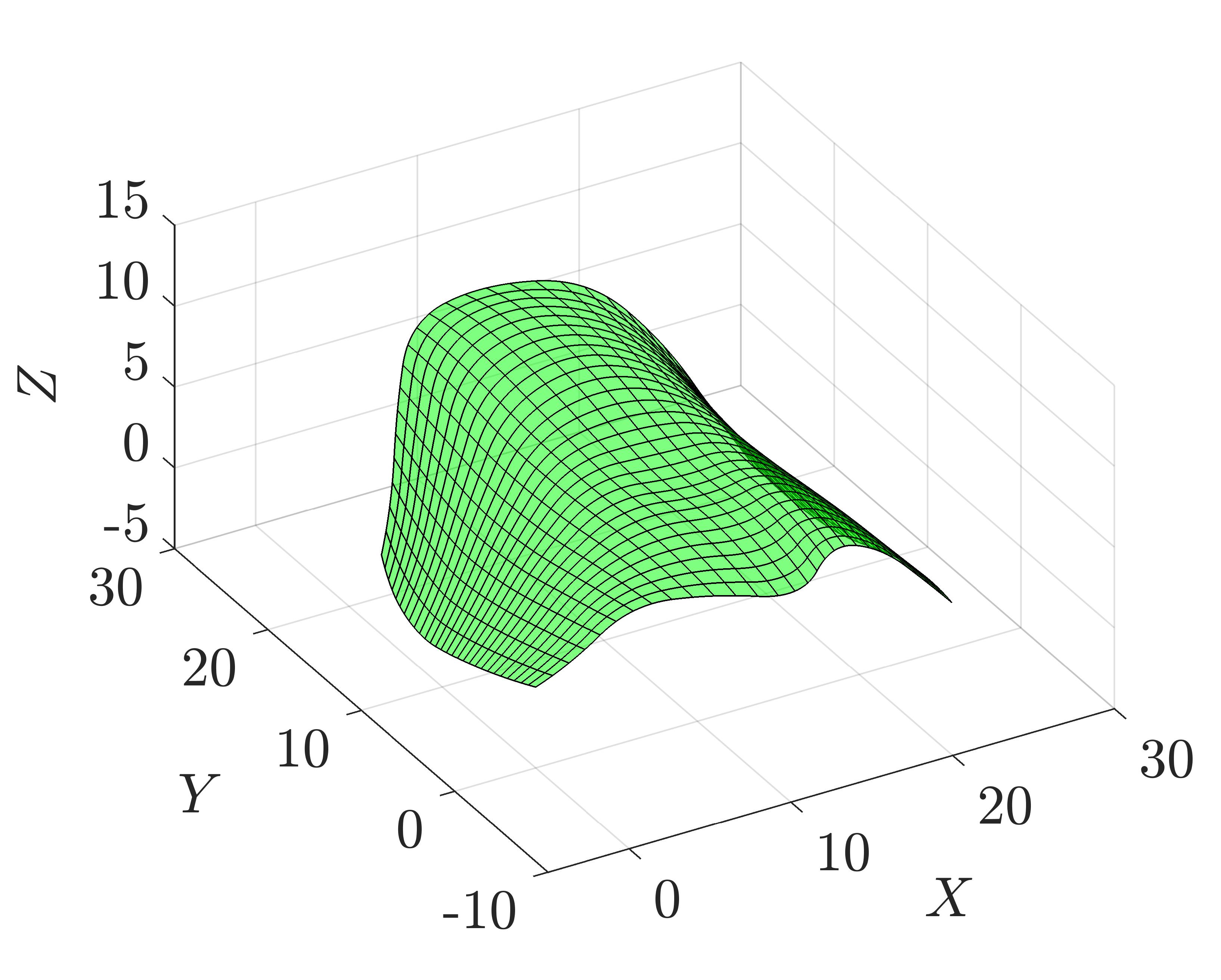}}
\put(-.15,6.8){\includegraphics[width=83mm]{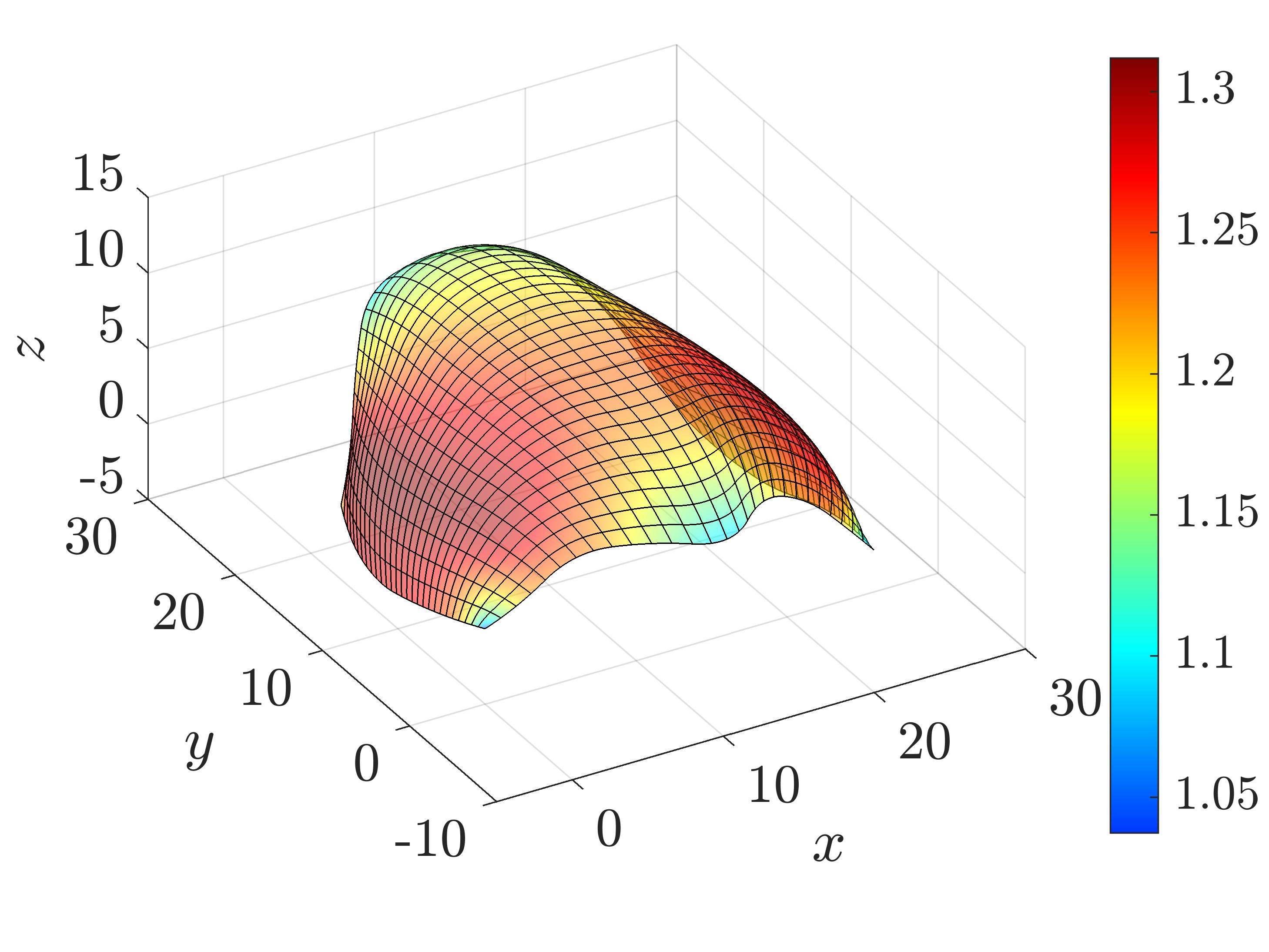}}
\put(-7.4,.7){\includegraphics[width=75mm]{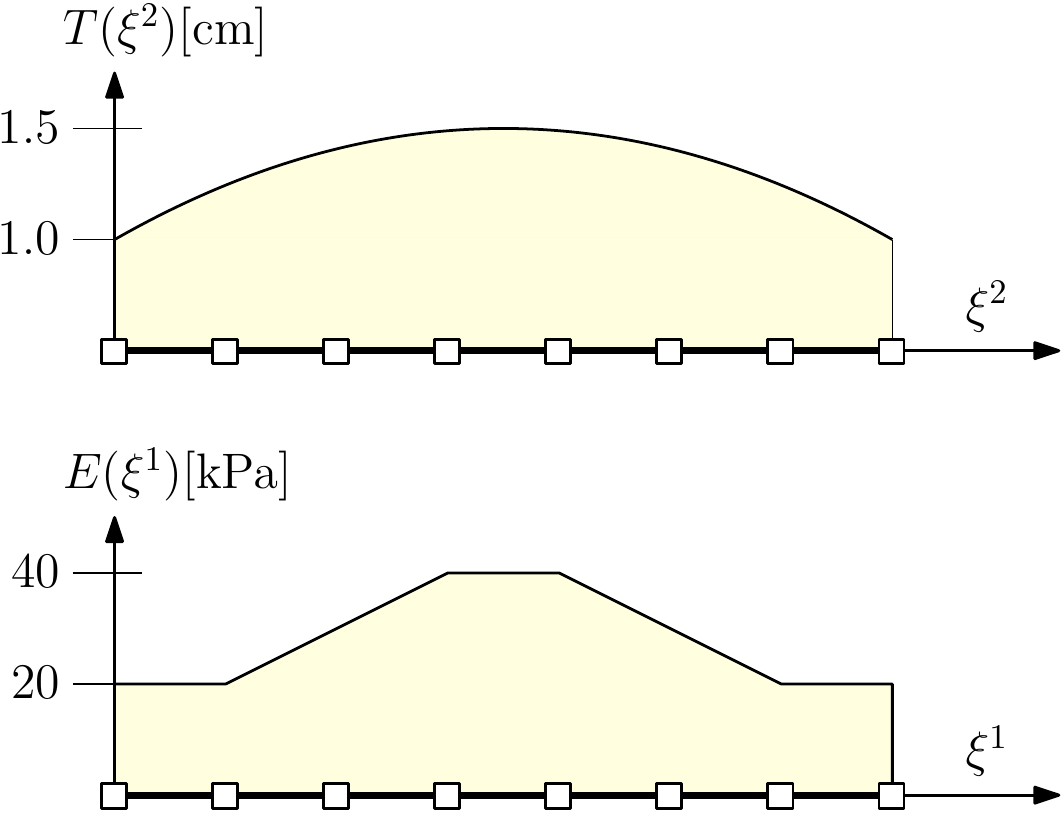}}
\put(1.5,.5){\includegraphics[width=58mm]{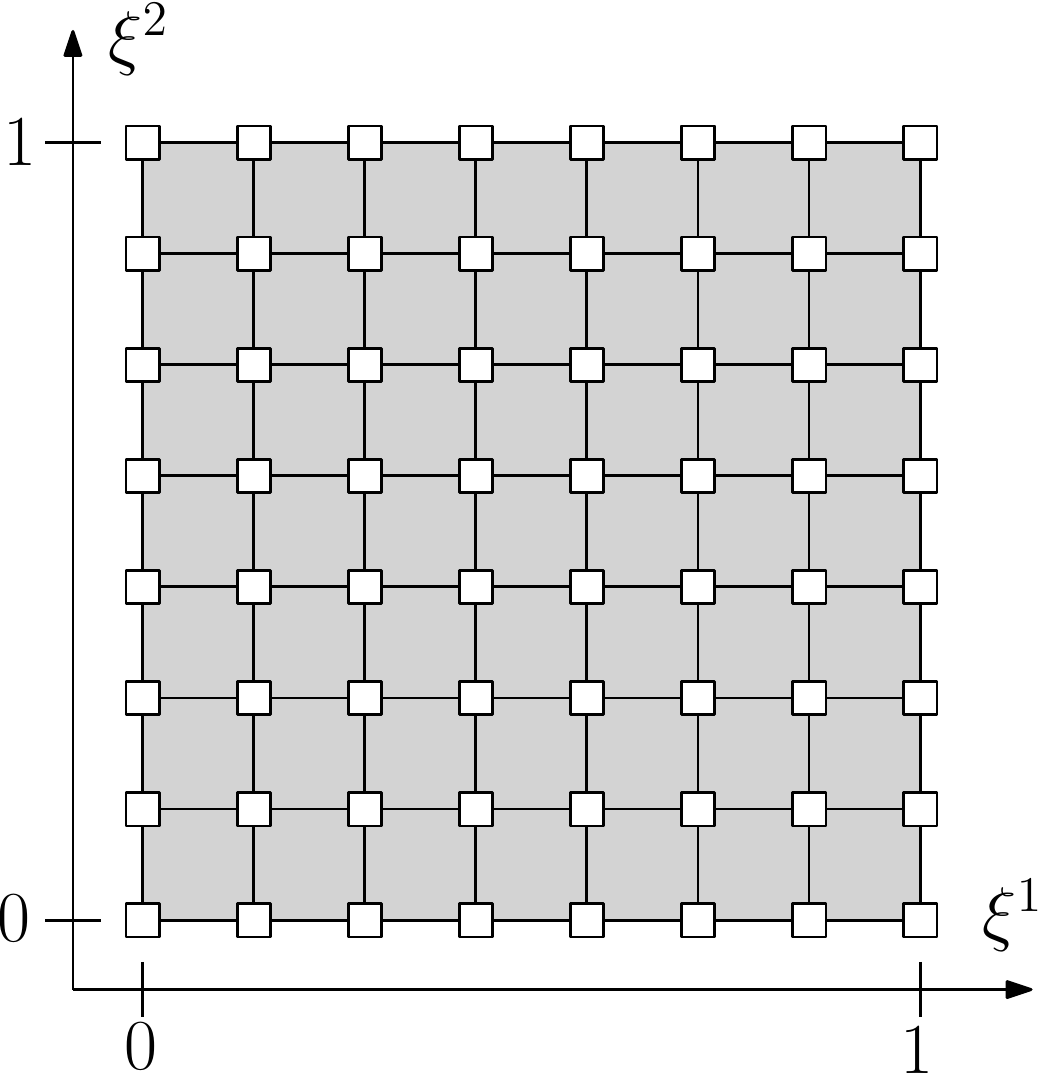}}
\put(-7.5,7){\small{a.}}
\put(1,7){\small{b.}}
\put(-7.5,0){\small{c.}}
\put(1,0){\small{d.}}
\end{picture}
\caption{Abdominal wall: a.~undeformed configuration; b.~deformed configuration colored by the surface stretch $J$, ranging between 1.04 and 1.31; c.~reference distribution for Young's modulus $E$ and thickness $T$; d.~material mesh.}
\label{fig:abdominal1}
\end{center}
\end{figure}

\begin{figure}[!ht]
\begin{center} \unitlength1cm
\begin{picture}(0,6)
\put(-8,0){\includegraphics[width=78mm]{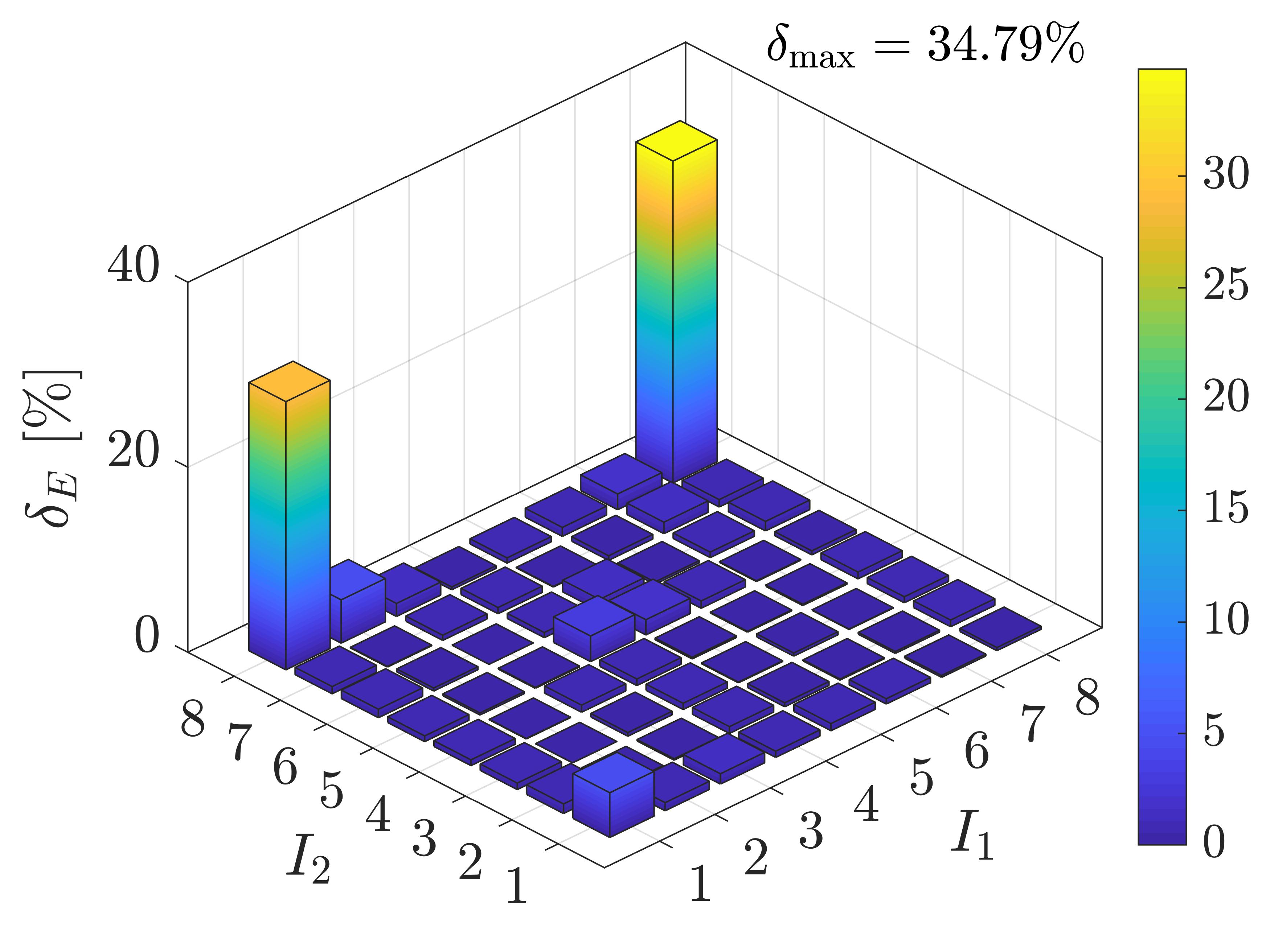}}
\put(0.3,0){\includegraphics[width=78mm]{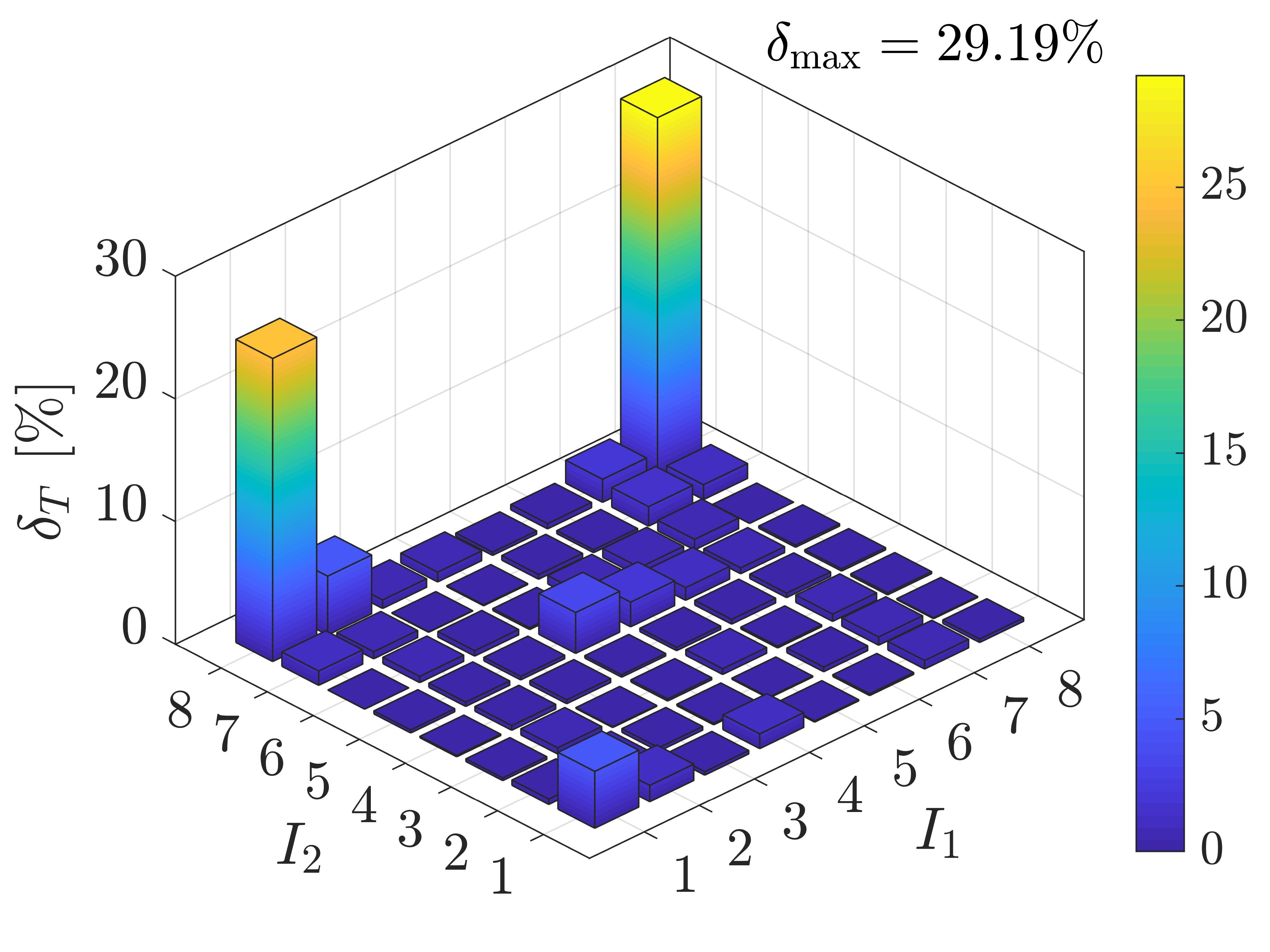}}
\put(-7.3,0){\small{a.}}
\put(.8,0){\small{b.}}
\end{picture}
\caption{Abdominal wall: Case 4.2, nodal error $\delta_I$ of: a.~Young's modulus $E$; b.~thickness $T$.}
\label{fig:abdominal_res}
\end{center}
\end{figure}

The last example considers the pressurization of the human abdominal wall in order to identify its Young's modulus and thickness distribution ($\bar d = 2$). A single-patch NURBS discretization is used \citep{borzeszkowski2020isogeometric}, which is based on the geometry model of \cite{lubowiecka2017membrane} and the measurement methodology of \cite{szymczak2012investigation}. The initially curved surface is pinned on all boundaries and  exposed to the uniform pressure $p=1.6\,\mathrm{kPa}$, which corresponds to the intra-abdominal pressure level (12 mmHg) in laparoscopic surgeries \citep{song2006mechanical,pachera2016numerical}, see Fig.~\ref{fig:abdominal1}a and~\ref{fig:abdominal1}b. 
$168\times168$ finite elements are used to generate experiment-like results for the deformation field, incorporating four load levels (at 25, 50, 75, 100 [\%] load). For the sake of demonstration, an analytic reference distribution of the material is chosen. It is defined in the domain $\boldsymbol{\xi}$ and mapped on the surface $\sS_0$. The Koiter material model (Sec.~\ref{sec:koiter}) is considered with varying Young's modulus in $\xi^1\mapsto X$ direction
\begin{equation}
   E(\xi^1) = \left\{ \begin{array}{rcl}
E_1 & \mbox{for} & \xi^1 \leq \frac{1}{7}\,  \vee \xi^1 \geq \frac{6}{7}\vspace{0.3em},\\
E_1+\frac{1}{2}(E_2-E_1) \cdot(7\xi^1-1) & \mbox{for} & \frac{1}{7} < \xi^1 < \frac{3}{7}\vspace{0.3em},\\
E_2 & \mbox{for} & \frac{3}{7} \leq \xi^1 \leq\frac{4}{7}\vspace{0.3em},\\
E_2-\frac{1}{2}(E_2-E_1)\cdot(7\xi^1-4) & \mbox{for} & \frac{4}{7}<\xi^1<\frac{6}{7},\\
\end{array}\right. \quad \xi^1 \in [0,1],
\end{equation}
and varying thickness in the $\xi^2 \mapsto Y$ direction
\begin{equation}
   T(\xi^2) = T_2  -  (T_2 - T_1) (2\,\xi^2 -1)^2\,, \quad \xi^2 \in [0,1],
\end{equation}
with $T_1 = 1\,$cm, $T_2 = 1.5\,$cm, $E_1=20\,$kPa, $E_2=40\,$kPa, see Fig.~\ref{fig:abdominal1}c. The Koiter parameters $\mu$ and $\Lambda$ are then calculated from \eqref{eq:muandlam}. $7\times7$ bilinear material elements are used to capture the material distribution (Fig.~\ref{fig:abdominal1}d). Objective minimization is performed with the lower bounds $[5\,\mathrm{kPa}, 0.5 \,\mathrm{cm}]$ and the upper bounds $[100\,\mathrm{kPa}, 5.0\,\mathrm{cm}]$. The constant initial estimates $E=6\,$kPa and $T=4\,$cm are considered. Results, for different FE meshes and noise level, are collected in Tab.~\ref{tab:5} and illustrated in Fig.~\ref{fig:abdominal_res}.\par
\begin{table}[!ht]
\centering
{\color{col1}\begin{tabulary}{\textwidth}{cccccccccc}
\toprule
Case & FE & mat. & mat. & $q(\boldsymbol{X})$  & exp. & load & noise & $\delta_\mathrm{{max}}$  & $\delta_{\mathrm{ave}}$ \\
     & $n_{\mathrm{el}}$ & $\bar{n}_{\mathrm{el}}$ & $n_{\mathrm{var}}$  &  & $n_{\mathrm{exp}}/n_{\mathrm{ll}}$ & $\mathrm{n_{ll}}$ & $[\%]$ & $[\%]$ & $[\%]$ \\
\midrule
4.1     & $28\times28$ & $7\times7$ & 128 & $E$ & $170^2$  & 4 & 0 & 53.07 (10.59) & 3.68 (2.18) \\
        &              &            &                      & $T$   &         &   &   & 43.01 (10.49) & 3.19 (1.97)\\
\midrule
4.2     & $56\times56$ & $7\times7$ & 128 & $E$ & $170^2$ & 4 & 0 & 34.79 (4.83) & 1.69 (0.72)  \\
        &              &            &                      & $T$   &         &   &   & 29.19 (4.79) & 1.49 (0.66) \\
\midrule
4.3     & $56\times56$ & $7\times7$ & 128 & $E$ & $170^2$ & 4 & 1 & $34.74\,\pm\,0.19$  & $1.74\,\pm\,0.029$ \\
        &              &            &                      & $T$   &         &   &   & $29.17\,\pm\,0.19$ & $1.53\,\pm\,0.02$ \\
        
\midrule
4.4     & $56\times56$ & $7\times7$ & 128 & $E$ & $170^2$ & 4 & 2 & $34.78\,\pm\,0.41$  & $1.83\,\pm\,0.054$ \\
        &              &            &                      & $T$   &         &   &   & $29.17\,\pm\,0.31$ & $1.61\,\pm\,0.05$ \\
        \midrule
4.5     & $56\times56$ & $7\times7$ & 128 & $E$ & $170^2$ & 4 & 4 & $34.6\,\pm\,0.85$  & $2.1\,\pm\,0.15$ \\
        &              &            &                      & $T$   &         &   &   & $29.15\,\pm\,0.79$ & $1.87\,\pm\,0.13$ \\
\bottomrule
\end{tabulary}}
\caption{Abdominal wall: Studied inverse analysis cases with their FE mesh, material mesh, design variables, experimental grid resolution, load levels, noise level, and resulting errors $\delta_\mathrm{max}$ and $\delta_{\mathrm{ave}}$. The values in brackets show the errors excluding the upper corner values. 25 repetitions were used for the statistical analysis of Cases 4.3--4.5.}
\label{tab:5}
\end{table}
High accuracy is met, except at the upper corner nodes, where high errors $\delta_{\mathrm{max}}$ and $\delta_{\mathrm{ave}}$ are observed (see Fig.~\ref{fig:abdominal_res}). FE refinement decreases the errors, as Cases 4.1 and 4.2 show, but corner errors remain still high. The poor corner accuracy indicates that the pressurization test is insensitive to the corner parameters of the considered geometry model, resulting in the inability to reconstruct those values accurately. The incorporation of additional tests, as well as the consideration of different geometry models, are expected to eliminate corner inaccuracies. The statistical analysis in Cases 4.3--4.5 shows that noise has a minor influence on the material reconstruction.
%
%
%
%
%
%
%-------------------------------------- CONCLUSIONS --------------------------------------
\section{Conclusions}\label{s:conclusions}
This work presents an inverse material identification procedure based on an isogeometric shell formulation and the FEMU. The procedure is very general and able to identify heterogeneous material distributions for challenging nonlinear inverse problems in the presence of measurement error. The theoretical framework accounts for large deformations and is based on a general material law definition with independent membrane and bending contributions that is suitable to model a large range of materials. The discretization of the governing equations and heterogeneous material parameter fields use efficient isogeometric shell elements and bilinear Lagrange elements, respectively. The former provide high continuity to accurately and efficiently describe deforming shells, while the latter is a good compromise for various materials that can be expected to contain both discontinuous and graded material properties. Moreover, the material discretization is independent from the analysis mesh, which allows to efficiently and accurately capture material discontinuities using adapted meshes, as they reduce the number of design variables and avoid overfitting. The minimization of the resulting least-squares objective is conducted with a trust region algorithm in-build to MATLAB. Using the analytical Jacobians results in a major speed-up compared to using finite difference approximations of the Jacobians. Various possible approximation errors and their influence on the inverse algorithm are discussed. Several numerical examples demonstrate the ability of the algorithm to contain all these errors. In order to mimic measurement errors in experiments, experiment-like data is produced by a highly-resolved forward problem with given analytical reference material distribution and the addition of random noise.\par
The biggest benefit of the proposed identification framework is that its building blocks can be easily substituted or extended. The trust-region algorithm can be replaced by a different optimization scheme to increase efficiency, assure global convergence and account for uncertainties. Another extension is the consideration of more sophisticated constitutive laws. The membrane and bending energy split favors combining and studying different material models, while the proposed discretization allows for local mesh refinement that can accurately account for possible material variations and discontinuities. To automate this process, adaptive meshing techniques need to be developed, which is planned for future work. Also, we plan to extend the abdominal wall study in Sec.~\ref{sec:abdominalwall} to the material identification of human abdominal wall tissue based on \textit{in vivo} measurements as reported for example in \citet{lubowiecka2021novel2}. Furthermore, thick and composite shell formulations with various constraints (contact, volume, area) can be used. Although we focus on material identification, additional parameters can be used as design variables. An important example relevant to experimental material reconstruction is the stiffness of the specimen fixations.\par
%
%
%
%
%
%
%-------------------------------------- Acknowledgements --------------------------------------
\section*{Acknowledgements}
We thank Dr.~Thang Duong and Karsten Paul for their comments and support. This work has been partially supported by the National Science Centre (Poland) under Grant No.~2017/27/B/ ST8/02518. Calculations have been carried out at the Academic Computing Centre in Gda{\'n}sk.
%
%
%
%
%
%
%-------------------------------------- APP --------------------------------------
\appendix
\section{Gradient and Hessian of the objective function} \label{s:appendix1}
Introducing the residual
\begin{equation}
    \bar{\textbf{U}}_\mrr(\mq) := \bar{\textbf{U}}_{\mathrm{exp}} - \bar{\textbf{U}}_\mathrm{FE}(\mq) ,
\end{equation}
with
\begin{equation}
    \bar{\textbf{U}}_{\mathrm{exp}} := \begin{bmatrix}
    \textbf{U}_{\mathrm{exp}} / \|  \textbf{U}_{\mathrm{exp}} \|  \\[0.2em]
    \textbf{R}_{\mathrm{exp}} / \|  \textbf{R}_{\mathrm{exp}} \| 
    \end{bmatrix}\,,\qquad \bar{\textbf{U}}_{\mathrm{FE}}(\mq) := \begin{bmatrix}
    \textbf{U}_{\mathrm{FE}}(\mq) / \|  \textbf{U}_{\mathrm{exp}} \|  \\[0.2em]
    \textbf{R}_{\mathrm{FE}}(\mq) / \|  \textbf{R}_{\mathrm{exp}} \| 
    \end{bmatrix}
\end{equation}
the objective function of \eqref{eq:minimization} simply becomes
\begin{equation}
    f = \frac{1}{2} \bar{\textbf{U}}_\mrr^{\mrT} \bar{\textbf{U}}_\mrr. 
\end{equation}
The gradient of $f(\textbf{q})$ w.r.t to the design variable vector $\textbf{q}$ is then given by
\begin{equation}\label{eq:gradapp}
    \textbf{g}(\textbf{q})=\frac{\partial f(\textbf{q})}{\partial \textbf{q}}=\textbf{J}(\textbf{q})^{\mathrm{T}}\bar{\textbf{U}}_\mrr(\mq)\ .
\end{equation}
where
\begin{equation}\label{eq:jacobian}
    \textbf{J}(\textbf{q})=\frac{\partial \bar{\textbf{U}}_\mrr(\mq)}{\partial \textbf{q}}=\begin{bmatrix}
    \mJ_\mathrm{U}\\
    \mJ_\mathrm{R}
    \end{bmatrix}\,,
\end{equation}
is the \textit{Jacobian} of the residual that contains the blocks
\begin{equation}\label{eq:jacobianU}
    \mJ_\mathrm{U} = -\frac{1}{\| \textbf{U}_{\mathrm{exp}} \|}\frac{\partial \textbf{U}_{\mathrm{FE}}(\mq)}{\partial \mq}\,,
\end{equation}
and
\begin{equation}\label{eq:jacobianR}
    \mJ_{\mathrm{R}} = -\frac{1}{\| \textbf{R}_\mathrm{exp} \|}\frac{\partial \textbf{R}_{\mathrm{FE}}(\mq)}{\partial \mq}\,.
\end{equation}
The \textit{Hessian} is the matrix of second derivatives of $f(\textbf{q})$, which now becomes (e.g.~see \citet{hansen2013least} for further details)
\begin{equation}\label{eq:hessian}
    \textbf{H}(\textbf{q})=\frac{\partial^2 f(\textbf{q})}{\partial \textbf{q}^2}=\textbf{J}^{\mathrm{T}}\textbf{J}+\bar{\textbf{U}}^{\mathrm{T}}_\mrr\frac{\partial ^2 \bar{\textbf{U}}_\mrr}{\partial \textbf{q}^2}\,.
\end{equation}
{\small \textbf{Remark 3}: One of the benefits of the least-squares form is that the first term in Eq. (\ref{eq:hessian}) can be already computed with the given Jacobian. Moreover, the second term is often neglected due to the residual $\bar{\textbf{U}}_\mrr$ approaching zero near the solution. This approximation is popular in various \textit{trust region} methods, as it allows for an evaluation of the Hessian matrix without extra computation of the second derivative of~$\bar{\textbf{U}}_\mrr$.}\par
In order to calculate the Jacobian in \eqref{eq:jacobianU}, we first need to expand
\begin{equation}\label{eq:dUFEdq}
    \frac{\partial\textbf{U}_{\mathrm{FE}}}{\partial\mq} = \frac{\partial\textbf{U}_{\mathrm{FE}}}{\partial \textbf{u}}\frac{\partial\textbf{u}}{\partial\textbf{q}}\,,
\end{equation}
where $\textbf{u}$ is the stacked vector of all $n_{\mathrm{no}}$ nodal displacements \eqref{eq:globalvector}, and $\partial\mU_\mathrm{FE}/\partial\muu$ is assembled from the $n_\mathrm{el}$ elemental contributions
\begin{equation}\label{eq:dUFE}
\frac{\partial\mU_\mathrm{FE}}{\partial\muu^e}=\begin{bmatrix}
\mN^e\big(\bx_1^{\mathrm{exp}}\big)\\[0.4em]
\mN^e\big(\bx_2^{\mathrm{exp}}\big)\\
\vdots\\[0.2em]
\mN^e\big(\bx_{n_\mathrm{exp}}^{\mathrm{exp}}\big)
\end{bmatrix},\qquad  e = 1, ...\,, n_\mathrm{el}
\end{equation}
that follow directly from \eqref{eq:UFE} and \eqref{eq:uh}. Further, $\partial\muu/\partial\mq$ is required in \eqref{eq:dUFEdq}. For the Dirichlet boundary nodes, $\muu$ is prescribed independently of $\mq$ and hence $\partial\muu/\partial\mq$ is zero. For the free nodes, $\partial\muu/\partial\mq$ follows from the FE element equilibrium equation \eqref{eq:solutionfem}, which depends on $\textbf{q}$ as follows
\begin{equation}\label{eq:balancemu}
\textbf{f}\big(\muu(\mq),\textbf{q}\big)=\textbf{f}_{\mathrm{int}}\big(\muu(\mq),\textbf{q}\big)-\textbf{f}_\mathrm{ext}\big(\muu\big)= \mathbf{0}\ .
\end{equation}
Differentiating Eq.~\eqref{eq:balancemu} w.r.t.~the design variable vector  $\textbf{q}$ then gives
\begin{equation}
\frac{\mrd\textbf{f}}{\mrd\textbf{q}}=\frac{\partial \textbf{f}_{\mathrm{int}}}{\partial \textbf{q}}+\frac{\partial\textbf{f}_{\mathrm{int}}}{\partial \muu}\frac{\partial \muu}{\partial \textbf{q}}-\frac{\partial \textbf{f}_{\mathrm{ext}}}{\partial \textbf{q}}-\frac{\partial\textbf{f}_{\mathrm{ext}}}{\partial \muu}\frac{\partial \muu}{\partial \textbf{q}}=\textbf{0}\,.
\end{equation}
Rewriting this equation gives
\begin{equation}\label{eq:derivativeofu}
\frac{\partial \muu}{\partial \textbf{q}}=-\left(\frac{\partial\textbf{f}_{\mathrm{int}}}{\partial \muu}-\frac{\partial\textbf{f}_{\mathrm{ext}}}{\partial \muu}\right)^{-1}\left(\frac{\partial\textbf{f}_{\mathrm{int}}}{\partial \textbf{q}}-\frac{\partial\textbf{f}_{\mathrm{ext}}}{\partial \textbf{q}}\right).
\end{equation}
Introducing the tangent stiffness matrix
\begin{equation}\label{eq:tangentstiffnessmatrix}
\textbf{K}=\frac{\partial\textbf{f}_{\mathrm{int}}}{\partial \muu}-\frac{\partial\textbf{f}_{\mathrm{ext}}}{\partial \muu}\,,
\end{equation}
and accounting for the fact that the external load vector $\textbf{f}_{\mathrm{ext}}$ does not depend on $\textbf{q}$, Eq.~(\ref{eq:derivativeofu}) can be expressed as
\begin{equation}\label{eq:dudq}
\frac{\partial \muu}{\partial \textbf{q}}=-\textbf{K}^{-1}\frac{\partial\textbf{f}_{\mathrm{int}}}{\partial \textbf{q}}\ ,
\end{equation}
where the $\partial\textbf{f}_{\mathrm{int}}/\partial \textbf{q}$ is the global sensitivity matrix $\textbf{S}$ (analogous to Eq.~\eqref{eq:sensitivitymatrixS}).\\In order to calculate the Jacobian in \eqref{eq:jacobianR}, we note that reaction force $\mR_{\mathrm{FE}}$ directly follows from equilibrium at the Dirichlet nodes, given by
\begin{equation}\label{eq:RFE}
    \mR_\mathrm{FE} = \mf^\mathrm{b}_\mathrm{int}\big(\muu(\mq),\mq\big) - \mf^\mathrm{b}_\mathrm{ext}\big(\muu\big)\,.
\end{equation}
Here superscript ``b" denotes the boundary forces, that are different from the forces of the free nodes given in \eqref{eq:balancemu}. From \eqref{eq:RFE} follows
\begin{equation}
    \frac{\partial\mR_\mathrm{FE}}{\partial\mq}=\frac{\partial \mf^\mathrm{b}_{\mathrm{int}}}{\partial \textbf{q}}+\frac{\partial\textbf{f}^\mathrm{\,b}_{\mathrm{int}}}{\partial \muu}\frac{\partial \muu}{\partial \textbf{q}}-\frac{\partial \textbf{f}^\mathrm{\,b}_{\mathrm{ext}}}{\partial \textbf{q}}-\frac{\partial\textbf{f}^\mathrm{\,b}_{\mathrm{ext}}}{\partial \muu}\frac{\partial \muu}{\partial \textbf{q}}\ .
\end{equation}
Introducing
\begin{equation}
    \mK^\mathrm{b} := \frac{\partial\textbf{f}^\mathrm{\,b}_{\mathrm{int}}}{\partial \muu}-\frac{\partial\textbf{f}^\mathrm{\,b}_{\mathrm{ext}}}{\partial \muu}\,,
\end{equation}
and using \eqref{eq:dudq} and $\partial\mf^\mathrm{b}_{\mathrm{ext}}/\partial \mq=\textbf{0}$, then gives
\begin{equation}
    \frac{\partial\mR_\mathrm{FE}}{\partial\mq} = \mS^\mathrm{b} - \mK^\mathrm{b} \frac{\partial \muu}{\partial \mq}\,.
\end{equation}
where
\begin{equation}
    \textbf{S}^\mathrm{b} := \frac{\partial \textbf{f}^\mathrm{\,b}_\mathrm{int}}{\partial \textbf{q}}  
\end{equation} 
is the sensitivity at the boundary.
With this, all pieces are given to evaluate $\mg$ in \eqref{eq:gradapp} and $\mH$ according to Remark 3.
\begin{small}
\bibliographystyle{apalike}
\bibliography{references}{}
\end{small}
\end{document}